\documentclass[a4paper,11pt]{article}
\pdfoutput=1

\usepackage{jcappub} 

\usepackage[T1]{fontenc}
\usepackage[utf8]{inputenc}

\usepackage{enumitem}
\usepackage{braket}
\usepackage{comment}
\usepackage{amsmath,amssymb,amsfonts,mathrsfs,amsthm}
\usepackage{booktabs}
\usepackage{mathtools}
\usepackage{bm}
\usepackage{dcolumn}
\usepackage{graphicx}   
\usepackage{latexsym}
\usepackage[dvipsnames]{xcolor}
\usepackage{colortbl}
\usepackage{soul}
\usepackage{manfnt,nicefrac}
\usepackage{scalerel, stackengine}
\usepackage{bbm}    
\usepackage{accents} 
\usepackage[makeroom]{cancel}
\usepackage[framemethod=tikz]{mdframed}
\usepackage[most]{tcolorbox}
\usepackage{cleveref}
\usepackage{color}
\usetikzlibrary{arrows}
\usepackage[acronym]{glossaries}

\usepackage[english]{babel}

\definecolor{LBlue}{rgb}{0.5,0.9,0.5}
\definecolor{LCyan}{rgb}{0.6,1,1}
\definecolor{LRed}{rgb}{0.9,0.6,0.7}
\definecolor{LGreen}{rgb}{0.98, 0.93, 0.36}

\newcommand\be{\begin{equation}}
\newcommand\ba{\begin{eqnarray}}
\newcommand\ee{\end{equation}}
\newcommand\ea{\end{eqnarray}}

\newcommand{\scri}{{\cal I}}
\newcommand{\RNum}[1]{\uppercase\expandafter{\romannumeral #1\relax}}

\newcommand{\dd}{\mathrm{d}}

\newcommand{\R}{\mathcal{R}}

\newcommand{\Tc}[1]{T^{(#1)}}
\newcommand{\tildeBox}{\tilde{\raisebox{0pt}[1.25\height]{$\Box$}}}
\newcommand{\orcid}[1]{\href{https://orcid.org/#1}{\includegraphics[width=8pt]{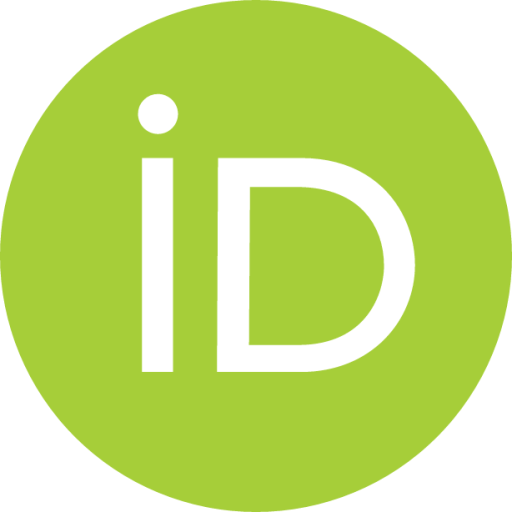}}}

\newcommand*{\rom}[1]{\expandafter\@slowromancap\romannumeral #1@}

\makeatletter
\newcommand{\dalembertian}{\mathop{\mathpalette\dalembertian@\relax}}
\newcommand{\dalembertian@}[2]{%
  \begingroup
  \sbox\z@{$\m@th#1\square$}%
  \dimen0=\fontdimen8
    \ifx#1\displaystyle\textfont\else
    \ifx#1\textstyle\textfont\else
    \ifx#1\scriptstyle\scriptfont\else
    \scriptscriptfont\fi\fi\fi3
  \makebox[\wd\z@]{%
    \hbox to \ht\z@{%
      \vrule width \dimen0
      \kern-\dimen0
      \vbox to \ht\z@{
        \hrule height \dimen0 width \ht\z@
        \vss
        \hrule height 2\dimen0
      }%
      \kern-2.5\dimen0
      \vrule width 2.5\dimen0
    }%
  }%
  \endgroup
}
\makeatother

\makeatletter
\newcommand*{\pgfunderleftarrow}{%
  \@ifstar
    {\let\ifpgf@depth\iftrue\mathpalette\@pgfunderleftarrow}
    {\let\ifpgf@depth\iffalse\mathpalette\@pgfunderleftarrow}%
}
\newcommand*{\@pgfunderleftarrow}[2]{%
  #2%
  \edef\pgf@math@fam{\the\fam}%
  \pgfpicture
    \pgfsetbaseline{0pt}
    \pgf@relevantforpicturesizefalse      
    \pgfsetroundcap                       
    \pgfsetarrowsend{left to}
    \pgfutil@tempdima=0.28pt%
    \advance\pgfutil@tempdima by.8\pgflinewidth%
    \pgfutil@tempdima-4\pgfutil@tempdima
    \sbox\pgfutil@tempboxa{$\m@th\fam\pgf@math@fam#1#2$}%
    \advance\pgfutil@tempdima-\dp\pgfutil@tempboxa
    \pgfutil@tempdimb\wd\pgfutil@tempboxa
    \pgfpathmoveto{\pgfqpoint{0pt}{\pgfutil@tempdima}}%
    \pgfpathlineto{\pgfqpoint{-\pgfutil@tempdimb}{\pgfutil@tempdima}}%
    \pgfusepath{stroke}
    \ifpgf@depth
      \pgf@relevantforpicturesizetrue
      \pgfpathmoveto{\pgfqpoint{0pt}{-\pgfutil@tempdimb}}%
      \pgfusepath{use as bounding box}%
    \fi
  \endpgfpicture
}
\makeatother

\makeatletter
\newcommand*{\pgfunderrightarrow}{%
  \@ifstar
    {\let\ifpgf@depth\iftrue\mathpalette\@pgfunderrightarrow}
    {\let\ifpgf@depth\iffalse\mathpalette\@pgfunderrightarrow}%
}
\newcommand*{\@pgfunderrightarrow}[2]{%
  #2%
  \edef\pgf@math@fam{\the\fam}%
  \pgfpicture
    \pgfsetbaseline{0pt}
    \pgf@relevantforpicturesizefalse      
    \pgfsetroundcap                       
    \pgfsetarrowsend{right to}
    \pgfutil@tempdima=0.28pt%
    \advance\pgfutil@tempdima by.8\pgflinewidth%
    \pgfutil@tempdima-4\pgfutil@tempdima
    \sbox\pgfutil@tempboxa{$\m@th\fam\pgf@math@fam#1#2$}%
    \advance\pgfutil@tempdima-\dp\pgfutil@tempboxa
    \pgfutil@tempdimb\wd\pgfutil@tempboxa
    \pgfpathmoveto{\pgfqpoint{-\pgfutil@tempdimb}{\pgfutil@tempdima}}%
    \pgfpathlineto{\pgfqpoint{0pt}{\pgfutil@tempdima}}%
    \pgfusepath{stroke}
    \ifpgf@depth
      \pgf@relevantforpicturesizetrue
      \pgfpathmoveto{\pgfqpoint{0pt}{-\pgfutil@tempdimb}}%
      \pgfusepath{use as bounding box}%
    \fi
  \endpgfpicture
}
\makeatother

\newacronym{gw}{GW}{Gravitational Wave}

\def\scri{\mathscr{I}}
\def\scrip{\scri^{+}}

\newacronym{gr}{GR}{general relativity}

\title{Asymptotic flatness beyond General Relativity}

\author[a]{David Maibach\orcid{0000-0002-5294-464X}}
\affiliation[a]{Institute for Theoretical Physics, University of Heidelberg,\\Philosophenweg 16, D-69120	Heidelberg, Germany}

\emailAdd{d.maibach@thphys.uni-heidelberg.de}

\abstract{The asymptotic symmetry group of asymptotically flat spacetimes gives rise to balance flux equations that constrain, fully non-perturbatively, the asymptotic strain measured by gravitational-wave detectors. Such constraints are sharp tools for identifying features such as the memory effect. As detector sensitivities improve, it becomes imperative to place the most promising beyond-GR candidates on the same footing. Whether the asymptotically flat framework applies to such theories at all is far from obvious and requires careful analysis of the additional degrees of freedom reaching future null infinity. In this work, we adress this question. We integrate the Bondi-Sachs hierarchy in the presence of an arbitrary stress-energy tensor and extract the falloff conditions its components must satisfy for the standard metric decay to close. We then feed the most general scalar-vector-tensor (SVT) theory with second-order equations of motion through this framework. Recasting the field equations in the effective Einstein form $G_{\mu\nu}=T^{\rm eff}_{\mu\nu}/G_4(\Phi,X)$ and evaluating every operator in the SVT Lagrangian against the falloff table, we condense the outcome into a constraint table for the coupling functionals and their derivatives at the asymptotic vacuum, sharpened by the scalar and vector equations of motion and vacuum stability. Remarkably few conditions survive. The scalar potential must vanish to cubic order at the asymptotic vacuum, the asymptotic Newton constant must be finite and positive, the scalar and vector modes must be canonically normalized, and the conformally coupled sector carries a frame subtlety. Every other coupling functional is protected by the structure of the theory itself. The constraint table thus provides a diagnostic for screening beyond-GR models against asymptotic flatness and establishes the BMS group as the asymptotic symmetry group across the entire admissible SVT class.}

\begin{document}
\maketitle
\flushbottom


\newcommand{\myhyperref}[1]{\hyperref[#1]{\ref{#1}}}




\section{Introduction}
\label{sec:intro}

The dawn of gravitational-wave astronomy has, for the first time, opened the strong-field, radiative regime of gravity to direct observational scrutiny. Almost a decade after the first direct detection of a binary black-hole coalescence \cite{PhysRevLett.116.061102}, the LIGO--Virgo--KAGRA network has compiled a catalog of $\sim$$10^2$ compact-binary mergers, while the next generation of ground-based instruments like the Einstein Telescope \cite{maggiore_science_2020} or Cosmic Explorer \cite{evans_cosmic_2023} and space-based observatories like LISA \cite{colpi_lisa_2024} are poised to extend this catalog by orders of magnitude and to push the sensitivity into regimes in which subtle radiative features become directly accessible. Among the many opportunities this represents, the prospect of testing General Relativity (GR) against viable deviations at the level of the gravitational waveform stands out as one of the most far-reaching. This concerns in particular modifications of GR, whether motivated by quantum-gravity considerations \cite{Cardoso:2016oxy,Cardoso:2017cqb,Cardoso:2017cfl,Deppe:2024fdo} or by dark-energy and dark-matter phenomenology \cite{Clifton:2011jh,Joyce:2014kja,CANTATA:2021ktz}, whose signatures would be smoking guns for many theoretical models of gravity and black holes.

For the full discriminative power of future measurements to truly unfold, it is imperative to have a precise description of \textit{all} waveform features inherent to a given theoretical model. For GR this requires a fully non-linear treatment, so that no non-linear (or other) features are inadvertently discarded. Famously, this has given rise, for instance, to the gravitational-wave (GW) memory \cite{christodoulouNonlinearNatureGravitation1991a, Strominger:2014pwa,Bieri:2015yia,Heisenberg:2023prj} and the asymptotic Bondi--Metzner--Sachs (BMS) symmetry group of null infinity \cite{Bondi:1962px,sachsGravitationalWavesGeneral1962,M_dler_2016}. These structures have been mostly explored in what we will refer to in this work as the asymptotic spacetime formalism: a spacetime model that stays agnostic with respect to the bulk dynamics but asymptotically settles to the Bondi--Sachs metric and eventually, at future null infinity, to flat Minkowski space \cite{M_dler_2016,DAmbrosio:2022clk,Maibach:2025iku}. It has since been applied as a tool for testing and assessing waveform accuracy \cite{Ashtekar:2019viz,Mitman:2020bjf,Borchers:2021vyw,Borchers_2023,DAmbrosio:2024zok} and studying the gravitational-wave memory \cite{Khera:2020mcz,Inchauspe:2024ibs,Cogez:2026frh,Zosso:2026czc}. In anticipation of new detector data, comparing features like the memory in beyond-GR theories with their GR counterparts on an equal footing therefore requires the same machinery in both settings. This includes in particular the asymptotic structure at $\scrip$, the Bondi--Sachs hierarchy, and the BMS group which can be converted into explicit constraint equations for instance through the Wald--Zoupas covariant phase space approach \cite{waldGeneralDefinitionConserved2000}. Before applying any phase space formalism, however, one has to answer the fundamental question: does the theory even admit an asymptotically flat solution at null infinity in the first place? While in the literature different beyond-GR models have been subject to the asymptotic spacetime formalism on a case-by-case basis \cite{tahura_brans-dicke_2021,hou_conserved_2021,Hou:2021bxz,hou_gravitational_2021,Tahura:2025ebb,Maibach:2026wpz}, a general analysis without restrictions from a concrete physical model \cite{Flanagan_2017} remains absent so far.

Applying the asymptotic spacetime formalism to a generic alternative theory of gravity is not automatic. In contrast to GR, where the vacuum Bondi--Sachs setup is canonical and the full structure of $\scrip$ has been worked out in various cases ( \cite{Bondi:1962px,sachsGravitationalWavesGeneral1962,Penrose:1964wq} as well as \cite{Barnich:2009se,Barnich:2010eb}), generic beyond-GR theories carry a host of free coupling functionals in their Lagrangians, i.e., potentials, curvature--scalar mixings, parity-odd Chern--Simons structures, double-dual Riemann couplings \cite{Horndeski:1974wa,Nicolis:2008in,Heisenberg:2014rta,Heisenberg:2018acv,Heisenberg:2023prj}, and may admit a much broader class of asymptotic behaviors, including non-trivial scalar hair \cite{Silva:2017uqg,Doneva:2021tvn,Doneva:2022ewd}, non-vanishing backgrounds \cite{Heisenberg:2018mxx,Heisenberg:2018acv}, modified peeling \cite{Kehrberger_2021} or simply slower falloffs that obstruct the standard BMS construction. The first task is therefore to identify which subset of the parameter space of beyond-GR theories admits an asymptotic structure compatible with the standard BMS arena. In other words, which choices of the coupling functionals render the spacetime ``asymptotically flat'' in the Bondi sense, and consequently admit the BMS group as the asymptotic isometry group.

A particularly powerful organizing framework for these questions is the scalar--vector--tensor (SVT) classification pioneered in \cite{Heisenberg:2018acv} and studied in the context of memory effects in \cite{Heisenberg:2023prj}. It is the most general action coupling a metric, a $U(1)$-gauge vector and a scalar that produces second-order equations of motion, derived through a decoupling limit of generalized Proca with massless gauge-invariance imposed via Stueckelberg fields. The SVT class encompasses Horndeski scalar--tensor gravity \cite{Horndeski:1974wa}, Galileon and Galileon-induced cosmologies \cite{Nicolis:2008in}, generalized Proca and beyond-generalized-Proca vector--tensor theories \cite{Heisenberg:2014rta}, double-dual Riemann gravity, Brans--Dicke (BD) theory in its Stueckelberg formulation, and various scalar--Gauss--Bonnet extensions as particular cases. As such, identifying the asymptotic-flatness boundary in the SVT parameter space outlined in \cite{Heisenberg:2018acv} simultaneously addresses a large portion of the beyond-GR phenomenological landscape.

The present work carries out an asymptotic analysis for the SVT family of theories classified in \cite{Heisenberg:2018acv}, i.e., we derive constraints on the theory's coupling functionals that guarantee the validity of the asymptotically flat metric parametrization tending towards Minkowski spacetime at null infinity. We proceed in two stages. First, in Sec.~\ref{sec:matter}, we derive the Bondi--Sachs hierarchy keeping a generic stress-energy tensor explicitly and read off the minimal falloff conditions on each component of $T_{ab}$ compatible with the asymptotically flat metric falloffs. We illustrate the framework on BD theory (Sec.~\ref{sec:matter-bd}), the simplest genuinely beyond-GR source. The complementary verifications on a free massless scalar and on Einstein--Maxwell theory are collected in App.~\ref{app:matter-scalar} and App.~\ref{app:matter-maxwell}. Second, in Sec.~\ref{sec:flatness_SVT} we recast the SVT field equations into the effective Einstein form $G_{\mu\nu}=T^{\rm eff}_{\mu\nu}/G_4(\Phi,X)$, compute the leading-order contributions of each operator in $L_2+L_3+L_4+L_5$ to $T^{\rm eff}_{\mu\nu}$, and demand that the resulting effective stress-energy satisfy the falloff table. This results in consistency conditions on the SVT coupling functionals $G_i$ at the asymptotic vacuum, which we then sharpen by demanding vacuum stability and well-defined scalar and vector equations of motion. The explicit per-operator computations behind these results are deferred to App.~\ref{app:svt-explicit}. We comment on the implications of the derived conditions for the asymptotic Killing vectors of SVT gravity in Sec.~\ref{sec:Killing_SVT}.




\section{Bondi--Sachs solution with a general stress-energy tensor}
\label{sec:matter}

The standard hierarchical integration of the Bondi--Sachs equations is carried out for vacuum, $T_{ab}=0$ (see \cite{M_dler_2016} for a detailed review), by inserting an asymptotic $1/r$ expansion of the metric functions and solving the Einstein equations and Bianchi identities order by order in $1/r$. In this section we repeat that integration while keeping a general stress-energy tensor $T_{ab}$. In the literature, similar computations can be found for particular types of matter, e.g.\ \cite{Flanagan_2017,hou_conserved_2021,tahura_brans-dicke_2021,hou_gravitational_2021,Hou:2021bxz,Tahura:2025ebb,Zosso:2024xgy}, but none of them stays completely agnostic. Note also that we carry out the derivation deliberately in a step-by-step manner, establishing a blueprint for future explorations of asymptotic flatness in more exotic theories. We assume only that each metric function and each covariant component of $T_{ab}$ admits an asymptotic expansion in powers of $1/r$ with coefficients that depend on $(u,x^A)$, and we leave those coefficients unrestricted, apart from the falloff conditions that are forced by the absence of logarithmic terms and by asymptotic flatness.\footnote{See \cite{Kehrberger_2021} and the discussion below.}

\subsection{Field equations and asymptotic expansion}
\label{sec:matter-setup}
We set the stage by introducing the Bondi--Sachs metric for asymptotically flat spacetimes where flatness refers to flatness at null infinity, $\scrip$. In the Bondi gauge, it reads \cite{M_dler_2016}
\be
\label{eq:bs-metric}
g_{ab}\dd x^a\dd x^b = -\frac{V}{r}e^{2\beta}\dd u^2 - 2e^{2\beta}\dd u\,\dd r + r^2 h_{AB}\big(\dd x^A - U^A\dd u\big)\big(\dd x^B - U^B\dd u\big),
\ee
with $\det[h_{AB}]=\det[q_{AB}]= q$, where $q_{AB}$ is the unit-sphere metric.\footnote{We note that the determinant condition is a choice, not a necessary requirement. In the literature, different choices have produced a large body of work on extensions of the BMS group simply by placing less restrictive constraints on the determinant \cite{Flanagan_2017}. The choice used here however establishes the most common gauge.} We denote the Einstein equations as
\be
\label{eq:einstein}
E_{ab}:=R_{ab}-\tfrac12 g_{ab}R^{c}{}_{c}-8\pi T_{ab}=0 ,
\ee
and we do not assume the divergence-free condition $\nabla_b T^{b}{}_{a}=0$ a priori. Its content is recovered below through the supplementary conditions.

The main equations split into three hypersurface equations, which determine $\beta$, $U^A$ and $V$ on a null cone $u=\text{const}$, and one evolution equation, which fixes $\partial_u h_{AB}$. Retaining the matter terms, the hypersurface equations are the $rr$, $rA$, and angular-trace projections of the Einstein equation \eqref{eq:einstein}. The $rr$ component, $E_{rr}=0$, gives
\be
\label{eq:hyp-beta}
\partial_r\beta = \frac{r}{16}\,h^{AC}h^{BD}(\partial_r h_{AB})(\partial_r h_{CD}) + 2\pi r\,T_{rr},
\ee
the $rA$ component, $E_{rA}=0$, gives
\be
\label{eq:hyp-U}
\partial_r\!\Big[r^4 e^{-2\beta}h_{AB}(\partial_r U^B)\Big] = 2r^4\partial_r\!\Big(\frac{1}{r^2}D_A\beta\Big) - r^2 h^{EF}D_E(\partial_r h_{AF}) + 16\pi r^2\,T_{rA},
\ee
and the angular trace, $h^{AB}E_{AB}=0$, gives
\ba
\label{eq:hyp-V}
2e^{-2\beta}(\partial_r V) &=& \R - 2h^{AB}\big[D_A D_B\beta + (D_A\beta)(D_B\beta)\big] + \frac{e^{-2\beta}}{r^2}D_A\big[\partial_r(r^4 U^A)\big] \nonumber\\
&& -\,\frac12 r^4 e^{-4\beta}h_{AB}(\partial_r U^A)(\partial_r U^B) + 8\pi\big[h^{AB}T_{AB} - r^2 T^{a}{}_{a}\big],
\ea
where $D_A$ is the covariant derivative of the conformal $2$-metric $h_{AB}$, $D^A=h^{AB}D_B$, $\R$ its Ricci scalar, and $\eth_A$ (with $\eth^A=q^{AB}\eth_B$ and $\eth^2=\eth_A\eth^A$) the covariant derivative of the unit-sphere metric $q_{AB}$. The two connections differ by the tensor
\be
\label{eq:connection}
D_A V^B = \eth_A V^B + \mathcal C^{B}{}_{AE}\,V^E ,\qquad \mathcal C^{C}{}_{AB} = \tfrac12 h^{CD}\big(\eth_A h_{DB}+\eth_B h_{DA} -\eth_D h_{AB}\big),
\ee
acting analogously on tensors of any rank. The evolution equation is the projection $m^A m^B E_{AB}=0$ onto the complex polarization dyad $m^A$ ($m^A m^B q_{AB}=0$),
\ba
\label{eq:evol}
m^A m^B\bigg\{ &&\!\!\, r\,\partial_r\!\big[r(\partial_u h_{AB})\big] - \tfrac12\partial_r\!\big[rV(\partial_r h_{AB})\big] - 2e^{\beta}D_A D_B e^{\beta} + h_{CA}D_B\big[\partial_r(r^2 U^C)\big] \nonumber\\
&&\!\! -\,\tfrac12 r^4 e^{-2\beta}h_{AC}h_{BD}(\partial_r U^C)(\partial_r U^D) + \tfrac{r^2}{2}(\partial_r h_{AB})(D_C U^C) + r^2 U^C D_C(\partial_r h_{AB}) \nonumber\\
&&\!\! -\,r^2(\partial_r h_{AC})h_{BE}(D_C U^E - D^E U^C) - 8\pi e^{2\beta}T_{AB} \bigg\} = 0 .
\ea
Finally, the supplementary conditions arise from the $a=r$ components of the contracted Bianchi identity. Using $E^{a}{}_{b}=R^{a}{}_{b}-\tfrac12\delta^{a}_{b}R^{c}{}_{c}-8\pi T^{a}{}_{b}$ they take the form
\be
\label{eq:supp}
\partial_r\big(r^2 e^{2\beta}E^{r}{}_{u}\big)=0 , \qquad \partial_r\big(r^2 e^{2\beta}E^{r}{}_{A}\big)=0 ,
\ee
so that $E^{r}{}_{u}$ and $E^{r}{}_{A}$ are satisfied everywhere once they hold at one value of $r$. In particular, this includes $\scrip$.

To solve the above equations, we need an ansatz for the a priori free metric functions $\beta,V,U^A,h_{AB}$. The most common choice in the literature is an asymptotic expansion in $1/r$, even though this technically invokes a regularity condition that has recently received some pushback from the mathematical community \cite{Kehrberger_2021}. Nevertheless, here we stick to this approach and utilize the following expansion. For the conformal $2$-metric we use, as in the vacuum case,
\be
\label{eq:h-exp}
h_{AB} = q_{AB} + \frac{c_{AB}}{r} + \frac{d_{AB}}{r^2} + \mathcal O(r^{-3}), \qquad h^{AB} = q^{AB} - \frac{c^{AB}}{r} - \frac{d^{AB}-q^{AC}c^{BD}c_{CD}}{r^2} + \mathcal O(r^{-3}),
\ee
with indices on $c_{AB},d_{AB}$ raised with $q^{AB}$. The determinant condition $\det[h_{AB}]= q$ then implies
\be
\label{eq:det-cond}
q^{AB}c_{AB}=0 ,\qquad q^{AB}d_{AB}=\tfrac12 c^{AB}c_{AB} ,\qquad q^{AB}\partial_u c_{AB}=0 ,
\ee
so that $c_{AB}$ is trace-free and only the trace-free part
\be
\label{eq:bAB-def}
b_{AB}:=d_{AB}-\tfrac12 q_{AB}\,q^{CD}d_{CD}
\ee
of $d_{AB}$ carries independent data. The radiative data of the metric are encoded in $c_{AB}$, with news tensor $N_{AB}=\tfrac12\partial_u c_{AB}$. With Eq.~\eqref{eq:h-exp} the connection difference Eq.~\eqref{eq:connection} expands as
\be
\label{eq:connection-exp}
\mathcal C^{C}{}_{AB} = \frac{1}{2r}\,q^{CD}\big(\eth_A c_{DB}+\eth_B c_{DA} -\eth_D c_{AB}\big) + \mathcal O(r^{-2}),
\ee
which controls the difference between $D_A$ and $\eth_A$ in the asymptotic expansion. Its trace vanishes identically, $\mathcal C^{A}{}_{AB} =\tfrac12 h^{AD}\eth_B h_{DA}=\tfrac12\,\eth_B\ln(\det h_{AB})=0$, because the determinant condition forces $\det h_{AB}=\det q_{AB}$. Consequently the divergence of a vector is connection-independent, $D_A W^A=\eth_A W^A$, a fact we use repeatedly below.

A remark on the underlying regularity assumption: as mentioned above, for generic Cauchy or characteristic data the gravitational field is expected to admit a polyhomogeneous expansion, in which logarithmic terms appear and the classical peeling property is weakened~\cite{Kehrberger_2021}. While we stress that our restriction to a pure $1/r$ expansion enforces the absence of logarithms in the metric functions' expansion, the analysis below shows that such terms are naturally generated by the Einstein equations once a general stress-energy expansion is admitted. Leaving the behavior of the latter fully undetermined would prevent any qualitative statement about its influence on the metric functions and is therefore not instructive in the context of this work. We emphasize, however, that the admissible $r$-dependence of the metric functions reacts very sensitively to changes in the behavior of the stress-energy tensor at large radii. We will not extend the analysis in this direction in this work.

For the matter we likewise expand every covariant component,
\be
\label{eq:T-exp}
T_{ab}(u,r,x^C)=\sum_{n}\frac{\Tc{n}_{ab}(u,x^C)}{r^{n}},
\ee
with unrestricted coefficients $\Tc{n}_{ab}$. The lowest powers that may appear are fixed by demanding that the asymptotic structure of the vacuum solution be preserved, i.e.\ that no logarithmic terms enter the expansion, and that the Bondi gauge conditions
\begin{align}
\label{eq:Bondi_gauge_cond}
\lim_{r\to\infty}\beta=\lim_{r\to\infty}U^A=0, && \lim_{r\to\infty}V/r=1, && \lim_{r\to\infty}h_{AB}=q_{AB}
\end{align}
continue to hold. As shown below these requirements enforce
\be
\label{eq:T-falloff}
\begin{aligned}
&T_{rr}=\mathcal O(r^{-3}),\quad T_{rA}=\mathcal O(r^{-2}),\quad T_{ur}=\mathcal O(r^{-3}),\\
&T_{uu}=\mathcal O(r^{-2}),\quad T_{uA}=\mathcal O(r^{-2}),\quad T_{AB}=\mathcal O(r^{-1}),
\end{aligned}
\ee
which are the slowest falloffs compatible with a smooth $1/r$ (non-polyhomogeneous) expansion and with asymptotic flatness. We emphasize that particular matter models may decay faster, where the rate is set by the falloff of the matter fields themselves as shown explicitly in App.~\ref{app:matter-scalar} and App.~\ref{app:matter-maxwell}. In what follows we keep the leading coefficients consistent with Eq.~\eqref{eq:T-falloff} and indicate at each step which component and which order is responsible for a given correction. Note that to properly determine the functional dependence of the metric functions on the matter content, one would compute the stress-energy tensor with the metric \eqref{eq:bs-metric}, insert it into the corresponding Einstein equation, and solve the resulting differential equation for each of the functions $\beta,U^A,V$ as well as for the expansion coefficients of $h_{AB}$. We will do so for BD theory in Sec.~\ref{sec:matter-bd} and, for a free scalar and Einstein--Maxwell, in App.~\ref{app:matter-scalar} and App.~\ref{app:matter-maxwell}. For the moment, however, we treat the matter tensor agnostically in order to carry out the computations in full generality.

\subsection{The $\beta$ hypersurface equation}
\label{sec:matter-beta}

Inserting Eq.~\eqref{eq:h-exp} into the geometric term of Eq.~\eqref{eq:hyp-beta} and using $\partial_r h_{AB}$$=-c_{AB}/r^2+\mathcal O(r^{-3})$ together with $h^{AC}h^{BD}(\partial_r h_{AB})\cdot$ $(\partial_r h_{CD})=c^{AB}c_{AB}/r^4+\mathcal O(r^{-5})$ gives
\be
\label{eq:beta-rhs}
\partial_r\beta = \frac{c^{AB}c_{AB}}{16\,r^3} + 2\pi r\,T_{rr} + \mathcal O(r^{-4}).
\ee
The matter source $2\pi r\,T_{rr}$ controls the leading falloff of $\beta$. Asymptotic flatness only requires $\lim_{r\to\infty}\beta=0$, so consistency of the ansatz requires
\be
\label{eq:beta-cond}
\Tc{2}_{rr}=0 ,
\ee
i.e.\ $T_{rr}=\mathcal O(r^{-3})$, while $\Tc{3}_{rr}$ remains unrestricted. With $T_{rr}=\Tc{3}_{rr}/r^3+\Tc{4}_{rr}/r^4+\mathcal O(r^{-5})$ the radial integration of Eq.~\eqref{eq:beta-rhs} subject to $\lim_{r\to\infty}\beta=0$ yields
\be
\label{eq:beta-sol}
\beta(u,r,x^A) = -\frac{2\pi\,\Tc{3}_{rr}}{r} - \frac{1}{r^2}\bigg(\frac{c^{AB}c_{AB}}{32}+\pi\,\Tc{4}_{rr}\bigg)+\mathcal O(r^{-3}) \;=:\; \frac{\beta_{(1)}}{r}+\frac{\beta_{(2)}}{r^2}+\mathcal O(r^{-3}).
\ee
A non-vanishing radial stress $\Tc{3}_{rr}$ therefore produces a genuine $\mathcal O(r^{-1})$ term in $\beta$, with $\beta_{(1)}=-2\pi\,\Tc{3}_{rr}$, while $\Tc{4}_{rr}$ shifts the $\mathcal O(r^{-2})$ coefficient $\beta_{(2)}=-\big(\tfrac{1}{32}c^{AB}c_{AB}+\pi\Tc{4}_{rr}\big)$. In vacuum $\beta_{(1)}=0$ and Eq.~\eqref{eq:beta-sol} reduces to the known result $\beta=-c^{AB}c_{AB}/(32\,r^2)+\mathcal O(r^{-3})$ \cite{M_dler_2016}. We keep $\beta_{(1)}$ explicit below.

\subsection{The $U^A$ hypersurface equation}
\label{sec:matter-U}

With the solution for $\beta$ and the ansatz for $h_{AB}$, we integrate Eq.~\eqref{eq:hyp-U} while keeping the full series $T_{rA}=\Tc{0}_{rA}+\Tc{1}_{rA}/r+\Tc{2}_{rA}/r^2+\Tc{3}_{rA}/r^3+\mathcal O(r^{-4})$, so that the matter source is $16\pi r^2 T_{rA}=16\pi\big(\Tc{0}_{rA}r^2+\Tc{1}_{rA}r+\Tc{2}_{rA} +\Tc{3}_{rA}/r\big)+\mathcal O(r^{-2})$. A further $\mathcal O(r^{0})$ contribution comes from the $\mathcal O(r^{-1})$ piece $\beta_{(1)}$ of $\beta$, i.e, the geometric term $2r^4\partial_r(r^{-2}D_A\beta)$, which in vacuum is $\mathcal O(r^{-1})$, now contributes $-6\,\eth_A\beta_{(1)}=12\pi\,\eth_A\Tc{3}_{rr}$ already at $\mathcal O(r^{0})$. Writing $F_A:=r^4 e^{-2\beta}h_{AB}\partial_r U^B$, the hypersurface equation reads
\be
\label{eq:U-rhs}
\partial_r F_A = 16\pi\Tc{0}_{rA}\,r^2 + 16\pi\Tc{1}_{rA}\,r + \big(\eth^E c_{AE}+16\pi\,\Tc{2}_{rA}+12\pi\,\eth_A\Tc{3}_{rr}\big) + \frac{1}{r}\big(S_A+16\pi\,\Tc{3}_{rA}\big) + \mathcal O(r^{-2}),
\ee
where $\eth_A$ is the covariant derivative on the unit sphere and
\be
\label{eq:SA-def}
S_A := \eth^B\!\big(2d_{AB}-q^{FG}c_{BG}c_{AF}\big) = 2\,\eth^B b_{AB}.
\ee
The second equality rests on two facts. First, $c_{AB}$ is symmetric and trace-free, and in two dimensions any such tensor obeys the algebraic identity
\be
\label{eq:2d-identity}
q^{FG}c_{BG}c_{AF}=c_{AF}\,c_{B}{}^{F}=\tfrac12\,q_{AB}\,c^{FG}c_{FG},
\ee
as is readily checked in an orthonormal frame. Second, the determinant condition Eq.~\eqref{eq:det-cond} gives $q^{CD}d_{CD}=\tfrac12 c^{FG}c_{FG}$, so that $\tfrac12 q_{AB}c^{FG}c_{FG}=q_{AB}\,q^{CD}d_{CD}$. Combining the two,
\be
\label{eq:2bAB}
2d_{AB}-q^{FG}c_{BG}c_{AF} = 2d_{AB}-q_{AB}\,q^{CD}d_{CD} = 2\Big(d_{AB}-\tfrac12 q_{AB}\,q^{CD}d_{CD}\Big) = 2\,b_{AB},
\ee
with $b_{AB}$ the trace-free part of $d_{AB}$, Eq.~\eqref{eq:bAB-def}. The nonlinear $c^2$ term has canceled against the trace of $d_{AB}$, leaving only $b_{AB}$. Radial integration, with $L_A(u,x^C)$ the integration constant, gives
\be
\label{eq:FA-full}
\begin{aligned}
F_A ={}& \tfrac{16\pi}{3}\Tc{0}_{rA}\,r^3 + 8\pi\Tc{1}_{rA}\,r^2 + \big(\eth^E c_{AE}+16\pi\,\Tc{2}_{rA}+12\pi\,\eth_A\Tc{3}_{rr}\big)\,r\\
&+ \big(S_A+16\pi\,\Tc{3}_{rA}\big)\ln r - 6L_A + \mathcal O(r^{-1}).
\end{aligned}
\ee

We can now read off the consequences of non-trivial $\Tc{0}_{rA}$ and $\Tc{1}_{rA}$ for $U^A$ and discard those incompatible with a Bondi frame. Since $\partial_r U^B=r^{-4}h^{BA}e^{2\beta}F_A$ with $h^{BA}e^{2\beta}=q^{BA}+\mathcal O(r^{-1})$, each leading power of Eq.~\eqref{eq:FA-full} fixes, after one further integration, a term of $U^A$:
\begin{itemize}
\item[(i)] $\Tc{0}_{rA}$ gives $F_A\sim r^3$, hence $\partial_r U^A\sim r^{-1}$ and $U^A\sim\tfrac{16\pi}{3}q^{AB}\Tc{0}_{rB}\,\ln r$. This diverges as $r\to\infty$ and violates the Bondi gauge condition $\lim_{r\to\infty}U^A=0$ outright. Hence $\Tc{0}_{rA}=0$.
\item[(ii)] $\Tc{1}_{rA}$ yields $F_A\sim r^2$, hence $\partial_r U^A\sim r^{-2}$ and $U^A\sim-8\pi q^{AB}\Tc{1}_{rB}/r$. This is finite and vanishes at infinity, yet it produces a metric cross term $g_{uA}=-r^2 h_{AB}U^B=8\pi\,r\,\Tc{1}_{rA}+\mathcal O(1)$ that grows linearly in $r$. The obstruction is seen by measuring this term in an orthonormal frame. Namely, since $g_{AB}=r^2 h_{AB}$, the unit angular legs scale as $e_{\hat A}=r^{-1}\hat e^{(q)}_{\hat A}$, so the dimensionless cross component is $g(\partial_u,e_{\hat A})\sim g_{uA}/r$, to be compared with $g_{uu}=\mathcal O(1)\to-1$. For $g_{uA}=\mathcal O(1)$ this tilt decays as $r^{-1}$ and $\partial_u$ becomes asymptotically orthogonal to the cuts of $\scrip$, as required of an inertial frame. For $g_{uA}=\mathcal O(r)$ it stays $\mathcal O(1)$, so $\partial_u$ retains a finite, non-decaying tilt and the coordinate basis never approaches a Minkowski basis. An asymptotically inertial Bondi frame therefore demands $g_{uA}=\mathcal O(1)$, i.e.\ $U^A=\mathcal O(r^{-2})$ in the integer $1/r$ expansion. As $\Tc{1}_{rA}$ is a genuine stress-tensor component and not pure gauge, it cannot be absorbed by a frame choice, and we conclude $\Tc{1}_{rA}=0$.
\end{itemize}
Thus the field equation, together with the requirement of an asymptotically inertial Bondi frame, derives the falloff $T_{rA}=\mathcal O(r^{-2})$. What remains is the $\ln r$ term in Eq.~\eqref{eq:FA-full} proportional to $S_A+16\pi\Tc{3}_{rA}$. Carried through $\partial_r U^B=r^{-4}h^{BA}e^{2\beta}F_A$ it gives $\partial_r U^A\sim r^{-4}\ln r$ and, after integration,
\be
\label{eq:Ulog}
U^A\;\supset\;-\tfrac13\,q^{AB}\big(S_B+16\pi\Tc{3}_{rB}\big)\, \frac{\ln r}{r^{3}} + \dots
\ee
This term does decay and therefore violates neither boundedness nor $\lim_{r\to\infty}U^A=0$. It is excluded instead by the standing regularity assumption of the formalism imposing that the metric functions are posited to admit a smooth, integer-power $1/r$ expansion, equivalently a smooth conformal completion at $\scrip$. In terms of the conformal factor $\Omega=1/r$ the offending term is $\propto\Omega^{3}\ln\Omega$, whose third $\Omega$-derivative diverges as $\Omega\to0$, rendering the conformal metric only $C^{2}$ at $\scrip$ (instead of $C^\infty$). This spoils the differentiable structure of null infinity and the peeling property of the Weyl tensor that the smooth $1/r$ ansatz is designed to guarantee. We thus demand a logarithm-free expansion and therefore set
\be
\label{eq:bAB-constraint}
S_A+16\pi\Tc{3}_{rA}=0 \quad\Longleftrightarrow\quad \eth^B b_{AB}=-8\pi\,\Tc{3}_{rA}.
\ee
In vacuum this is $S_A=0$, which by the $\eth$-calculus forces the two independent components of $b_{AB}$ to vanish. With matter it instead ties $b_{AB}$ to the $\mathcal O(r^{-3})$ part of $T_{rA}$, so that the trace-free part of the subleading angular metric $d_{AB}$ no longer vanishes. If one relaxes the regularity assumption and admits a polyhomogeneous (logarithmic) expansion, this constraint is dropped one obtains only a partial peeling property at $\scrip$.\footnote{Compare to the results of \cite{Kehrberger_2021}.}

Imposing $\Tc{0}_{rA}=\Tc{1}_{rA}=0$ and the log-removal condition, Eq.~\eqref{eq:FA-full} reduces to
\be
\label{eq:FA-sol}
r^4 e^{-2\beta}h_{AB}\,\partial_r U^B = \big(\eth^E c_{AE}+16\pi\,\Tc{2}_{rA}+12\pi\,\eth_A\Tc{3}_{rr}\big)\,r - 6L_A + \mathcal O(r^{-1}).
\ee
Solving for $\partial_r U^B$ with $h^{AB}e^{2\beta}=q^{AB}+\big(2\beta_{(1)}q^{AB}-c^{AB}\big)/r+\mathcal O(r^{-2})$ and integrating subject to $\lim_{r\to\infty}U^A=0$ yields
\be
\label{eq:U-sol}
\begin{split}
U^A &= -\frac{1}{2r^2}\Big(\eth_B c^{AB}+16\pi\,q^{AB}\Tc{2}_{rB} +12\pi\,\eth^A\Tc{3}_{rr}\Big)\\
&\quad + \frac{1}{r^3}\bigg[\,2L^A + \tfrac13\,c^{AE}\eth^F c_{EF} + \tfrac{16\pi}{3}\,c^{AB}\Tc{2}_{rB}\\
&\qquad\quad + 4\pi\,c^{AB}\eth_B\Tc{3}_{rr} + \tfrac{4\pi}{3}\,\Tc{3}_{rr}\,\eth_E c^{AE}\\
&\qquad\quad + \tfrac{64\pi^2}{3}\,\Tc{3}_{rr}\,q^{AB}\Tc{2}_{rB} + 16\pi^2\,\Tc{3}_{rr}\,\eth^A\Tc{3}_{rr}\,\bigg] + \mathcal O(r^{-4}).
\end{split}
\ee
The matter modifies $U^A$ already at $\mathcal O(r^{-2})$ through the leading radial-angular momentum flux $\Tc{2}_{rA}$ and through the radial stress $\Tc{3}_{rr}$ inherited from $\beta_{(1)}$. At $\mathcal O(r^{-3})$ it enters both linearly, via $\Tc{2}_{rA}$ and $\Tc{3}_{rr}$ contracted with the shear $c_{AB}$, i.e.\ the terms $\tfrac{16\pi}{3}c^{AB}\Tc{2}_{rB}$, $4\pi c^{AB}\eth_B\Tc{3}_{rr}$ and $\tfrac{4\pi}{3}\Tc{3}_{rr}\eth_E c^{AE}$, and quadratically, via the products $\Tc{3}_{rr}\Tc{2}_{rA}$ and $\Tc{3}_{rr}\eth_A\Tc{3}_{rr}$ generated by the $2\beta_{(1)}q^{AB}/r$ piece of $h^{AB}e^{2\beta}$. Eq. \eqref{eq:U-sol} is to be compared against the vacuum solution $U^A=-\eth_B c^{AB}/(2r^2)+r^{-3}\big(2L^A+\tfrac13 c^{AE}\eth^F c_{EF}\big) +\mathcal O(r^{-4})$ \cite{M_dler_2016}.

\subsection{The $V$ hypersurface equation and the mass aspect}
\label{sec:matter-V}

In Eq.~\eqref{eq:hyp-V} every term except the matter source already has its radial order fixed by the solutions of Secs.~\ref{sec:matter-beta} and \ref{sec:matter-U}. Using $\beta=\mathcal O(r^{-1})$, $U^A=\mathcal O(r^{-2})$ and $h_{AB}=q_{AB}+\mathcal O(r^{-1})$ (so that $r^4 U^A=\mathcal O(r^2)$ and $\partial_r U^A=\mathcal O(r^{-3})$), the geometric pieces are
\be
\label{eq:V-orders}
\begin{aligned}
&2e^{-2\beta}\partial_r V = 2\,\partial_r V + \mathcal O(r^{-1}), \qquad \R = \R[q] + \mathcal O(r^{-1}) = 2 + \mathcal O(r^{-1}),\\[2pt]
&2h^{AB}\big[D_A D_B\beta+(D_A\beta)(D_B\beta)\big] = \mathcal O(r^{-1}), \qquad \frac{e^{-2\beta}}{r^2}D_A\big[\partial_r(r^4 U^A)\big] = \mathcal O(r^{-1}),\\[2pt]
&\tfrac12 r^4 e^{-4\beta}h_{AB}(\partial_r U^A)(\partial_r U^B) = \mathcal O(r^{-2}).
\end{aligned}
\ee
Hence the entire geometric content of Eq.~\eqref{eq:hyp-V} reduces to $2\,\partial_r V = 2 + \mathcal O(r^{-1})$ wherever the leading $\mathcal O(r^0)$ balance fixes $V\sim r$, while every geometric correction is $\mathcal O(r^{-1})$ or smaller. The only term whose order is not yet pinned, tracing back to the falloff of $T_{ur}$ not been constrained so far, is the matter source $8\pi[h^{AB}T_{AB}-r^2 T^{a}{}_{a}]$.

Let us carfully analyze this piece now: The matter source of Eq.~\eqref{eq:hyp-V} simplifies considerably once the trace $T^{a}{}_{a}=g^{ab}T_{ab}$ is written out with the contravariant metric components $g^{ur}=-e^{-2\beta}$, $g^{rr}=\tfrac{V}{r}e^{-2\beta}$, $g^{rA}=-U^A e^{-2\beta}$, $g^{AB}=r^{-2}h^{AB}$. The angular contraction $h^{AB}T_{AB}$ then cancels identically so that $T_{AB}$ drops out of the $V$ equation altogether leaving
\be
\label{eq:V-matter}
8\pi\big[h^{AB}T_{AB}-r^2 T^{a}{}_{a}\big] = 8\pi e^{-2\beta}\Big[2r^2 T_{ur} - rV\,T_{rr} + 2r^2 U^A T_{rA}\Big].
\ee
We estimate the three terms using only results already established. By Sec.~\ref{sec:matter-U}, $U^A=\mathcal O(r^{-2})$ and $T_{rA}=\mathcal O(r^{-2})$, so $2r^2 U^A T_{rA}=\mathcal O(r^{-2})$ is subleading. By Sec.~\ref{sec:matter-beta}, $T_{rr}=\mathcal O(r^{-3})$. Hence for any leading power $V=\mathcal O(r^{p})$ the term $rV\,T_{rr}=\mathcal O(r^{p-2})$ lies one order below $\partial_r V=\mathcal O(r^{p-1})$ and can never balance the left-hand side. Note that this term carries the unknown $V$ itself, so Eq.~\eqref{eq:hyp-V} is strictly a linear first-order differential equation,
\be
\label{eq:V-ode}
\partial_r V + 4\pi r\,T_{rr}\,V = Q, \qquad Q := \tfrac12 e^{2\beta}\,(\text{geometric RHS}) + 8\pi r^2\big(T_{ur}+U^A T_{rA}\big),
\ee
rather than a pure quadrature. Because $T_{rr}=\mathcal O(r^{-3})$, the coefficient $4\pi r\,T_{rr}=\mathcal O(r^{-2})$ has integrating factor $\mu=\exp\!\int^{r}\!4\pi r'T_{rr}\,\dd r'=1+\mathcal O(r^{-1})\to1$, so the homogeneous term only shifts the $\mathcal O(r^{-1})$ and higher coefficients of $V$ and leaves both the leading balance and the free integration constant $M$ untouched. The only term able to compete with the leading geometry is thus $2r^2 T_{ur}$, whose falloff we have not yet constrained.

The geometric right-hand side of Eq.~\eqref{eq:hyp-V} tends to $\R\to\R[q]=2$ and the left-hand side to $2e^{-2\beta}\partial_r V\to2\partial_r V$. Writing $V=v_1 r+v_0+\mathcal O(r^{-1})$ and reading off the consequence of each leading $T_{ur}$ coefficient (with $8\pi e^{-2\beta}\cdot 2r^2 T_{ur}$ the relevant source), we find:
\begin{itemize}
\item[(i)] $\Tc{0}_{ur}$ adds $16\pi\Tc{0}_{ur}r^2$ to the source, forcing $\partial_r V\sim 8\pi\Tc{0}_{ur}r^2$ and $V\sim\tfrac{8\pi}{3}\Tc{0}_{ur}r^3$, so $V/r\to\infty$. This violates asymptotic flatness $\lim_{r\to\infty}V/r=1$. Hence $\Tc{0}_{ur}=0$.
\item[(ii)] $\Tc{1}_{ur}$ adds $16\pi\Tc{1}_{ur}r$, forcing $V\sim 4\pi\Tc{1}_{ur}r^2$ and again $V/r\to\infty$. Hence $\Tc{1}_{ur}=0$. \item[(iii)] $\Tc{2}_{ur}$ adds $16\pi\Tc{2}_{ur}$ at $\mathcal O(r^0)$, so the leading balance gives $2v_1=2+16\pi\Tc{2}_{ur}$, i.e.\ $V/r\to v_1=1+8\pi\Tc{2}_{ur}$. Asymptotic flatness $v_1=1$ then forces $\Tc{2}_{ur}=0$.
\end{itemize}
The $V$ hypersurface equation therefore derives
\be
\label{eq:V-cond}
\Tc{0}_{ur}=\Tc{1}_{ur}=\Tc{2}_{ur}=0 \quad\Longleftrightarrow\quad T_{ur}=\mathcal O(r^{-3}).
\ee
No condition on $T_{AB}$ arises, as it canceled in Eq.~\eqref{eq:V-matter}. With $T_{ur}=\Tc{3}_{ur}/r^3+\mathcal O(r^{-4})$ and the $rV\,T_{rr}$ piece of Eq.~\eqref{eq:V-matter} (now $\mathcal O(r^{-1})$, since $V\sim r$), the matter source enters first at $\mathcal O(r^{-1})$,
\be
\label{eq:V-matter-sub}
8\pi\big[h^{AB}T_{AB}-r^2 T^{a}{}_{a}\big] = \frac{8\pi\big(2\Tc{3}_{ur}-\Tc{3}_{rr}\big)}{r} + \mathcal O(r^{-2}).
\ee

A pure power series $V=v_1 r+v_0+v_{-1}/r+\dots$ has $\partial_r V=v_1-v_{-1}/r^2+\dots$ with no $\mathcal O(r^{-1})$ term, so any $\mathcal O(r^{-1})$ contribution to $\partial_r V$ would integrate to a $\ln r$ in $V$, which we exclude on the same smoothness/peeling grounds invoked in Sec.~\ref{sec:matter-U}. To isolate that contribution we solve the $V$ hypersurface equation Eq.~\eqref{eq:hyp-V} for the radial derivative,
\begin{align}
\label{eq:V-drV}
\partial_r V = \tfrac12 e^{2\beta}\bigg\{ &\R - 2h^{AB}\big[D_A D_B\beta+(D_A\beta)(D_B\beta)\big] + \frac{e^{-2\beta}}{r^2}D_A\big[\partial_r(r^4 U^A)\big] \notag\\
&- \tfrac12 r^4 e^{-4\beta}h_{AB}(\partial_r U^A)(\partial_r U^B) + 8\pi\big(h^{AB}T_{AB}-r^2 T^{a}{}_{a}\big)\bigg\},
\end{align}
and expand the brace order by order with the solutions established above. Its geometric terms (those without stress-energy components) contribute at $\mathcal O(r^{-1})$
\be
\label{eq:V-geom-log}
\begin{aligned}
&\R = 2+\frac{\eth_A\eth_B c^{AB}}{r}+\mathcal O(r^{-2}),\qquad -2h^{AB}D_A D_B\beta = -\frac{2\,\eth^2\beta_{(1)}}{r}+\mathcal O(r^{-2}),\\[2pt]
&\frac{e^{-2\beta}}{r^2}D_A\!\big[\partial_r(r^4 U^A)\big] = \frac{2\,\eth_A u^A_{(2)}}{r}+\mathcal O(r^{-2}),
\end{aligned}
\ee
with $\eth^2:=\eth_A\eth^A$ and $u^A_{(2)}=-\tfrac12\big(\eth_B c^{AB}+16\pi q^{AB}\Tc{2}_{rB} +12\pi\eth^A\Tc{3}_{rr}\big)$ is the leading coefficient of $U^A$ from Eq.~\eqref{eq:U-sol}. The two quadratic terms in the brace are $\mathcal O(r^{-2})$ and do not enter at this order. The remaining $\mathcal O(r^{-1})$ contributions are the matter source Eq.~\eqref{eq:V-matter-sub} and, from the $e^{2\beta}$ prefactor, the cross term $e^{2\beta}\R[q]=2+4\beta_{(1)}/r+\mathcal O(r^{-2})$ between the $\mathcal O(r^{-1})$ part of $e^{2\beta}$ and the leading geometric value $\R[q]=2$. Collecting these, the first order in $1/r$ of $\partial_r V$, i.e.\ the coefficient of $r^{-1}$, which we denote $[\partial_r V]_{r^{-1}}$, is then
\be
\label{eq:V-rhs1}
[\partial_r V]_{r^{-1}} = \tfrac12\bigg[ \underbrace{\big(\eth_A\eth_B c^{AB}+2\eth_A u^A_{(2)}\big)} _{=\,-16\pi\eth_AT^A_r{}^{(2)}-12\pi\eth^2\Tc{3}_{rr}} \;-\;2\eth^2\beta_{(1)}+4\beta_{(1)} +8\pi\big(2\Tc{3}_{ur}-\Tc{3}_{rr}\big)\bigg].
\ee
Inserting $\beta_{(1)}=-2\pi\Tc{3}_{rr}$ and demanding $[\partial_r V]_{r^{-1}}=0$ yields the constraint
\be
\label{eq:V-logcond}
2\,\Tc{3}_{ur} = \big(2+\eth^2\big)\Tc{3}_{rr}+2\,\eth_A\Tc{2}_r{}^A, \qquad T^A_r{}^{(2)}:=q^{AB}\Tc{2}_{rB}.
\ee
Like Eq.~\eqref{eq:bAB-constraint} of the $U^A$ sector, Eq.~\eqref{eq:V-logcond} is a constraint that ties the matter coefficients together and forms part of the conservation content carried by the field equations.

With the logarithm thereby removed, integrating Eq.~\eqref{eq:hyp-V}, where the mass aspect $M(u,x^A)$ entering as the integration constant $M:=-\tfrac12\lim_{r\to\infty}(V-r)$, gives
\be
\label{eq:V-sol}
V(u,r,x^A) = r - 2M(u,x^A) + \mathcal O(r^{-1}).
\ee
Thus, $M$ remains free Cauchy data on the initial null cone, exactly as in vacuum, while the matter populates the $\mathcal O(r^{-1})$ and higher coefficients subject to the log-free constraint Eq.~\eqref{eq:V-logcond}. Carrying the integration one order further, the next-to-leading coefficient is obtained from the $\mathcal O(r^{-2})$ part of Eq.~\eqref{eq:V-drV} and, using the log condition $[\partial_r V]_{r^{-1}}=0$, simplifies to
\be
\label{eq:V-1overr}
V = r - 2M + \frac{V_{(1)}}{r} + \mathcal O(r^{-2}), \qquad V_{(1)} = 2\beta_{(1)}^2 - 2\beta_{(2)} - \tfrac12\,\Sigma_{(2)},
\ee
with $\beta_{(1)}=-2\pi\Tc{3}_{rr}$, $\beta_{(2)}=-\tfrac{1}{32}c^{AB}c_{AB} -\pi\Tc{4}_{rr}$ from Eq.~\eqref{eq:beta-sol}, and $\Sigma_{(2)}$ the coefficient of $r^{-2}$ in the right-hand side of Eq.~\eqref{eq:hyp-V}. As discussed above, the $\Sigma_{(1)}$ part is forbidden by the assumption of smoothness of all metric functions at $\scrip$. The right-hand side of Eq.~\eqref{eq:hyp-V} separates into the explicit matter source $8\pi[h^{AB}T_{AB}-r^2 T^{a}{}_{a}]$ and the remaining, purely geometric terms ($\R$, the $\beta$-Hessian $-2h^{AB}[D_AD_B\beta+(D_A\beta)(D_B\beta)]$, the $U^A$-divergence $\tfrac{e^{-2\beta}}{r^2}D_A[\partial_r(r^4U^A)]$ and the $-\tfrac12 r^4e^{-4\beta}h_{AB}(\partial_rU^A)(\partial_rU^B)$ term). Accordingly the $\mathcal O(r^{-2})$ coefficient splits as
\be
\label{eq:Sigma-split}
\Sigma_{(2)} = \Sigma_{(2)}^{\rm(mat)} + \Sigma_{(2)}^{\rm(geom)},
\ee
where $\Sigma_{(2)}^{\rm(mat)}$ is the $\mathcal O(r^{-2})$ coefficient of the matter source and $\Sigma_{(2)}^{\rm(geom)}$ that of the geometric terms. The matter part is given by
\be
\label{eq:Sigma-matter}
\Sigma_{(2)}^{\rm(mat)} = 8\pi\Big[\,2\Tc{4}_{ur}-\Tc{4}_{rr} + 2M\,\Tc{3}_{rr} + 2\,u^A_{(2)}\Tc{2}_{rA} - 2\beta_{(1)}\big(2\Tc{3}_{ur}-\Tc{3}_{rr}\big)\Big],
\ee
where $u^A_{(2)}$ is the leading coefficient of $U^A$ in Eq.~\eqref{eq:U-sol}. The mass--stress coupling $2M\,\Tc{3}_{rr}$ originates from the $-rV\,T_{rr}$ piece of Eq.~\eqref{eq:V-matter} evaluated on $V\simeq r-2M$.

For the geometric part $\Sigma_{(2)}^{\rm(geom)}$, the leading ingredient is the $\mathcal O(r^{-2})$ term of the angular Ricci scalar, which for $h_{AB}=q_{AB}+c_{AB}/r+d_{AB}/r^2$ reads
\be
\label{eq:R2}
\mathcal R_{(2)} = \eth_A\eth_B d^{AB} - \tfrac12\,\eth^2\!\big(c_{CD}c^{CD}\big) + \tfrac12\,c_{CD}c^{CD} + \tfrac14\,\eth_A c_{BC}\,\eth^A c^{BC} - \tfrac12\,\big(\eth_B c^{AB}\big)\big(\eth^C c_{AC}\big),
\ee
where the determinant condition $q^{AB}d_{AB}=\tfrac12 c_{CD}c^{CD}$ has been used and only the first term carries $d_{AB}$. Adding the $\mathcal O(r^{-2})$ parts of the $\beta$-Hessian, the $U^A$-divergence (for which $\det h_{AB}=\det q_{AB}$ gives $D_A W^A=\eth_A W^A$) and the $(\partial_r U)^2$ term gives the closed form
\ba
\label{eq:Sigma-geom}
\Sigma_{(2)}^{\rm(geom)} &=& \mathcal R_{(2)} - 2\,\eth^2\beta_{(2)} + 2\big(\eth_A c^{AB}\big)\eth_B\beta_{(1)} + 2\,c^{AB}\,\eth_A\eth_B\beta_{(1)} - 2\,\eth_A\beta_{(1)}\,\eth^A\beta_{(1)} \nonumber\\
&& +\ \eth_A u^A_{(3)} - 4\beta_{(1)}\,\eth_A u^A_{(2)} - 2\,q_{AB}\,u^A_{(2)} u^B_{(2)} ,
\ea
with $\beta_{(1)},\beta_{(2)}$ from Eq.~\eqref{eq:beta-sol} and $u^A_{(2)},u^A_{(3)}$ the $\mathcal O(r^{-2}),\mathcal O(r^{-3})$ coefficients of $U^A$ in Eq.~\eqref{eq:U-sol}. In vacuum ($\beta_{(1)}=0$, $\beta_{(2)}=-\tfrac1{32} c_{CD}c^{CD}$, $u^A_{(2)}=-\tfrac12\eth_B c^{AB}$, $u^A_{(3)}=2L^A+\tfrac13 c^{AE}\eth^F c_{EF}$) it collapses to the standard vacuum coefficient.

\subsection{The evolution equation}
\label{sec:matter-evol}

In the evolution equation \eqref{eq:evol} the radial orders are set by the solutions found above, namely $\beta=\mathcal O(r^{-1})$, $U^A=\mathcal O(r^{-2})$, $V=r-2M+\mathcal O(r^{-1})$ and $T_{AB}=\Tc{1}_{AB}/r+\mathcal O(r^{-2})$. We find three terms of $\mathcal O(r^{-1})$, the time-derivative term $r\partial_r(r\,\partial_u h_{AB})=-\partial_u d_{AB}/r+\mathcal O(r^{-2})$, the matter term $-8\pi e^{2\beta}T_{AB}=-8\pi\Tc{1}_{AB}/r+\mathcal O(r^{-2})$, and, crucially, because $\beta$ now decays as $r^{-1}$ rather than the vacuum $r^{-2}$, the curvature term $-2e^{\beta}D_AD_B e^{\beta}=-2\,\eth_A\eth_B\beta_{(1)}/r+\mathcal O(r^{-2})$, which in vacuum would be subleading. All other terms are of $\mathcal O(r^{-2})$. After contraction with the dyad $m^A m^B$, we have
\be
\label{eq:evol-lead}
\begin{aligned}
&m^A m^B\Big[\,r\partial_r\big(r\,\partial_u h_{AB}\big) - 2e^{\beta}D_AD_B e^{\beta} - 8\pi e^{2\beta}T_{AB}\,\Big]\\
&\qquad = -\frac{m^A m^B}{r}\Big[\,\partial_u d_{AB} + 2\,\eth_A\eth_B\beta_{(1)} + 8\pi\,\Tc{1}_{AB}\,\Big] + \mathcal O(r^{-2}).
\end{aligned}
\ee
The news $N_{AB}=\tfrac12\partial_u c_{AB}$ enters Eq.~\eqref{eq:evol} only through the $r$-independent piece of $r\,\partial_u h_{AB}=\partial_u c_{AB}+\mathcal O(r^{-1})$, which the outer radial derivative in $r\partial_r(r\,\partial_u h_{AB})$ annihilates. No other term of Eq.~\eqref{eq:evol} is $\mathcal O(r^{0})$. The evolution equation therefore places no condition on $\partial_u c_{AB}$ and the news is freely specifiable characteristic data. Eq.~\eqref{eq:evol} instead fixes the retarded-time derivative of the next expansion coefficient.

We nevertheless continue with the $1/r$ analysis. The contraction with $m^A m^B$ isolates the trace-free sector. Because $m^A$ is null, $m^A m_A=0$, the unit-sphere metric $q_{AB}=2\,m_{(A}\bar m_{B)}$ (normalization $m_A\bar m^A=1$) satisfies $m^A m^B q_{AB}=0$. Hence $m^A m^B$ annihilates the pure-trace part $\tfrac12 q_{AB}q^{CD}S_{CD}$ of any symmetric $S_{AB}$ and returns only the spin-weight-$2$ component of its trace-free part. The single complex equation $m^A m^B(\cdots)=0$, together with its conjugate $\bar m^A\bar m^B(\cdots)=0$, is therefore equivalent to the vanishing of the full real trace-free symmetric tensor. At $\mathcal O(r^{-1})$ this turns Eq.~\eqref{eq:evol-lead} into a trace-free tensor equation for $b_{AB}=d_{AB}-\tfrac12 q_{AB}q^{CD}d_{CD}$, that is
\be
\label{eq:bAB-evol}
\partial_u b_{AB} = -2\,\big(\eth_A\eth_B\beta_{(1)}\big)^{\mathrm{TF}} - 8\pi\,\big(\Tc{1}_{AB}\big)^{\mathrm{TF}} = 4\pi\,\big(\eth_A\eth_B\Tc{3}_{rr}\big)^{\mathrm{TF}} - 8\pi\,\big(\Tc{1}_{AB}\big)^{\mathrm{TF}},
\ee
where $(\,\cdot\,)^{\mathrm{TF}}$ denotes the trace-free part with respect to $q_{AB}$ and $\beta_{(1)}=-2\pi\Tc{3}_{rr}$ from Eq.~\eqref{eq:beta-sol}. In vacuum this reduces to $\partial_u b_{AB}=0$, equivalently $m^A m^B\partial_u d_{AB}=0$. The complementary trace of $\partial_u d_{AB}$ is not free but fixed by the determinant condition Eq.~\eqref{eq:det-cond}, $q^{AB}\partial_u d_{AB}=c^{AB}\partial_u c_{AB}$. Note in particular that $q^{AB}\partial_u d_{AB}\neq0$. Eqs.~\eqref{eq:bAB-constraint} and \eqref{eq:bAB-evol} constrain the same tensor $b_{AB}$ in two independent ways: Eq.~\eqref{eq:bAB-constraint} fixes its angular divergence $\eth^B b_{AB}$ in terms of the $\mathcal O(r^{-3})$ radial-angular flux $\Tc{3}_{rA}$, while Eq.~\eqref{eq:bAB-evol} fixes its retarded-time derivative $\partial_u b_{AB}$ in terms of $\Tc{3}_{rr}$ (through $\beta_{(1)}$) and the angular stress $\Tc{1}_{AB}$. Since $\eth_A$ is $u$-independent, $\partial_u$ and $\eth^B$ commute, so $\partial_u\!\big(\eth^B b_{AB}\big)=\eth^B\!\big(\partial_u b_{AB}\big)$ may be evaluated from either equation. From Eq.~\eqref{eq:bAB-constraint}, $\partial_u(\eth^B b_{AB})=-8\pi\,\partial_u\Tc{3}_{rA}$, while from Eq.~\eqref{eq:bAB-evol}, $\eth^B(\partial_u b_{AB})=4\pi\,\eth^B(\eth_A\eth_B\Tc{3}_{rr})^{\mathrm{TF}} -8\pi\,\eth^B(\Tc{1}_{AB})^{\mathrm{TF}}$. Equating the two yields an integrability condition purely among the matter coefficients,
\be
\label{eq:integrability}
\partial_u\Tc{3}_{rA} = \eth^B\big(\Tc{1}_{AB}\big)^{\mathrm{TF}} - \tfrac12\,\eth^B\big(\eth_A\eth_B\Tc{3}_{rr}\big)^{\mathrm{TF}} .
\ee
This is a constraint-propagation condition stating that the constraint Eq.~\eqref{eq:bAB-constraint} imposed on each cut is preserved by the evolution Eq.~\eqref{eq:bAB-evol}. Since Eqs.~\eqref{eq:bAB-constraint} and \eqref{eq:bAB-evol} are different projections of the same Einstein equations $G_{ab}=8\pi T_{ab}$, the contracted Bianchi identity $\nabla_b G^{b}{}_{a}\equiv0$ guarantees the system is consistent. Equivalently, it reassures that the matter obeys $\nabla_b T^{b}{}_{a}=0$. Thus, Eq.~\eqref{eq:integrability} holds automatically for any matter coupled through Einstein's equations.

We highlight that Eq.~\eqref{eq:integrability} is not a single component of $\nabla_b T^{b}{}_{a}=0$. Writing $\nabla_b T^{b}{}_{A}=\tfrac{1}{\sqrt{-g}}\partial_b(\sqrt{-g}\,T^{b}{}_{A}) -\tfrac12(\partial_A g_{cd})T^{cd}$ with $\sqrt{-g}=e^{2\beta}r^2\sqrt q$, the mixed flux $T^{b}{}_{A}=g^{bc}T_{cA}$ never carries $T_{rr}$, so $T_{rr}$ enters the $a=A$ conservation law only through the source term $\partial_A g_{cd}\,T^{cd}$ via $T^{cd}\supset g^{cr}g^{dr}T_{rr}$. With $T_{rr}=\mathcal O(r^{-3})$ this contributes at $\mathcal O(r^{-2})$. The $\mathcal O(r^{-1})$ part of $\nabla_b T^{b}{}_{A}=0$ therefore relates $\partial_u\Tc{3}_{rA}$ to $\eth^B\Tc{1}_{AB}$ and the lower-order fluxes, with no $\Tc{3}_{rr}$ term. The $\Tc{3}_{rr}$ contribution to Eq.~\eqref{eq:integrability} arises instead from $\beta_{(1)}=-2\pi\Tc{3}_{rr}$ in the evolution Eq.~\eqref{eq:bAB-evol}, i.e.\ from a field equation. Thus Eq.~\eqref{eq:integrability} follows from the field equations and conservation together while conservation alone does not produce it.

At $\mathcal O(r^{-2})$ the structure changes qualitatively. Now, all nine terms of Eq.~\eqref{eq:evol} contribute in contrast to the three of Eq.~\eqref{eq:evol-lead}. Writing $h_{AB}=q_{AB}+c_{AB}/r+d_{AB}/r^2+e_{AB}/r^3+\dots$, the time-derivative term gives $r\partial_r(r\,\partial_u h_{AB})=-\partial_u d_{AB}/r-2\partial_u e_{AB}/r^2 +\dots$, so the dyad projection now determines the retarded-time derivative of the trace-free part of the next coefficient $e_{AB}$. Collecting the $\mathcal O(r^{-2})$ pieces of the nine terms,
\ba
\label{eq:evol-NLO}
2\,m^A m^B\partial_u e_{AB} = m^A m^B\Big[\, && (M c_{AB}-d_{AB}) - 2\,\mathcal H^{(2)}_{AB} - \eth_B u^{(3)}_{A} - 2\,u^{(2)}_{A}u^{(2)}_{B} \nonumber\\
&& -\,\tfrac12 c_{AB}\,\eth_C u_{(2)}^{C} - u_{(2)}^{C}\,\eth_C c_{AB} + c_{A}{}^{C}\big(\eth_C u^{(2)}_{B}-\eth_B u^{(2)}_{C}\big) \nonumber\\
&& -\,8\pi\big(\Tc{2}_{AB}+2\beta_{(1)}\Tc{1}_{AB}\big)\,\Big],
\ea
where $u^{(n)}_A:=q_{AB}u^B_{(n)}$, the last line collects the matter, and the $\beta$-Hessian coefficient denotes the $\mathcal O(r^{-2})$ part of $D_AD_B\beta+(D_A\beta)(D_B\beta)$. It is
\be
\label{eq:Hessian2}
\mathcal H^{(2)}_{AB} := \eth_A\eth_B\beta_{(2)} -\tfrac12 q^{EF}\big(\eth_A c_{FB}+\eth_B c_{FA}-\eth_F c_{AB}\big)\eth_E\beta_{(1)} +\eth_A\beta_{(1)}\,\eth_B\beta_{(1)} +2\beta_{(1)}\,\eth_A\eth_B\beta_{(1)} .
\ee
Eq.~\eqref{eq:evol-NLO} fixes $\partial_u$ of the trace-free part of $e_{AB}$ in terms of the mass aspect $M$, the lower coefficients $c_{AB},d_{AB},u^A_{(2)},$ $u^A_{(3)},\beta_{(1)},\beta_{(2)}$ and the matter coefficients $\Tc{1}_{AB},\Tc{2}_{AB}$. Relative to Eq.~\eqref{eq:bAB-evol} the new matter inputs are $\Tc{2}_{AB}$ and the cross term $\beta_{(1)}\Tc{1}_{AB}$. In vacuum it reduces to the standard subleading Bondi evolution.

\subsection{Supplementary conditions: mass and angular-momentum aspect}
\label{sec:matter-supp}

By Eq.~\eqref{eq:supp}, $r^2 e^{2\beta}E^{r}{}_{a}$ is independent of $r$, so the supplementary equations $E^{r}{}_{a}=0$ need only be imposed at one radius. Evaluating them as $r\to\infty$ turns each into a flux-balance law at $\scrip$. With $E^{r}{}_{a}=R^{r}{}_{a}-8\pi T^{r}{}_{a}$ (the term $\tfrac12\delta^r_a R$ vanishes for $a=u,A$) and $e^{2\beta}\to1$, the condition $\lim_{r\to\infty}(r^2 E^{r}{}_{a})=0$ reads
\be
\label{eq:flux-balance}
\lim_{r\to\infty}\big(r^2 R^{r}{}_{a}\big) = 8\pi\lim_{r\to\infty}\big(r^2 T^{r}{}_{a}\big) ,
\ee
the left side being the asymptotic curvature and the right the net radial flux of energy ($a=u$) and angular momentum ($a=A$) at $\scrip$.\footnote{ Here, by ``net'' we refer to the fact that $T^{r}{}_{a}$ may also carry total-$u$-derivative pieces that only shift the instantaneous mass-aspect $M$/flux split, so that it is the finiteness of the balance, not either side separately, that is physical (cf.\ Sec.~\ref{sec:matter-bd}).} With $g^{ru}=-e^{-2\beta}$, $g^{rr}=\tfrac{V}{r}e^{-2\beta}$, $g^{rB}=-U^B e^{-2\beta}$ the mixed flux $T^{r}{}_{a}=g^{rc}T_{ca}$ is
\ba
\label{eq:Tru}
T^{r}{}_{u} &=& e^{-2\beta}\Big(-T_{uu}+\tfrac{V}{r}T_{ur}-U^B T_{uB}\Big),\\
\label{eq:TrA}
T^{r}{}_{A} &=& e^{-2\beta}\Big(-T_{uA}+\tfrac{V}{r}T_{rA}-U^B T_{BA}\Big).
\ea

The components $T_{uu}$ and $T_{uA}$ enter the field equations only here. They are unconstrained by the hypersurface and evolution equations. The left-hand side of Eq.~\eqref{eq:flux-balance}, $\lim_{r\to\infty}(r^2 R^{r}{}_{a})$, is built entirely from the asymptotic metric data ($M$, $N_{AB}$, $c_{AB}$, $L_A$) of the preceding sections and therefore can be safely assumed to be finite. Consequently, the same should hold for the matter flux $\lim_{r\to\infty}(r^2 T^{r}{}_{a})$. We highlight here that the order of the analysis of the two components matters, because only some of the components in Eqs.~\eqref{eq:Tru}--\eqref{eq:TrA} are yet constrained.

Before turning to the fluxes we settle the one component that the integration hierarchy leaves open, i.e., the angular trace $q^{AB}T_{AB}$, since it feeds the angular flux $T^{r}{}_{A}$. It is pinned by none of the hierarchy equations as it cancels in the $V$-equation source, and the evolution equation retains only the trace-free part of $T_{AB}$. Thus, its value is instead determined by the angular trace of the field equations, $g^{AB}E_{AB}=0$, i.e.\ (using $g^{AB}g_{AB}=2$)
\be
\label{eq:angular-trace}
8\pi\,g^{AB}T_{AB} = g^{AB}R_{AB} - R^{c}{}_{c} ,
\ee
which fixes the angular pressure in terms of the curvature already built from the other components. This is, however, not an independent equation. Once the main equations hold, the $a=r$ component of the contracted Bianchi identity $\nabla_b E^{b}{}_{a}=0$ reduces to $(\partial_r g^{AB})E_{AB}=-\tfrac{2}{r}\,g^{AB}E_{AB}=0$, so $g^{AB}E_{AB}=0$ is satisfied automatically. To extract the falloff without assuming a matter model, we evaluate the right-hand side of Eq.~\eqref{eq:angular-trace} directly on the asymptotic solution, where the metric already carries the established falloffs ($\beta=\mathcal O(r^{-1})$, $U^A=\mathcal O(r^{-2})$, $V=r-2M+\dots$, $h_{AB}=q_{AB}+\dots$). A computation of the Bondi--Sachs Ricci tensor gives, independently of the shear,
\be
\label{eq:angtrace-curv}
g^{AB}R_{AB}-R^{c}{}_{c}=\frac{4\,\partial_u\beta_{(1)}}{r^2}+\mathcal O(r^{-3}),
\ee
so that, with $g^{AB}=r^{-2}h^{AB}$ and $\beta_{(1)}=-2\pi\Tc{3}_{rr}$,
\be
\label{eq:angtrace-gen}
h^{AB}T_{AB}=r^2\,g^{AB}T_{AB}=\frac{\partial_u\beta_{(1)}}{2\pi}+\mathcal O(r^{-1}) =-\,\partial_u\Tc{3}_{rr}+\mathcal O(r^{-1}).
\ee
Thus the angular trace does not decay in general. Instead, it carries a non-vanishing $\mathcal O(1)$ piece $-\partial_u\Tc{3}_{rr}$ and reduces to $\mathcal O(r^{-1})$ precisely when $\Tc{3}_{rr}$ is $u$-independent, in particular when $\Tc{3}_{rr}=0$. This already sharpens the entry $T_{AB}=\mathcal O(r^{-1})$ of the falloff table Eq.~\eqref{eq:T-falloff} in the sense that the trace-free part is $\mathcal O(r^{-1})$ unconditionally (fixed by the evolution equation), but the trace is $\mathcal O(1)$ unless $\partial_u\Tc{3}_{rr}=0$.

With the angular trace in hand we turn to the fluxes. Take $T^{r}{}_{A}$, Eq.~\eqref{eq:TrA}, first, as it involves only quantities already fixed. With $T_{rA}=\mathcal O(r^{-2})$, $U^A=\mathcal O(r^{-2})$ (Sec.~\ref{sec:matter-U}) and the trace-free part of $T_{AB}$ at $\mathcal O(r^{-1})$ (evolution equation, Sec.~\ref{sec:matter-evol}), one has $\tfrac{V}{r}T_{rA}\to\Tc{2}_{rA}/r^2$, while the cross term carries the explicit coefficient $u^{B}_{(2)}$, writing $U^B=u^{B}_{(2)}/r^2+\mathcal O(r^{-3})$. The angular trace contributes through the $\mathcal O(1)$ part $\Tc{0}_{BA}=\tfrac12 q_{BA}h^{CD}T_{CD}$, which by Eq.~\eqref{eq:angtrace-gen} equals $-\tfrac12 q_{BA}\partial_u\Tc{3}_{rr}$, so that
\be
\label{eq:UTBA}
r^2\,U^B T_{BA}=u^{B}_{(2)}\,\Tc{0}_{BA}+\mathcal O(r^{-1}) \;\xrightarrow[r\to\infty]{}\;-\tfrac12\,u_{(2)A}\,\partial_u\Tc{3}_{rr},
\ee
which vanishes for matter with $u$-independent radial stress and is finite in every case. Hence $\lim_{r\to\infty}(r^2 T^{r}{}_{A})=\Tc{2}_{rA}-\lim_{r\to\infty}(r^2 T_{uA}) +\tfrac12 u_{(2)A}\partial_u\Tc{3}_{rr}$, which irrespective of that finite trace term is finite only if $T_{uA}=\mathcal O(r^{-2})$.

Only now, with $T_{uA}$ constrained, can $T^{r}{}_{u}$ of Eq.~\eqref{eq:Tru} be controlled. Its cross term $U^B T_{uB}$ involves $T_{uB}=T_{uA}$, which was not yet bounded above. Using $T_{ur}=\mathcal O(r^{-3})$, i.e. Eq.~\eqref{eq:V-cond}, and the freshly-derived $T_{uB}=\mathcal O(r^{-2})$,
\be
\label{eq:UTuB}
r^2\,\tfrac{V}{r}T_{ur}=\mathcal O(r^{-1}),\qquad r^2\,U^B T_{uB}=u^{B}_{(2)}\,\Tc{2}_{uB}/r^2+\mathcal O(r^{-3})\;\xrightarrow[r\to\infty]{}\;0 ,
\ee
so $\lim_{r\to\infty}(r^2 T^{r}{}_{u})=-\lim_{r\to\infty}(r^2 T_{uu})$, finite only if $T_{uu}=\mathcal O(r^{-2})$. In sum,
\be
\label{eq:uu-uA-falloff}
T_{uu}=\mathcal O(r^{-2}),\qquad T_{uA}=\mathcal O(r^{-2}).
\ee
A slower one would make $\lim_{r\to\infty}(r^2 T^{r}{}_{a})$ diverge and is excluded, while a faster one is allowed and merely sets the corresponding limit to zero. Denoting the surviving coefficients $\Tc{2}_{uu}=\lim_{r\to\infty}(r^2 T_{uu})$ and $\Tc{2}_{uA}=\lim_{r\to\infty}(r^2 T_{uA})$,
\be
\label{eq:flux-limits}
\lim_{r\to\infty}\!\big(r^2 T^{r}{}_{u}\big)=-\,\Tc{2}_{uu}, \qquad \lim_{r\to\infty}\!\big(r^2 T^{r}{}_{A}\big)=\Tc{2}_{rA}-\Tc{2}_{uA} +\tfrac12\,u_{(2)A}\,\partial_u\Tc{3}_{rr},
\ee
the last term being the angular-trace contribution Eq.~\eqref{eq:UTBA}, which is present only for $u$-dependent radial stress (see Eq.~\eqref{eq:angtrace-gen}).

Let us now continue with Eq.~\eqref{eq:flux-balance}. For $a=u$ the geometric side of the latter is the $\mathcal O(r^{-2})$ coefficient of $R^{r}{}_{u}$ on the asymptotic solution. Evaluating $R^{r}{}_{u}$ directly for the metric Eq.~\eqref{eq:bs-metric} (keeping the slow $\mathcal O(r^{-1})$ part of $\beta$) gives
\be
\label{eq:Rru-geom}
\lim_{r\to\infty}\big(r^2 R^{r}{}_{u}\big) = 2\,\partial_u M + \partial_u\!\big(\eth_A u^{A}_{(2)}\big) + 2\,\partial_u\beta_{(1)} + N_{AB}N^{AB},
\ee
where $u^{A}_{(2)},\beta_{(1)}$ are the leading coefficients of $U^A,\beta$. In vacuum $\beta_{(1)}=0$ and $u^{A}_{(2)}=-\tfrac12\eth_B c^{AB}$, so $\partial_u(\eth_A u^{A}_{(2)})=-\eth_A\eth_B N^{AB}$ and Eq.~\eqref{eq:Rru-geom} collapses to the familiar $2\partial_u M-\eth_A\eth_B N^{AB}+N_{AB}N^{AB}$. With matter, $\beta_{(1)}=-2\pi\Tc{3}_{rr}\neq0$ and $u^{A}_{(2)}$ carries the extra pieces $-8\pi q^{AB}\Tc{2}_{rB}-6\pi\eth^A\Tc{3}_{rr}$ from Eqs.~\eqref{eq:U-sol},\eqref{eq:beta-sol}. Inserting these together with the energy flux $8\pi\lim(r^2 T^{r}{}_{u})=-8\pi\Tc{2}_{uu}$ into Eq.~\eqref{eq:flux-balance} gives the mass-aspect evolution
\be
\label{eq:M-evol}
2\,\partial_u M = \eth_A\eth_B N^{AB} - N_{AB}N^{AB} - 8\pi\,\Tc{2}_{uu} + 8\pi\,\partial_u\eth^B\Tc{2}_{rB} + \big(6\pi\,\eth^2 + 4\pi\big)\partial_u\Tc{3}_{rr} .
\ee
The last two groups are matter corrections beyond the energy flux. The $\Tc{3}_{rr}$ terms come from the slow falloff of $\beta$, directly via $2\partial_u\beta_{(1)}$ and indirectly through $u^{A}_{(2)}$. All of them vanish for matter with $\Tc{2}_{rA}=\Tc{3}_{rr}=0$, recovering the standard law. Integrating over the unit sphere ($m=\tfrac{1}{4\pi}\oint M\,\dd\Omega$), the divergence terms $\eth_A\eth_B N^{AB}$, $\eth^B\Tc{2}_{rB}$ and $\eth^2\Tc{3}_{rr}$ drop, leaving the Bondi mass-loss formula
\be
\label{eq:massloss}
\frac{\dd m}{\dd u} = -\frac{1}{8\pi}\oint N_{AB}N^{AB}\,\dd\Omega - \oint \Tc{2}_{uu}\,\dd\Omega + \frac12\,\partial_u\!\oint \Tc{3}_{rr}\,\dd\Omega .
\ee
The first term is the loss to gravitational radiation. The second is the matter energy flux ($\Tc{2}_{uu}\ge0$ under the null energy condition, so matter further decreases the Bondi mass). The last is a total retarded-time derivative, absorbable into a redefinition of the Bondi mass.

For $a=A$ the geometric side of Eq.~\eqref{eq:flux-balance}, i.e. the $\mathcal O(r^{-2})$ coefficient of $R^{r}{}_{A}$, is the corresponding, lengthier standard expression \cite{M_dler_2016}. Equating it through Eq.~\eqref{eq:flux-balance} to the matter angular-momentum flux $8\pi\lim_{r\to\infty}(r^2 T^{r}{}_{A})$, evaluated in Eq.~\eqref{eq:flux-limits} (including the $\mathcal O(1)$-trace term for $u$-dependent radial stress), gives the angular-momentum-aspect evolution
\ba
\label{eq:L-evol}
-3\,\partial_u L_A &=& \eth_A M - \tfrac14\,\eth^E\big(\eth_E\eth^F c_{AF} - \eth_A\eth^F c_{EF}\big) + \tfrac18\,\eth_A\big(c_{EF}N^{EF}\big) \nonumber\\
&& -\,\eth_C\big(c^{CF}N_{FA}\big) + \tfrac12\,c^{EF}\big(\eth_A N_{EF}\big) + 8\pi\lim_{r\to\infty}\!\big(r^2 T^{r}{}_{A}\big).
\ea
In Eq.~\eqref{eq:L-evol} the geometric side is written in its vacuum form (the terms preceding the flux). As for the mass aspect, however, the matter content is not exhausted by $8\pi\lim(r^2 T^{r}{}_{A})$. A direct evaluation of $R^{r}{}_{A}$ shows that its $\mathcal O(r^{-2})$ coefficient also carries $\partial_u$-derivatives of the subleading coefficients $\beta_{(1)},\beta_{(2)},u^{A}_{(2)},u^{A}_{(3)}$, whose field-equation values Eqs.~\eqref{eq:beta-sol},\eqref{eq:U-sol} inject further $T_{ab}$-dependence. We record these terms explicitly. With $\beta_{(1)}=-2\pi\Tc{3}_{rr}$, $\beta_{(2)}=-\big(\tfrac1{32}c^{EF}c_{EF} +\pi\Tc{4}_{rr}\big)$, $u^{A}_{(2)}=-\tfrac12\eth_B c^{AB}-8\pi q^{AB}\Tc{2}_{rB}-6\pi\eth^A\Tc{3}_{rr}$ and $u^{A}_{(3)}$ from Eq.~\eqref{eq:U-sol}, the computed $\mathcal O(r^{-2})$ coefficient of $R^{r}{}_{A}$, equated through Eq.~\eqref{eq:flux-balance} to $8\pi\lim_{r\to\infty}(r^2 T^{r}{}_{A})$ of Eq.~\eqref{eq:flux-limits}, gives
\be
\label{eq:L-evol-matter}
\begin{split}
-3\,\partial_u L_A &= \mathcal G_A - 8\pi\,\Tc{2}_{uA} + \pi\,\partial_u\eth_A\Tc{4}_{rr} - 2\pi\,(\partial_u\Tc{3}_{rr})\,\eth^B c_{AB}\\
&\quad - 32\pi^2\,\partial_u\!\big(\Tc{3}_{rr}\,\Tc{2}_{rA}\big) - 16\pi^2\,\Tc{3}_{rr}\,\partial_u\eth_A\Tc{3}_{rr}\\
&\quad - 24\pi^2\,(\eth_A\Tc{3}_{rr})(\partial_u\Tc{3}_{rr}) ,
\end{split}
\ee
where $\mathcal G_A$ denotes the vacuum geometric side of Eq.~\eqref{eq:L-evol} ($\eth_A M$ together with its shear terms). Eq.~\eqref{eq:L-evol-matter} is the complete matter content of the $\mathcal O(r^{-2})$ balance. A direct evaluation of $R^{r}{}_{A}$ shows that $\lim_{r\to\infty}(r^2 R^{r}{}_{A})$ is at most quadratic in the asymptotic data. Consequently, the matter enters at most quadratically in $T_{ab}$ and couples to the shear only through the single bilinear term $-2\pi(\partial_u\Tc{3}_{rr})\,\eth^B c_{AB}$. The purely gravitational $\mathcal G_A$ does carry the standard shear-quadratic structure of Eq.~\eqref{eq:L-evol}, but no shear-quadratic term multiplies $T_{ab}$.

Several features stand out. First, the leading matter source is the angular-momentum flux $-8\pi\Tc{2}_{uA}$, the $A$-analog of the energy flux $-8\pi\Tc{2}_{uu}$ in Eq.~\eqref{eq:M-evol}. Second, the radial flux $\Tc{2}_{rA}$ does not enter linearly, instead the $+8\pi\Tc{2}_{rA}$ of Eq.~\eqref{eq:flux-limits} is canceled by the $-u^{A}_{(2)}$ contribution of the geometric side, so $\Tc{2}_{rA}$ survives only quadratically in $\partial_u(\Tc{3}_{rr}\Tc{2}_{rA})$. Third, $\pi\,\partial_u\eth_A\Tc{4}_{rr}$ descends from $\beta_{(2)}$ and the matter quadratics from $u^{A}_{(3)}$. Fourth, the entire shear--matter coupling collapses to the single term $-2\pi(\partial_u\Tc{3}_{rr})\eth^B c_{AB}$. The many candidate cross terms from the $c^{AB}\Tc{2}_{rB}$, $c^{AB}\eth_B\Tc{3}_{rr}$ and $\Tc{3}_{rr}\,\eth_E c^{AE}$ pieces of Eq.~\eqref{eq:U-sol} cancel against the geometric-side and flux contributions, leaving only this radial-stress--shear term.

\subsection{Summary of the derived $T_{ab}$ falloffs.}
Compiling the results of the previous section together with those of Secs.~\ref{sec:matter-beta}--\ref{sec:matter-evol}, every covariant component of the stress-energy tensor has its first allowed order in $1/r$ fixed by a specific field-equation component, with the slower (forbidden) orders set to zero by an explicit asymptotic-flatness or finiteness requirement. All in all, we find:

\begin{itemize}\setlength{\itemsep}{0.25em}
\item $T_{rr}=\Tc{3}_{rr}/r^3+\mathcal O(r^{-4})$ determined by the $\beta$-equation \eqref{eq:beta-cond} forcing $\Tc{0}_{rr}=\Tc{1}_{rr}=\Tc{2}_{rr}=0$ via $\lim_{r\to\infty}\beta=0$ (which excludes $r^2$, $r$ and $\ln r$ growth in $\beta$).
\item $T_{rA}=\Tc{2}_{rA}/r^2+\mathcal O(r^{-3})$ determined by the $U^A$-equation \eqref{eq:FA-sol} which forbids $\Tc{0}_{rA}\neq0$ (it would produce $U^A\sim\ln r$, divergent) and $\Tc{1}_{rA}\neq0$ (incompatible with $\lim_{r\to\infty}U^A=0$).
\item $T_{ur}=\Tc{3}_{ur}/r^3+\mathcal O(r^{-4})$ determined by the $V$-equation \eqref{eq:V-cond} forces $\Tc{0}_{ur}=\Tc{1}_{ur}=\Tc{2}_{ur}=0$ via $\lim_{r\to\infty}V/r=1$.
\item $T_{AB}^{\mathrm{TF}}=\mathcal O(r^{-1})$ (trace-free part) determined by the evolution equation \eqref{eq:evol}. The leading $\mathcal O(r^0)$ trace-free contribution is incompatible with finite news.
\item $h^{AB}T_{AB}=-\partial_u\Tc{3}_{rr}+\mathcal O(r^{-1})$ (angular trace) determined by the angular-trace Einstein equation \eqref{eq:angular-trace}. Generically $\mathcal O(1)$, reducing to $\mathcal O(r^{-1})$ iff $\partial_u\Tc{3}_{rr}=0$.
\item $T_{uu}=\Tc{2}_{uu}/r^2+\mathcal O(r^{-3})$ determined by the supplementary balance Eq.~\eqref{eq:flux-balance} on $T^{r}{}_{u}$ 
\item $T_{uA}=\Tc{2}_{uA}/r^2+\mathcal O(r^{-3})$ determined by the supplementary balance Eq.~\eqref{eq:flux-balance} on $T^{r}{}_{A}$ 
\end{itemize}

This summary sharpens the schematic entries of Eq.~\eqref{eq:T-falloff}: each row establishes the slowest decay consistent with the corresponding constraint, where a faster falloff is always admissible and simply sets the listed leading coefficient to zero. The trace of $T_{AB}$ is the only entry whose generic order is $\mathcal O(1)$ rather than a strict power of $r$. It reduces to $\mathcal O(r^{-1})$ when the radial stress $\Tc{3}_{rr}$ is $u$-independent.

Before moving on, the framework should prove itself on concrete matter models. For a free massless scalar and for Einstein--Maxwell theory this verification is carried out in App.~\ref{app:matter-scalar} and App.~\ref{app:matter-maxwell}. In both cases the field equations reproduce the metric falloffs, i.e., every stress-energy component lands on or below its bound in Eq.~\eqref{eq:T-falloff}, and the standard Bondi--Sachs structure emerges without further restriction. In the following subsection we present instead the example that genuinely tests the framework beyond GR.

\subsection{Brans--Dicke theory as a beyond-GR example}
\label{sec:matter-bd}

We run the construction presented in this section in a generalized format explicitly for BD gravity \cite{tahura_brans-dicke_2021}, again treating every metric function as an unconstrained smooth $1/r$ series and imposing only $\lim_{r\to\infty}\beta=0$, $\lim_{r\to\infty}U^A=0$, $\lim_{r\to\infty}V/r=1$, $h_{AB}\to q_{AB}$. In vacuum the Jordan-frame BD action is
\be
\label{eq:bd-action}
S=\frac{1}{16\pi}\int\!\dd^4x\,\sqrt{-g}\,\Big[\phi R -\frac{\omega}{\phi}\,g^{ab}\partial_a\phi\,\partial_b\phi\Big],
\ee
with constant BD parameter $\omega$.\footnote{In the language of the SVT class analyzed in Sec.~\ref{sec:flatness_SVT} \cite{Heisenberg:2018acv}, the action Eq.~\eqref{eq:bd-action} is the point $G_4(\Phi)=\Phi$, $G_2(\Phi,X)=2\omega X/\Phi$ of the parameter space, with all remaining coupling functionals switched off (cf.\ Eqs.~\eqref{eq:svt-L2-full}--\eqref{eq:svt-L5-full}).} Varying $\phi$ gives
\be
\label{eq:bd-scalar-eom}
\frac{2\omega}{\phi}\,\Box\phi +R-\frac{\omega}{\phi^2}\,(\partial\phi)^2=0 .
\ee
Eliminating $R$ through the trace of the metric field equation below, $\phi R=\tfrac{\omega}{\phi}(\partial\phi)^2+3\,\Box\phi$, collapses Eq.~\eqref{eq:bd-scalar-eom} to
\be
\label{eq:bd-wave}
(2\omega+3)\,\Box\phi=0 ,
\ee
so that on shell the vacuum BD scalar obeys the free wave equation $\Box\phi=0$ for $2\omega+3\neq0$ (the degenerate value $\omega=-3/2$, for which the scalar is non-dynamical, is excluded throughout). Varying $g_{ab}$ yields, after moving the $\phi R$ prefactor to the right, the effective Einstein equation $G_{ab}=8\pi T_{ab}$ with
\be
\label{eq:bd-Teff}
8\pi T_{ab}=\frac{1}{\phi}\big(\nabla_a\nabla_b\phi-g_{ab}\Box\phi\big) +\frac{\omega}{\phi^2}\big(\partial_a\phi\,\partial_b\phi -\tfrac12 g_{ab}(\partial\phi)^2\big).
\ee
On shell, by Eq.~\eqref{eq:bd-wave}, the $g_{ab}\Box\phi$ term drops and $T_{ab}$ plays the role of the stress-energy of Secs.~\ref{sec:matter-beta}--\ref{sec:matter-supp}. The first, linear piece is the non-minimal ``improvement'' term absent for the free scalar. It is this term that makes BD a nontrivial test of the framework. The prefactor $\phi$ can be seen as a modification to the asymptotic Newton constant, $G_N\propto\phi_0^{-1}$, and is finite iff $\phi_0\neq0$.

Asymptotic flatness requires the BD field to settle to a finite, non-zero constant $\phi_0$, i.e., a non-vanishing background, unlike the massless scalar of App.~\ref{app:matter-scalar}. We therefore adopt the smooth expansion
\be
\label{eq:bd-phi-ansatz}
\phi(u,r,x^A)=\phi_0+\frac{\varphi_1(u,x^A)}{r}+\frac{\varphi_2(u,x^A)}{r^2} +\mathcal O(r^{-3}),\qquad \phi_0=\text{const}\neq0 ,
\ee
so that $\partial_r\phi=-\varphi_1/r^2-2\varphi_2/r^3+\mathcal O(r^{-4})$ and $\partial_u\phi=\dot\varphi_1/r+\mathcal O(r^{-2})$. For the metric, we use the same smooth $1/r$ expansions as in the general analysis, in particular the angular block retains the ansatz Eq.~\eqref{eq:h-exp} with the determinant conditions Eq.~\eqref{eq:det-cond}, and we keep the BD field massless, i.e., potential-free.

In Bondi gauge the Christoffel symbols entering the scalar Hessian obey the exact relations
\be
\label{eq:bd-christoffels}
\Gamma^u_{rr}=\Gamma^u_{rA}=\Gamma^u_{ur}=0,\quad \Gamma^r_{rr}=2\,\partial_r\beta,\quad \Gamma^u_{AB}=\tfrac12 e^{-2\beta}\partial_r\!\big(r^2h_{AB}\big),\quad \Gamma^B_{rA}=\tfrac1r\,\delta^B_A+\tfrac12 h^{BC}\partial_r h_{CA},
\ee
so that the components of Eq.~\eqref{eq:bd-Teff} read, on shell and prior to any expansion,
\be
\label{eq:bd-Tab-closed}
\begin{aligned}
8\pi T_{rr}&=\frac{1}{\phi}\big(\partial_r^2\phi-2\,\partial_r\beta\,\partial_r\phi\big) +\frac{\omega}{\phi^2}(\partial_r\phi)^2,\\
8\pi T_{rA}&=\frac{1}{\phi}\Big(\partial_r\partial_A\phi-\tfrac1r\,\partial_A\phi -\tfrac12 h^{BC}\partial_rh_{CA}\,\partial_B\phi-\Gamma^r_{rA}\,\partial_r\phi\Big) +\frac{\omega}{\phi^2}\,\partial_r\phi\,\partial_A\phi,\\
8\pi T_{AB}&=\frac{1}{\phi}\Big(\partial_A\partial_B\phi-\Gamma^C_{AB}\partial_C\phi -\tfrac12 e^{-2\beta}\partial_r(r^2h_{AB})\,\partial_u\phi -\Gamma^r_{AB}\,\partial_r\phi\Big)\\
&\quad +\frac{\omega}{\phi^2}\Big(\partial_A\phi\,\partial_B\phi -\tfrac12 r^2h_{AB}(\partial\phi)^2\Big),\\
8\pi T_{ur}&=\frac{1}{\phi}\Big(\partial_u\partial_r\phi -\Gamma^r_{ur}\,\partial_r\phi-\Gamma^A_{ur}\,\partial_A\phi\Big) +\frac{\omega}{\phi^2}\Big(\partial_u\phi\,\partial_r\phi +\tfrac12 e^{2\beta}(\partial\phi)^2\Big),
\end{aligned}
\ee
and analogously for $T_{uu}$, $T_{uA}$, with $(\partial\phi)^2$ as given below Eq.~\eqref{eq:scalar-Tab-closed} of App.~\ref{app:matter-scalar}. In contrast to the minimally coupled scalar, the BD stress depends on the unsolved metric functions. $T_{rr}$ contains $\partial_r\beta$, and $T_{AB}$ contains $\partial_r(r^2h_{AB})$ and $\Gamma^r_{AB}$. Thus, each hypersurface equation below is a genuine differential equation in which geometry and matter must be solved together. Expanding the Hessian entries with Eqs.~\eqref{eq:bd-phi-ansatz} and \eqref{eq:h-exp} gives, at leading order,
\be
\label{eq:bd-Tab-orders}
\begin{aligned}
\nabla_r\nabla_r\phi&=\frac{2\varphi_1}{r^3}+\mathcal O(r^{-4}), &\nabla_r\nabla_A\phi&=-\frac{2\,\partial_A\varphi_1}{r^2}+\mathcal O(r^{-3}),\\
\nabla_A\nabla_B\phi&=-q_{AB}\,\dot\varphi_1+\mathcal O(r^{-1}), &\nabla_u\nabla_r\phi&=-\frac{\dot\varphi_1}{r^2}+\mathcal O(r^{-3}),\\
\nabla_u\nabla_u\phi&=\frac{\ddot\varphi_1}{r}+\mathcal O(r^{-2}), &\nabla_u\nabla_A\phi&=\frac{\partial_A\dot\varphi_1}{r}+\mathcal O(r^{-2}),
\end{aligned}
\ee
the $AB$ entry being the non-decaying $\Phi_{AB}=\mathcal O(r^0)$ Hessian generated by $\Gamma^u_{AB}=r\,q_{AB}+\mathcal O(r^0)$. The on-shell wave equation Eq.~\eqref{eq:bd-wave} has the free-wave structure $\Box\phi=(2\partial_u\varphi_2+\eth^2\varphi_1)/r^3+\mathcal O(r^{-4})$. It leaves $\varphi_1(u,x^A)$ freely specifiable and fixes $2\partial_u\varphi_2=-\eth^2\varphi_1$ at leading order.

\subsubsection{$\beta$-equation.}
Inserting $8\pi T_{rr}$ of Eq.~\eqref{eq:bd-Tab-closed} into Eq.~\eqref{eq:hyp-beta} and collecting the $\partial_r\beta$ terms on the left turns the $\beta$-equation into the linear ordinary differential equation
\be
\label{eq:bd-beta-ode}
\Big(1+\frac{r}{2\phi}\,\partial_r\phi\Big)\partial_r\beta =\frac{r}{16}\,h^{AC}h^{BD}(\partial_rh_{AB})(\partial_rh_{CD}) +\frac{r}{4}\bigg(\frac{\partial_r^2\phi}{\phi} +\frac{\omega}{\phi^2}\,(\partial_r\phi)^2\bigg).
\ee
The bracket on the left is $1-\varphi_1/(2\phi_0 r)+\mathcal O(r^{-2})$, so the self-coupling enters only beyond the orders displayed. Expanding the right-hand side with Eqs.~\eqref{eq:bd-phi-ansatz},\eqref{eq:h-exp},
\be
\partial_r\beta=\frac{\varphi_1}{2\phi_0}\,\frac{1}{r^{2}} +\bigg[\frac{c^{AB}c_{AB}}{16}+\frac{3\varphi_2}{2\phi_0} +\frac{(\omega-1)\,\varphi_1^2}{4\phi_0^2}\bigg]\frac{1}{r^{3}}+\mathcal O(r^{-4}),
\ee
and radial integration subject to $\lim_{r\to\infty}\beta=0$ yields
\be
\label{eq:bd-beta}
\beta=-\frac{\varphi_1}{2\phi_0}\,\frac1r -\bigg[\frac{c^{AB}c_{AB}}{32}+\frac{3\varphi_2}{4\phi_0} +\frac{(\omega-1)\,\varphi_1^2}{8\phi_0^2}\bigg]\frac{1}{r^{2}}+\mathcal O(r^{-3}) .
\ee
Thus $\beta_{(1)}=-\varphi_1/(2\phi_0)\neq0$. Re-inserting Eq.~\eqref{eq:bd-beta} into $8\pi T_{rr}$ of Eq.~\eqref{eq:bd-Tab-closed} gives $\Tc{3}_{rr}=\varphi_1/(4\pi\phi_0)$ and $8\pi\Tc{4}_{rr}=6\varphi_2/\phi_0+(\omega-1)\varphi_1^2/\phi_0^2$, so Eq.~\eqref{eq:bd-beta} coincides with the general solution Eq.~\eqref{eq:beta-sol}, $\beta_{(1)}=-2\pi\Tc{3}_{rr}$, $\beta_{(2)}=-\big(\tfrac{1}{32}c^{AB}c_{AB}+\pi\Tc{4}_{rr}\big)$.

\subsubsection{$U^A$-equation.}
With $\beta$ and $h_{AB}$ known, the matter source of Eq.~\eqref{eq:hyp-U} expands (using $\Gamma^r_{rA}=\big(\partial_A\beta_{(1)}-u_{(2)A}\big)/r +\mathcal O(r^{-2})$, the $-u_{(2)A}$ piece arising from the $g^{rB}\partial_rg_{BA}$ term of $\Gamma^r_{rA}$) to
\be
\label{eq:bd-TrA}
16\pi r^2T_{rA}=-\frac{4\,\partial_A\varphi_1}{\phi_0} +\frac{2}{\phi_0}\bigg[-3\,\partial_A\varphi_2 +\tfrac12\,c^B{}_A\partial_B\varphi_1 -\varphi_1\,u_{(2)A} +\frac{3-2\omega}{2\phi_0}\,\varphi_1\partial_A\varphi_1\bigg]\frac1r +\mathcal O(r^{-2}) ,
\ee
where the appearance of the (yet to be determined) coefficient $u_{(2)A}$ in the source is another instance of the geometry--matter coupling of the BD hierarchy. It is harmless, since $u_{(2)A}$ is fixed at the previous order of this same equation. The absence of $r^2$- and $r$-pieces means the two general obstructions of Sec.~\ref{sec:matter-U} ($\Tc{0}_{rA},\Tc{1}_{rA}\neq0$) are absent. The $\mathcal O(r^0)$ piece feeds the linear-in-$r$ coefficient of $F_A=r^4e^{-2\beta}h_{AB}\partial_rU^B$ in Eq.~\eqref{eq:FA-sol} through $\Tc{2}_{rA}=-\partial_A\varphi_1/(4\pi\phi_0)$. The two radial integrations of the $U^A$ equation subject to $\lim_{r\to\infty}U^A=0$ then give
\be
\label{eq:bd-U}
u^A_{(2)}=-\tfrac12\,\eth_B c^{AB}+\frac{1}{2\phi_0}\,\eth^A\varphi_1 ,
\ee
in agreement with the general solution Eq.~\eqref{eq:U-sol} evaluated on the BD coefficients (the pieces $16\pi q^{AB}\Tc{2}_{rB}=-\tfrac{4}{\phi_0}\eth^A \varphi_1$ and $12\pi\eth^A\Tc{3}_{rr}=\tfrac{3}{\phi_0}\eth^A\varphi_1$ combining to $-\tfrac{1}{\phi_0}\eth^A\varphi_1$). The $\mathcal O(r^{-3})$ coefficient follows likewise from Eq.~\eqref{eq:U-sol}. The $\mathcal O(r^{-1})$ piece of Eq.~\eqref{eq:bd-TrA} then enters the log-removal constraint Eq.~\eqref{eq:bAB-constraint}, which ties $b_{AB}$ to the BD data. Substituting the solved $u_{(2)A}$ of Eq.~\eqref{eq:bd-U},
\be
\label{eq:bd-bAB}
\eth^B b_{AB}=-8\pi\,\Tc{3}_{rA} =\frac{1}{\phi_0}\bigg[3\,\partial_A\varphi_2 -\tfrac12\,c^B{}_A\partial_B\varphi_1 -\tfrac12\,\varphi_1\,\eth^Bc_{AB} -\frac{1-\omega}{\phi_0}\,\varphi_1\partial_A\varphi_1\bigg] .
\ee
Every $\beta_{(1)}$- and $\Tc{3}_{rr}$-dependent term of the general analysis is therefore active here.

\subsubsection{$V$-equation: failure of the standard boundary condition.}
The solved functions determine every term of Eq.~\eqref{eq:hyp-V} at $\mathcal O(r^0)$. Namely, the geometric terms contribute $\R\to2$ (the $\beta$-Hessian, $U^A$-divergence and quadratic terms are all $\mathcal O(r^{-1})$, as in Sec.~\ref{sec:matter-V}), while the matter source, evaluated with $\Gamma^u_{AB}$ of Eq.~\eqref{eq:bd-christoffels} and $(\partial\phi)^2=2\varphi_1\dot\varphi_1/r^3+\mathcal O(r^{-4})$, is
\be
\label{eq:bd-V-source}
8\pi\big[h^{AB}T_{AB}-r^2T^{a}{}_{a}\big] =-\frac{2\dot\varphi_1}{\phi_0}+\mathcal O(r^{-1}),\qquad 8\pi\,r^2T^{a}{}_{a} =-\frac{2\omega\,\varphi_1\dot\varphi_1}{\phi_0^2\,r}+\mathcal O(r^{-2}) .
\ee
Radial integration of $2e^{-2\beta}\partial_rV=2-2\dot\varphi_1/\phi_0+\mathcal O(r^{-1})$ then yields
\be
\label{eq:bd-Vdefect}
V=\Big(1-\frac{\dot\varphi_1}{\phi_0}\Big)r-2M(u,x^A)+\dots, \qquad \frac{V}{r}\;\xrightarrow[\;r\to\infty\;]{}\;1-\frac{\dot\varphi_1}{\phi_0} \;\neq\;1 
\ee
which clearly fails to satisfy the Bondi boundary condition $\lim_{r\to\infty}V/r=1$ for $\dot\varphi_1\neq0$, by a $u$-dependent $\mathcal O(1)$ slope that no choice of integration constant can repair. Equivalently, in the language of the general analysis, $\Tc{2}_{ur}=-\dot\varphi_1/(8\pi\phi_0)$ read off from Eq.~\eqref{eq:bd-Tab-closed} violates $T_{ur}=\mathcal O(r^{-3})$ and triggers item (iii) of Sec.~\ref{sec:matter-V}, $V/r\to1+8\pi\Tc{2}_{ur}$. Before addressing this issue further, let us address the remaining equations of the theory.

\subsubsection{Evolution equation.}
Evaluating the trace-free $\mathcal O(r^{-1})$ part of $T_{AB}$ on the solved metric ($\Gamma^u_{AB}=rq_{AB}+\tfrac12c_{AB}-2\beta_{(1)}q_{AB}+\mathcal O(r^{-1})$ and $\Gamma^r_{AB}=r(N_{AB}-q_{AB})+\mathcal O(1)$) gives
\be
\label{eq:bd-T1AB}
8\pi\big(\Tc{1}_{AB}\big)^{\rm TF} =\frac{1}{\phi_0}\Big[\big(\eth_A\eth_B\varphi_1\big)^{\rm TF} -\tfrac12\,\dot\varphi_1\,c_{AB}+\varphi_1\,N_{AB}\Big],
\ee
while the geometric source of Eq.~\eqref{eq:evol-lead} is $2\eth_A\eth_B\beta_{(1)}=-\eth_A\eth_B\varphi_1/\phi_0$. The Hessian pieces cancel between the two, leaving
\be
\label{eq:bd-evol}
m^Am^B\Big[\partial_u d_{AB} +\frac{1}{\phi_0}\Big(\varphi_1\,N_{AB}-\tfrac12\,\dot\varphi_1\,c_{AB}\Big) \Big]=0.
\ee
Unlike for the free scalar, the trace-free part of $d_{AB}$ is no longer conserved but is sourced by the product of the scalar background and the news. The source decouples as $\varphi_1\to0$.

\subsubsection{Re-inserting the solutions: the BD stress falloffs.}
Substituting the solved $\beta$, $U^A$, $V$, $h_{AB}$ back into Eq.~\eqref{eq:bd-Tab-closed} yields the scaling behavior of every covariant component,
\be
\label{eq:bd-Tab-summary}
\begin{aligned}
T_{rr}&=\frac{\varphi_1}{4\pi\phi_0\,r^{3}}+\mathcal O(r^{-4}), &T_{rA}&=-\frac{\partial_A\varphi_1}{4\pi\phi_0\,r^{2}}+\mathcal O(r^{-3}), &T_{AB}&=-\frac{q_{AB}\,\dot\varphi_1}{8\pi\phi_0}+\mathcal O(r^{-1}),\\
T_{ur}&=-\frac{\dot\varphi_1}{8\pi\phi_0\,r^{2}}+\mathcal O(r^{-3}), &T_{uu}&=\frac{\ddot\varphi_1}{8\pi\phi_0\,r}+\mathcal O(r^{-2}), &T_{uA}&=\frac{\partial_A\dot\varphi_1}{8\pi\phi_0\,r}+\mathcal O(r^{-2}).
\end{aligned}
\ee
Compared with Eq.~\eqref{eq:T-falloff} we thus find that $T_{rr}$ and $T_{rA}$ saturate their bounds, whereas the free scalar over-satisfied them. The $T_{AB}$ trace sits exactly at the marginal allowance $-\tfrac12 q_{AB}\partial_u\Tc{3}_{rr}$ of the angular-trace analysis. Finally, $T_{ur}$, $T_{uu}$, $T_{uA}$ violate their bounds by one power each. All three violating coefficients are total retarded-time derivatives of the radial stress,
\be
\label{eq:bd-total-derivatives}
\Tc{2}_{ur}=-\tfrac12\,\partial_u\Tc{3}_{rr},\qquad \Tc{1}_{uu}=\tfrac12\,\partial_u^2\Tc{3}_{rr},\qquad \Tc{1}_{uA}=\tfrac12\,\partial_u\partial_A\Tc{3}_{rr},
\ee
precisely the total-derivative structures that the general analysis of Secs.~\ref{sec:matter-V}--\ref{sec:matter-supp} absorbs into redefinitions of the mass aspect rather than into genuine flux. Anticipating Sec.~\ref{sec:svt-flatness}, this pattern is the universal signature of a scalar coupled non-minimally to the curvature. Note here that nothing physical diverges here. If one nevertheless inserts Eq.~\eqref{eq:bd-Tab-summary} into the standard-falloff formulas, e.g.\ the energy flux $\lim_{r\to\infty}r^2T^{r}{}_{u}$, which with $T_{uu}=\mathcal O(r^{-1})$ appears to grow linearly in $r$, one obtains ill-defined expressions that merely flag that the standard Bondi--Sachs identifications presuppose falloffs that fail here, cf.\ Eq.~\eqref{eq:bd-Vdefect}. We highlight that a consistent Bondi expansion exists, but requires modified falloffs and a correspondingly redefined mass aspect (which can be achieved by switching to the Einstein-frame), as carried out for scalar--tensor gravity in \cite{tahura_brans-dicke_2021,hou_conserved_2021,Hou:2021bxz}.

\subsubsection{Resolution: the Einstein frame.}

The standard cure is the conformal rescaling $\tilde g_{ab}=\Omega^2 g_{ab}$ with $\Omega^2=\phi/\phi_0$. Using, in $n=4$, $\tilde R=\Omega^{-2}[R-6\Box\ln\Omega -6(\nabla\ln\Omega)^2]$ and $\sqrt{-g}=\Omega^{-4}\sqrt{-\tilde g}$, together with the identity $\Omega^{-2}\Box\ln\Omega=\tildeBox\ln\Omega-2(\tilde\nabla\ln\Omega)^2$, the gravitational term becomes $\sqrt{-g}\,\phi R=\phi_0\sqrt{-\tilde g}\,[\tilde R +6\tildeBox\ln\Omega-6(\tilde\nabla\ln\Omega)^2]$. Discarding the total-derivative $\tildeBox\ln\Omega$ and combining with the transformed kinetic term $-\tfrac{\omega}{\phi}\sqrt{-g}(\nabla\phi)^2=-\omega\phi_0\sqrt{-\tilde g}(\tilde\nabla\ln\phi)^2$ yields the Einstein-frame action of GR minimally coupled to a free massless scalar,
\be
\label{eq:bd-conformal}
S=\int\!\dd^4x\,\sqrt{-\tilde g}\,\Big[\frac{\phi_0}{16\pi}\,\tilde R -\tfrac12\,\tilde g^{ab}\partial_a\tilde\phi\,\partial_b\tilde\phi\Big],\qquad \tilde\phi=\sqrt{\frac{(2\omega+3)\,\phi_0}{16\pi}}\;\ln\!\frac{\phi}{\phi_0} \quad(2\omega+3>0),
\ee
whose metric field equation is
\be
\label{eq:bd-einstein-frame}
\tilde G_{ab}=\frac{8\pi}{\phi_0}\,\tilde T^{(\tilde\phi)}_{ab},\qquad \tilde T^{(\tilde\phi)}_{ab}=\partial_a\tilde\phi\,\partial_b\tilde\phi -\tfrac12\tilde g_{ab}(\tilde\partial\tilde\phi)^2 ,
\ee
i.e.\ an Einstein-frame Newton constant $\tilde G_N=\phi_0^{-1}$. The rescaling used here is the special case $G(\Phi)=\Phi$ of $\tilde g_{ab}=[G(\Phi)/G(\phi_0)]\,g_{ab}$ where $G$ denotes the non-minimal coupling functional. We emphasize that rescaling does not apply to an arbitrary coupling functional. Since $\phi/\phi_0=1+\varphi_1/(\phi_0 r)+\mathcal O(r^{-2})\to1$, the conformal factor is $1+\mathcal O(r^{-1})$. It thus leaves the leading Minkowski structure and the conformal boundary $\scrip$ untouched, so the two frames share the same $\scrip$ and the same BMS group. In the Einstein frame the analysis is identical to that of App.~\ref{app:matter-scalar} with $\tilde\phi_1$ in place of $\phi_1$ and $8\pi\mapsto8\pi/\phi_0$ giving $\tilde\beta_{(1)}=0$, the finite energy flux $\tilde T_{uu}=(\partial_u\tilde\phi_1)^2/r^2$, and the mass-loss and angular-momentum-aspect laws Eqs.~\eqref{eq:scalar-M-evol}--\eqref{eq:scalar-L-evol}. The Jordan-frame $\beta_{(1)}=-\varphi_1/(2\phi_0)\neq0$ of Eq.~\eqref{eq:bd-beta} is precisely the imprint of the $\mathcal O(r^{-1})$ conformal factor relating the frames. Indeed $g_{ur}=\Omega^{-2}\tilde g_{ur}$ reproduces it exactly and the falloff defect Eq.~\eqref{eq:bd-Vdefect} is the statement that the standard Bondi--Sachs identifications apply to $\tilde g_{ab}$ rather than to $g_{ab}$.

The conformal rescaling sitting between Jordan- and Einstein-frame is an invertible redefinition of the field variables $(g_{ab},\phi)\mapsto(\tilde g_{ab},\tilde\phi)$, regular wherever $\phi>0$ (i.e.\ for $2\omega+3>0$ around the vacuum $\phi_0>0$). Substituting it into the action yields the identical functional up to a total derivative (irrelevant for the field equations, though not automatically for covariant-phase-space boundary terms), so on this domain the classical solution spaces are in one-to-one correspondence and in vacuum BD is GR plus a free massless scalar written in different variables. A physical distinction between the frames arises only once a matter coupling is specified and held fixed: BD matter couples minimally to $g_{ab}$, i.e. test bodies follow $g$-geodesics, so a detector measures the $g$-frame strain, including the scalar-induced breathing polarization, and in the Einstein variables this same coupling reappears as a $\tilde\phi$-dependent matter sector rather than as metric strain. Frame-invariant observables at $\scrip$ such as energy flux, memory offsets, charges agree either way once the dictionary is applied. Then, the finite Einstein-frame flux, $(\partial_u\tilde\phi_1)^2=\tfrac{2\omega+3}{16\pi\phi_0}\,\dot\varphi_1^2$, coincides with the flux of the modified-falloff Jordan-frame analyses \cite{tahura_brans-dicke_2021,Hou:2021bxz}. For the vacuum questions addressed here, i.e., the existence of $\scrip$, the asymptotic symmetry group, finiteness of fluxes, the identification is exact, and the Einstein frame is simply the variable choice in which the standard falloff table and the integration of Secs.~\ref{sec:matter-beta}--\ref{sec:matter-supp} apply.

For waveform observables, by contrast, the Einstein frame is only the intermediate computational arena. A detector is built from matter, and BD matter follows geodesics of the Jordan metric. The measured strain is the $g$-frame geodesic deviation, to which one must translate back at the end of any Einstein-frame computation. The dictionary is finite and explicit: the conformal rescaling $\Omega^2=1+\varphi_1/(\phi_0 r)+\mathcal O(r^{-2})$, the radius shift $\tilde r=r+\varphi_1/(2\phi_0)$, and an angle-dependent shift $u\to u+T(x^A)$ of the retarded time, acting exactly as the supertranslations of Eq.~\eqref{eq:BMS-KV}. Its physical effect is a reshuffling of the polarization content. The transverse-traceless strain $\tilde c_{AB}$ carries over up to the terms generated by this time shift, while the scalar $\varphi_1/\phi_0$, invisible in the Einstein-frame metric, enters the Jordan-frame geodesic deviation directly as the breathing polarization characteristic of scalar--tensor gravity \cite{Eardley:1973zuo}, together with the associated scalar (breathing) memory \cite{tahura_brans-dicke_2021,Zosso:2024xgy,heisenberg2025unifyingordinarynullmemory}. Waveform predictions quoted in the Einstein frame without this translation would miss precisely the extra polarization that makes BD observationally distinguishable from GR.

Importantly, translating back to the Jordan frame does not re-import the would-be divergences encountered above. First, the strain a detector registers is built from local curvature at large but finite $r$, and every term the dictionary adds is of order $1/r$. The breathing contribution $\propto\varphi_1/\phi_0$ and the terms generated by the time shift $u\to u+T(x^A)$ enter the waveform at the same $1/r$ order as the transverse-traceless strain itself. Even the $\mathcal O(1)$ defect in $g_{uu}$ contributes to the physical tidal field only at $\mathcal O(1/r)$, since for a detector of fixed proper size $s$ the associated geodesic-deviation terms scale as $\partial_A\partial_B\dot\varphi_1\,(s/r)$. Second, the violating coefficients of Eq.~\eqref{eq:bd-Tab-summary} are total retarded-time derivatives, Eq.~\eqref{eq:bd-total-derivatives}. For any radiation burst with $\dot\varphi_1\to0$ as $u\to\pm\infty$ their $u$-integrals vanish identically, e.g.\ $\int\ddot\varphi_1\,\dd u= [\dot\varphi_1]^{+\infty}_{-\infty}=0$. They therefore contribute zero net radiated energy and no secular waveform feature. Energy reversibly borrowed by the wave zone during the burst and returned afterwards, affecting only the instantaneous split between mass aspect and flux. That split is precisely what the redefined Jordan-frame mass aspect of \cite{tahura_brans-dicke_2021,Hou:2021bxz} renders finite order by order. In short, the back-translation re-imports the bookkeeping subtleties of the Jordan frame, the adapted mass aspect and time normalization, but never divergences in observables. The apparent divergence of the naive Jordan-frame flux integral is an artifact of the frame-presentation, not an obstruction to asymptotic flatness.

\subsubsection{Conclusion.}

Among the theories analyzed in this work, BD gravity is the prominent exception to the falloff table Eq.~\eqref{eq:T-falloff}. Taken at face value, the theory does not satisfy it. By the logic of Secs.~\ref{sec:matter-V}--\ref{sec:matter-supp} this would ordinarily be conclusive: a non-vanishing $\Tc{2}_{ur}$ is precisely the obstruction of item (iii) of Sec.~\ref{sec:matter-V}, and the Jordan metric $g_{ab}$ indeed fails the standard Bondi--Sachs falloffs whenever the scalar radiates. What rescues the theory is that the violation can be transformed away. The conformal rescaling $\tilde g_{ab}=(\phi/\phi_0)\,g_{ab}$ leading to Eq.~\eqref{eq:bd-conformal} removes the improvement term $\nabla_a\nabla_b\phi-g_{ab}\Box\phi$ responsible for the excess altogether, with a factor $\Omega^2=1+\mathcal O(r^{-1})$ that leaves $\scrip$ and the BMS group untouched. In the Einstein-frame variables every component of the stress lands on or below its bound and the hierarchy of Secs.~\ref{sec:matter-beta}--\ref{sec:matter-supp} applies verbatim. This escape route, however, is not generic. It exists because the coupling multiplying the curvature depends on the scalar alone, $G_4=G_4(\Phi)$ with $G_4=\Phi$ for BD, for which the required field redefinition is a conformal rescaling of the metric. For a coupling with genuine dependence on the kinetic invariant $X$, the corresponding transformation is disformal and maps $G_4(\Phi,X)R$ onto another Horndeski/DHOST theory rather than onto Einstein gravity. No choice of metric variable then restores the standard table, and one is left with the weaker option of a $u$-dependent mass aspect with modified falloffs, made possible by the total-derivative structure Eq.~\eqref{eq:bd-total-derivatives} of the violating coefficients.

The lesson is that the falloff table is a statement about a metric variable, not about a theory. For any non-minimally coupled theory the choice of field variables matters for the asymptotic bookkeeping, even though, at the classical level and on the domain where the conformal rescaling is regular ($\phi>0$, $2\omega+3>0$), the two frames are related by an invertible field redefinition with solution spaces in one-to-one correspondence.\footnote{Our statements are purely classical. At the quantum level the equivalence of conformal frames is a subtler and partly open question, see e.g.~\cite{Kamenshchik:2014waa}.} The BD example makes this concrete: the non-vanishing $\beta_{(1)}$ of Eq.~\eqref{eq:bd-beta} and the $\Phi_{AB}=\mathcal O(r^0)$ Hessian of Eq.~\eqref{eq:bd-Tab-orders}, both anticipated by the general framework, are precisely the Jordan-frame imprint of the $\mathcal O(r^{-1})$ conformal factor. Only one variable choice satisfies the standard falloff table, so which metric carries the standard Bondi--Sachs asymptotics is a derived property of the theory, and the GR-form identifications of mass and flux apply verbatim only in that frame. A violation of the table by total retarded-time derivatives in a non-minimally coupled theory is accordingly not an obstruction to asymptotic flatness but an instruction to look for the variables in which the table holds, and whether such variables exist is decided by the coupling functional.




\section{Asymptotic flatness in SVT theory}
\label{sec:flatness_SVT}

In Secs.~\ref{sec:matter-setup}--\ref{sec:matter-supp} we constructed the Bondi--Sachs hierarchy with a generic stress-energy tensor obeying only smooth asymptotic-flat falloffs, and in Sec.~\ref{sec:matter-bd} and Apps.~\ref{app:matter-scalar}--\ref{app:matter-maxwell} we analyzed particular instances of theories with extra fields and (non-)minimal coupling. We now adopt the complementary, agnostic viewpoint. Rather than fixing a particular matter model, we ask under what conditions each operator in the most general SVT effective Lagrangian \cite{Heisenberg:2018acv} preserves asymptotic flatness. The full SVT action (\cite{Heisenberg:2018acv}, also analyzed in \cite{Heisenberg:2023prj}) contains an arbitrary kinetic sector ($L_2,L_3$) together with two sectors ($L_4,L_5$) in which the scalar and vector couple non-minimally to the curvature itself. To handle these on equal footing we recast the metric equations of motion (EOM) as Einstein's equations sourced by an effective stress-energy $T^{\rm eff}_{\mu\nu}/G_4$ that absorbs all non-Einstein-tensor pieces, with $G_4$ one of the theory's coupling functionals, and then apply the falloff check to this $T^{\rm eff}_{\mu\nu}$. Note that we thereby assume that the metric functions obey their flatness constraints. All computations relevant to this section are explicitly shown in App.~\ref{app:svt-explicit}.

\subsection{Why asymptotic flatness?}
\label{sec:svt-motivation}

Before entering the analysis, we address an apparent tension in its very setup. The SVT class \cite{Heisenberg:2018acv}, like its Horndeski and generalized-Proca subclasses, is motivated primarily by dark-energy and dark-matter phenomenology \cite{Clifton:2011jh,Joyce:2014kja,CANTATA:2021ktz,Kase:2018aps,Frusciante:2019xia}, whose natural arena is a cosmological background with $\Lambda\neq0$ and spacelike conformal boundary, not Minkowski space. Why, then, is the asymptotically flat limit the right setting for the questions posed here? The motivation is threefold.

First, the generation of gravitational waves in a compact binary and their propagation through the local wave zone occupy scales negligible compared to the Hubble radius. For ground-based signals $\lambda_{\rm GW}/H_0^{-1}\sim10^{-20}$ \cite{ligoscientificcollaborationLALSuiteLIGOScientific2020}, and even for LISA sources $\lambda_{\rm GW}/H_0^{-1}\lesssim10^{-14}$ \cite{LISA}. Standard gravitational-wave practice in GR therefore models the source and its radiation in an asymptotically flat framework and accounts for the cosmological propagation entirely through redshifted masses and the luminosity distance \cite{poisson2014gravity,maggiore2008gravitational}. Corrections to the waveform from $\Lambda$ enter suppressed by powers of $\sqrt{\Lambda}\,r\ll1$ across the entire wave zone. The same logic applies verbatim to beyond-GR candidates. 
Furthermore, for the additional SVT fields to be phenomenologically viable in the first place, their fifth forces must be suppressed in local environments via the chameleon \cite{Khoury:2003aq,Khoury:2003rn} or Vainshtein \cite{Vainshtein:1972sx,Babichev:2013usa} mechanisms, or related constructions \cite{Brax:2013ida,Joyce:2014kja}. Precisely in the screened regime the extra fields settle to (approximately) constant environmental values with radiative $1/r$ perturbations on top, which is exactly the asymptotic ansatz employed in this work, cf.\ the BD example of Sec.~\ref{sec:matter-bd} with its constant background $\phi_0$. The asymptotically flat framework is thus not an idealization despite the dark-energy motivation but the correct effective description of the near-to-wave-zone geometry of a screened theory.
Finally, note that the multi-messenger constraint from GW170817 \cite{LIGOScientific:2017zic} on the tensor propagation speed has sharply narrowed the dark-energy-viable corner of the Horndeski/SVT landscape \cite{Creminelli:2017sry,Ezquiaga:2017ekz,Baker:2017hug,Sakstein:2017xjx,Langlois:2017dyl}. The surviving operator content lies within the class analyzed below, so the present analysis directly addresses the phenomenologically relevant parameter space.

We highlight that, nonetheless, the rigorous formulation of gravitational radiation with a positive cosmological constant remains under active development. For $\Lambda>0$ the conformal boundary is spacelike, no canonical analog of the Bondi news is available, and the asymptotic symmetry analysis is considerably subtler \cite{Ashtekar:2014zfa,Compere:2019bua}. 

\subsection{The full SVT action and the effective Einstein equation}
\label{sec:svt-action}

The full SVT Lagrangian \cite{Heisenberg:2018acv,Heisenberg:2023prj} decomposes as $L=L_2+L_3+ L_4+L_5$, where each piece reads, in exact correspondence with eqs.~(39)--(42) of \cite{Heisenberg:2023prj},
\begin{align}
\label{eq:svt-L2-full}
L_2 &= G_2(\Phi,X,Y,F,\tilde F),\\[2pt]
\label{eq:svt-L3-full}
L_3 &= -G_3(\Phi,X)\,\Box\Phi + \bigl[\hat G_3(\Phi,X)\,g_{\alpha\beta} + \mathring{\hat G}_3(\Phi,X)\,\Phi_\alpha\Phi_\beta\bigr]\, \tilde F^{\mu\alpha}F^{\nu\beta}\,\Phi_{\mu\nu},\\[2pt]
\label{eq:svt-L4-full}
L_4 &= G_4(\Phi,X)\,R + G_{4,X}\bigl[(\Box\Phi)^2 - \Phi^{\mu\nu}\Phi_{\mu\nu}\bigr] + \hat G_4(\Phi,X)\,L^{\mu\nu\alpha\beta}F_{\mu\nu}F_{\alpha\beta}\notag\\
&\quad + \bigl[\mathring{\hat G}_4(\Phi) + \tfrac12 \hat G_{4,X}(\Phi,X)\bigr]\,\tilde F^{\mu\alpha}\tilde F^{\nu\beta} \,\Phi_{\mu\nu}\Phi_{\alpha\beta},\\[2pt]
\label{eq:svt-L5-full}
L_5 &= G_5(\Phi,X)\,G^{\mu\nu}\Phi_{\mu\nu} - \tfrac{G_{5,X}}{6}\bigl[(\Box\Phi)^3 - 3\,\Box\Phi\,\Phi^{\mu\nu}\Phi_{\mu\nu} + 2\,\Phi_{\mu\nu}\Phi^{\nu\lambda}\Phi_\lambda{}^\mu\bigr],
\end{align}
with $\Phi_\mu\equiv\nabla_\mu\Phi$, $\Phi_{\mu\nu}\equiv\nabla_\mu\nabla_\nu \Phi$, the four SVT scalar invariants
\be
\label{eq:svt-building-blocks}
X \equiv -\tfrac12\Phi_\mu\Phi^\mu,\;\; Y \equiv \Phi_\mu\Phi_\nu F^{\mu\alpha}F^\nu{}_\alpha,\;\; F \equiv -\tfrac14 F^{\mu\nu}F_{\mu\nu},\;\; \tilde F \equiv F^{\mu\nu}\tilde F_{\mu\nu},
\ee
built from the scalar $\Phi$ and the field strength $F_{\mu\nu}=\nabla_\mu A_\nu-\nabla_\nu A_\mu$ together with its Hodge dual $\tilde F^{\mu\nu}= \tfrac12\epsilon^{\mu\nu\alpha\beta}F_{\alpha\beta}$, and the double-dual Riemann tensor
\be
\label{eq:doubledual-def}
L^{\mu\nu\alpha\beta} \equiv \tfrac14\epsilon^{\mu\nu\rho\sigma} \epsilon^{\alpha\beta\gamma\delta}R_{\rho\sigma\gamma\delta}.
\ee
The coupling functionals $G_2,G_3,\hat G_3,\mathring{\hat G}_3,G_4,\hat G_4,\mathring{\hat G}_4,G_5$ are otherwise arbitrary. Eqs.~\eqref{eq:svt-L2-full}--\eqref{eq:svt-L5-full} constitute the most general massless SVT theory with second-order EOM, and $L_5$ contains no further SVT mixings beyond the Horndeski quintic \cite{Heisenberg:2018acv}.

For the per-term variation of the action it is useful to record the algebraic dependence of each SVT scalar invariant on the inverse metric, namely
\begin{subequations}
\label{eq:bb-variations}
\begin{align}
\label{eq:bb-Phi}
\frac{\partial \Phi}{\partial g^{\mu\nu}} &= 0,\\
\label{eq:bb-X}
\frac{\partial X}{\partial g^{\mu\nu}} &= -\tfrac12\,\Phi_\mu\Phi_\nu, \\
\label{eq:bb-F}
\frac{\partial F}{\partial g^{\mu\nu}} &= -\tfrac12\,F_\mu{}^\alpha F_{\nu \alpha},\\
\label{eq:bb-Ftilde}
\frac{\partial \tilde F}{\partial g^{\mu\nu}} &= \tfrac12\,g_{\mu\nu}\,\tilde F, \\
\label{eq:bb-Y}
\frac{\partial Y}{\partial g^{\mu\nu}} &= 2\,\Phi_{(\mu}F_{\nu)}{}^\alpha F_\alpha{}^\beta\Phi_\beta + \Phi^\alpha\Phi^\beta F_{\mu\alpha}F_{\nu\beta}.
\end{align}
\end{subequations}
The scalar $\Phi$ has no algebraic metric dependence. The combination $\sqrt{-g}\,\tilde F$ is parity-odd and metric-independent, which fixes Eq.~\eqref{eq:bb-Ftilde} from $\delta(\sqrt{-g}\,\tilde F)=0$ combined with $\delta\sqrt{-g}/\sqrt{-g}=-\tfrac12 g_{\mu\nu}\delta g^{\mu\nu}$. The algebraic variations Eq.~\eqref{eq:bb-variations} combine, via the standard identity
\be
\label{eq:Tmunu-rule}
T_{\mu\nu} = -\frac{2}{\sqrt{-g}}\frac{\delta(\sqrt{-g}\,L)}{\delta g^{\mu\nu}} = -2\,\frac{\partial L}{\partial g^{\mu\nu}} + g_{\mu\nu}\,L \quad(\text{algebraic piece only}),
\ee
with the Palatini identity for the curvature factors $\delta R_{\mu\nu\alpha \beta}/\delta g^{\rho\sigma}$ and the Christoffel variation of the second-derivative tensor
\be
\label{eq:dPhi-Christoffel}
\delta\Phi_{\mu\nu}\big|_{\delta g} = -\,\delta\Gamma^\lambda_{\mu\nu}\, \Phi_\lambda,\quad \delta\Gamma^\lambda_{\mu\nu} = \tfrac12 g^{\lambda\rho}\bigl(\nabla_\mu \delta g_{\nu\rho} + \nabla_\nu\delta g_{\mu\rho} - \nabla_\rho\delta g_{\mu \nu}\bigr),
\ee
giving the stress-energy contribution of each Lagrangian term. Note that the matter fields $\Phi$, $A_\mu$, $F_{\mu\nu}$ are not varied.

Every operator in Eqs.~\eqref{eq:svt-L2-full}--\eqref{eq:svt-L5-full} that contains a $\Phi_{\mu\nu}$ factor can be written in the form $\mathcal L = K^{\mu\nu}\, \Phi_{\mu\nu}$ for some symmetric tensor $K^{\mu\nu}$ built from the matter--field content. Inserting Eq.~\eqref{eq:dPhi-Christoffel} into $\delta(\sqrt{-g}\,K^{\mu\nu}\Phi_{\mu\nu})$ and integrating by parts produces
\be
\label{eq:Palatini-master}
\;\mathcal D\bigl[K^{\alpha\beta}\bigr]_{\mu\nu} \;\equiv\; 2\,\nabla_\lambda\!\bigl(K^\lambda{}_{(\mu}\,\Phi_{\nu)}\bigr) - \nabla_\lambda\!\bigl(K_{\mu\nu}\,\Phi^\lambda\bigr),\;
\ee
with $K^{\mu\nu}$ implicitly symmetrized in $\mu\!\leftrightarrow\!\nu$ (only the symmetric part contracts with $\Phi_{\mu\nu}$). Eq.~\eqref{eq:Palatini-master} reproduces $\mathcal D[g^{\alpha\beta}]_{\mu\nu}= 2\Phi_{\mu\nu}-g_{\mu\nu}\Box\Phi$ as a consistency check, since for $K^{\mu \nu}=g^{\mu\nu}$ one has $\mathcal L=\Box\Phi$ and this $\mathcal D$ combines with the algebraic pieces $-2\delta g^{\mu\nu}\Phi_{\mu\nu}/\delta g^{\mu\nu}$ and $g_{\mu\nu}\Box\Phi$ to give vanishing total stress-energy, as required for a total divergence.

Applied operator by operator to $L_2+L_3+L_4+L_5$, these two rules, i.e. the algebraic variation Eq.~\eqref{eq:Tmunu-rule} and the Palatini integration-by-parts master formula Eq.~\eqref{eq:Palatini-master}, produce one stress-energy contribution per Lagrangian structure, i.e., an algebraic piece built directly from the field content, plus, whenever the operator carries a $\Phi_{\mu\nu}$ or curvature factor, a derivative remainder of the form Eq.~\eqref{eq:Palatini-master}. The resulting tensors $T^{(2)}_{\mu\nu}$, $T^{(3,\Box)}_{\mu\nu}$, $T^{(3,\hat G_3)}_{\mu\nu}$, $T^{(3,\mathring{\hat G}_3)}_{\mu\nu}$, $T^{(4,\rm kin)}_{\mu\nu}$, $T^{(4,\hat G_4)}_{\mu\nu}$, $T^{(4,\mathring{\hat G}_4)}_{\mu\nu}$, $T^{(5)}_{\mu\nu}$ and the curvature--scalar tensor $\mathcal E^{(5)}_{\mu\nu}$ are recorded in closed form in App.~\ref{app:svt-stress}. The falloff analysis below requires only their schematic structure, and we will quote individual components from the appendix where needed.

Collecting all sectoral contributions, the effective stress-energy is
\begin{align}
\label{eq:Teff-def}
T^{\rm eff}_{\mu\nu} &= T^{(2)}_{\mu\nu} + T^{(3,\Box)}_{\mu\nu} + T^{(3,\hat G_3)}_{\mu\nu} + T^{(3,\mathring{\hat G}_3)}_{\mu\nu}
 + T^{(4,\rm kin)}_{\mu\nu} + T^{(4,\hat G_4)}_{\mu\nu} + T^{(4,\mathring{\hat G}_4)}_{\mu\nu} + T^{(5)}_{\mu\nu}\notag\\
&\quad - \bigl(g_{\mu\nu}\Box - \nabla_\mu\nabla_\nu\bigr)G_4 - \mathcal E^{(5)}_{\mu\nu},
\end{align}
with the explicit forms recorded in App.~\ref{app:svt-stress} (Eqs.~\eqref{eq:T2-explicit}, \eqref{eq:T3-Hor-explicit}--\eqref{eq:T3-mix-explicit}, \eqref{eq:T4kin-explicit}--\eqref{eq:T4mathringG4-explicit} and \eqref{eq:T5-explicit}, each $T^{(\cdot)}_{\mu\nu}$ containing its algebraic piece plus the Palatini derivative remainder Eqs.~\eqref{eq:D3-explicit}, \eqref{eq:D4kin-explicit}, \eqref{eq:D4hatG4-explicit}, or \eqref{eq:D4mGhat4-explicit}, and the $\delta G^{\mu\nu}$ curvature--scalar piece given by Eq.~\eqref{eq:E5-explicit}). The metric field equation then takes the Einstein form
\be
\label{eq:svt-Einstein}
{\;G_{\mu\nu} \;=\; \frac{1}{G_4(\Phi,X)}\,T^{\rm eff}_{\mu\nu}.\;}
\ee
Invertibility of Eq.~\eqref{eq:svt-Einstein} requires $G_4(\Phi,X)\neq 0$ throughout the spacetime. For the flatness analysis, which refers entirely to $\scrip$ where the building blocks satisfy $(\Phi,X)\to(\phi_0,0)$, this reduces to the boundary condition
\be
\label{eq:G4-asymp-nonzero}
G_4(\phi_0,0) \;\neq\; 0 .
\ee
The SVT action \cite{Heisenberg:2018acv,Heisenberg:2023prj} carries an overall prefactor $1/(2\kappa_0)$ with $\kappa_0\equiv 8\pi G_0$ and $G_0$ the bare gravitational constant, so the coefficient of $R$ in the Lagrangian is $G_4(\Phi,X)/(2\kappa_0)$. Comparing with Einstein--Hilbert $R/(16\pi G_N)$, $G_4(\Phi,X)$ is a dimensionless rescaling of the inverse bare coupling as it multiplies the inverse Newton constant. At $\scrip$ the field-independent limit $G_4(\phi_0,0)$ therefore sets the asymptotic Newton constant measured by a distant observer,
\be
\label{eq:GN-asymp}
G_N^{\rm asymp} \;=\; \frac{G_0}{G_4(\phi_0,0)}.
\ee
Positivity $G_4(\phi_0,0)>0$ ensures positive asymptotic kinetic energy for the graviton (no ghost). The value $G_4(\phi_0,0)=1$ recovers Einstein gravity at $\scrip$ with $G_N^{\rm asymp}=G_0$. Finally, Eq.~\eqref{eq:GN-asymp} calibrates the mass aspect and angular-momentum aspect to physical units. Bulk variations of $G_4(\Phi,X)$ near matter sources rescale gravity locally but do not affect the asymptotic identification. Under Eq.~\eqref{eq:G4-asymp-nonzero} the Bondi--Sachs hierarchy of Secs.~\ref{sec:matter-beta}--\ref{sec:matter-supp} applies verbatim to the SVT theory after the substitution
\be
\label{eq:Teff-substitution}
T_{\mu\nu} \;\longmapsto\; \frac{T^{\rm eff}_{\mu\nu}}{G_4(\Phi,X)}
\ee
throughout. A remark on normalization is in order here. The substitution Eq.~\eqref{eq:Teff-substitution} identifies the source of the effective Einstein equation \eqref{eq:svt-Einstein}, which carries no factor of $8\pi$, with the source $8\pi\,T_{\mu\nu}$ of Eq.~\eqref{eq:einstein}. Accordingly, throughout this section and App.~\ref{app:svt-explicit} the coefficients $\Tc{n}_{ab}$ denote the $1/r^{n}$ expansion coefficients of $T^{\rm eff}_{ab}/G_4$ itself and differ from their counterparts in Sec.~\ref{sec:matter} by a factor of $8\pi$. For BD, for instance, $\bigl[T^{\rm eff}_{rr}/G_4\bigr]^{(3)}=2\varphi_1/\phi_0 =8\pi\,\Tc{3}_{rr}$, with $\Tc{3}_{rr}=\varphi_1/(4\pi\phi_0)$ from Eq.~\eqref{eq:bd-Tab-summary}. Since the falloff table and every relation quoted below are linear in the source, this rescaling affects neither the falloff orders nor the conditions derived from them. The asymptotic-flatness conditions on the SVT coupling functionals now follow by demanding that the rescaled effective stress-energy respect the falloff table Eq.~\eqref{eq:T-falloff}, which we put to a test sector by sector.

\subsection{Asymptotic-flatness conditions on the SVT coupling functionals}
\label{sec:svt-flatness}

We apply a self-consistency check: we insert the leading asymptotic-flat metric values
\begin{align}
\label{eq:svt-leading-metric}
&g_{uu} = g_{ur} = -1,\quad g_{rr}=g_{uA}=g_{rA}=0,\quad g_{AB}=r^2 q_{AB}, \notag\\
&g^{uu}=g^{uA}=g^{rA}=0,\;\; g^{ur}=-1,\;\; g^{rr}=1,\;\; g^{AB}=q^{AB}/r^2,
\end{align}
together with the radiative ansatz $\Phi=\phi_0+\phi_1(u,x^C)/r+\mathcal O(r^{-2})$, $A_u=q(u,x^C)/r+\mathcal O(r^{-2})$, $A_A=\mathcal A_A^{(0)}+\mathcal A_A^{(1)}/r+ \dots$, and verify that $T^{\rm eff}_{\mu\nu}/G_4$ respects Eq.~\eqref{eq:T-falloff}. Here $\phi_0$ is the asymptotic scalar value. Since the SVT coupling functionals depend on $\Phi$ undifferentiated it is not pure gauge (in contrast to the free scalar of App.~\ref{app:matter-scalar}), and we keep it general, $\phi_0\neq0$. From the radiative expansion, it follows immediately that the four SVT scalar invariants Eq.~\eqref{eq:svt-building-blocks} have leading decays
\be
\label{eq:svt-buildingblocks-orders}
\Phi-\phi_0=\mathcal O(r^{-1}),\;\; X=\mathcal O(r^{-3}),\;\; F=\mathcal O(r^{-4}),\;\; \tilde F=\mathcal O(r^{-4}),\;\; Y=\mathcal O(r^{-6}),
\ee
while the curvature blocks acquire their slowest orders from the radiative shear $c_{AB}$ at $\scrip$,
\be
\label{eq:svt-curvature-orders}
R_{\mu\nu\alpha\beta},\,G_{\mu\nu},\,L^{\mu\nu\alpha\beta} = \mathcal O(r^{-1}),
\ee
sourced by the news $N_{AB}=\tfrac12\partial_u c_{AB}$. Crucially, the leading decays Eq.~\eqref{eq:svt-buildingblocks-orders} are {independent of $\phi_0$} as every building block $X,\Phi_{\mu\nu},F,\tilde F,Y$ is a derivative of $\Phi$, so the constant $\phi_0$ drops out identically. The asymptotic value enters the effective stress only through the arguments of the coupling functionals. We accordingly expand each $G_i$ of Eqs.~\eqref{eq:svt-L2-full}--\eqref{eq:svt-L5-full} in a regular Taylor series about the asymptotic vacuum
\be
\label{eq:svt-vacuum}
\bigl(\Phi,X,Y,F,\tilde F\bigr)\big|_{\scrip}=(\phi_0,0,0,0,0),\qquad \phi_0\neq0\ \text{arbitrary},
\ee
and write a subscript $0$ for evaluation there, i.e.\ $G_i|_0\equiv G_i(\phi_0,0,0,0,0)$ and $\partial_\Phi G_i|_0\equiv\partial_\Phi G_i(\phi_0,0,0,0,0)$, and so on. Equivalently, the parenthetical forms $G_i(\phi_0)$, $G_{i,\Phi}(\phi_0)$, $G_4(\phi_0,0)$ used below denote the same vacuum values, with only the leading arguments displayed and all remaining invariants set to their vanishing vacuum values. At any given order in $1/r$ only finitely many such coefficients enter. Because the building-block decays do not see $\phi_0$, every falloff condition derived below holds verbatim for arbitrary $\phi_0$, reducing to the special case $\phi_0=0$ only for a sector that happens to be minimally coupled. For BD, $G_4=\Phi$ gives $G_4|_0=\phi_0$, so a non-trivial vacuum $\phi_0\neq0$ is exactly the finite-Newton-constant requirement of Sec.~\ref{sec:matter-bd}.

Let us now state the working assumptions on the coupling functionals explicitly. Throughout this subsection each $G_i$ is taken to be (i) a fixed function of the invariants $(\Phi,X,Y,F,\tilde F)$ alone, with no explicit spacetime dependence, as diffeomorphism invariance requires, and (ii) regular at the asymptotic vacuum Eq.~\eqref{eq:svt-vacuum}, i.e.\ analytic there, so that the Taylor expansion above exists with finite coefficients and, at any given order in $1/r$, only finitely many of them enter. Beyond (i) and (ii), and the invertibility condition $G_4(\phi_0,0)\neq0$ of Sec.~\ref{sec:svt-action}, no conditions are imposed. Away from the vacuum the coupling functionals remain completely arbitrary. Assumption (ii) is provisional and relaxed in Secs.~\ref{sec:growth-metric}--\ref{sec:growth-eom}, where singular power-law behavior at the vacuum is admitted.

We note a further ingredient recurring throughout the analysis, i.e., the covariant Hessian of the scalar, computed directly from $\Phi_{\mu\nu}=\partial_\mu\partial_\nu\Phi-\Gamma^\lambda_{\mu\nu} \partial_\lambda\Phi$ on the leading metric Eq.~\eqref{eq:svt-leading-metric},
\be
\label{eq:Phi-uv-orders}
\begin{aligned}
&\Phi_{uu}=\ddot\phi_1/r,\;\;\Phi_{uA}=\partial_A\dot\phi_1/r,\;\; \Phi_{ur}=-\dot\phi_1/r^2,\\
&\Phi_{rA}=-2\partial_A\phi_1/r^2,\;\;\Phi_{rr}=2\phi_1/r^3,\;\; \Phi_{AB}=-q_{AB}\dot\phi_1+\mathcal O(r^{-1}),
\end{aligned}
\ee
i.e.\ $\Phi_{uu},\Phi_{uA}=\mathcal O(r^{-1})$, $\Phi_{ur},\Phi_{rA}=\mathcal O(r^{-2})$, $\Phi_{rr}=\mathcal O(r^{-3})$, and crucially $\Phi_{AB}=\mathcal O(r^0)$, with leading coefficient $-q_{AB}\dot\phi_1$ supplied by the Christoffel correction $\Gamma^u_{AB}\partial_u\Phi=r q_{AB}\dot\phi_1/r$. The trace $g^{AB}\Phi_{AB}=-2\dot\phi_1/r^2$ exactly cancels $2g^{ur}\Phi_{ur}= 2\dot\phi_1/r^2$ in $\Box\Phi$, leaving $\Box\Phi=2\phi_1/r^3=\mathcal O(r^{-3})$ as expected for a free outgoing wave.

We start our analysis with the $1/G_4$ prefactor. Taylor-expanding the gravitational coupling functional,
\be
\label{eq:G4-expansion}
G_4 = G_4(\phi_0) + G_{4,\Phi}(\phi_0)\,\frac{\phi_1}{r} + G_{4,X}(\phi_0)\,X + \dots = G_4(\phi_0) + \mathcal O(r^{-1}),
\ee
gives $$ \frac{1}{G_4} = \frac{1}{G_4(\phi_0)}\Bigl[1 - \frac{G_{4,\Phi}(\phi_0)}{G_4(\phi_0)} \frac{\phi_1}{r} + \mathcal O(r^{-2})\Bigr]. $$ The prefactor therefore contributes only sub-leading corrections to each component of $T^{\rm eff}_{\mu\nu}/G_4$ and does not alter the falloff orders of $T^{\rm eff}_{\mu\nu}$ itself.

\subsubsection{$L_2$ sector.}
The trace piece $g_{\mu\nu}(G_2-G_{2,\tilde F}\tilde F)$ produces $T^{\rm eff}_{ur}\supset -(G_2-G_{2,\tilde F}\tilde F)$ at leading order. The strictest table entry $T_{ur}=\mathcal O(r^{-3})$ then forces $G_2$ to vanish to cubic order in the deviation $\Phi-\phi_0=\mathcal O(r^{-1})$ from the vacuum,
\be
\label{eq:flat-G2}
\;G_2\big|_0=0,\qquad \partial_\Phi G_2\big|_0=0,\qquad \partial_\Phi^2 G_2\big|_0=0.\;
\ee
These three conditions are the standard absence of a cosmological constant, a tadpole, and a mass term in the scalar sector, evaluated at the vacuum Eq.~\eqref{eq:svt-vacuum}. The tadpole condition $\partial_\Phi G_2|_0=0$ is, equivalently, the statement that $\Phi=\phi_0$ is a critical point of the scalar potential i.e.\ the true asymptotic vacuum of the theory rather than merely a convenient expansion point. A non-zero $\partial_\Phi G_2|_0$ would source the scalar EOM at $\Phi=\phi_0$ and make the radiative ansatz $\Phi=\phi_0+\phi_1/r+\mathcal O(r^{-2})$ dynamically inconsistent. The avoidance of the mass term simply follows from the requirement of a well-defined $1/r$ expansion.

\subsubsection{$L_3$ sector.}
Integration by parts gives $\sqrt{-g}\,(-G_3\,\Box\Phi) = \sqrt{-g}\,\nabla^\mu G_3\,\nabla_\mu\Phi + \text{total divergence}$, so $G_3(\phi_0,0)$ itself never enters $T_{\mu\nu}$. Only the derivative coefficients $G_{3,\Phi}(\phi_0,0)$ and $G_{3,X}(\phi_0,0)$ do, via Eq.~\eqref{eq:T3-Hor-explicit}. With $\Phi_\mu\Phi_\nu = \mathcal O(r^{-2})$ and $Xg_{\mu\nu}=\mathcal O(r^{-3})$,
\begin{align}
T^{(3,\Box)}_{uu} &\supset -2G_{3,\Phi}\bigl(\Phi_u^2 + X g_{uu}\bigr) = -2G_{3,\Phi}\,\dot\phi_1^2/r^2 + \mathcal O(r^{-3}) = \mathcal O(r^{-2}),\notag\\
T^{(3,\Box)}_{ur} &\supset -2G_{3,\Phi}\bigl(\Phi_u\Phi_r + X g_{ur}\bigr) = \mathcal O(r^{-4}),
\end{align}
where the second line collapses to $\mathcal O(r^{-4})$ because $\Phi_u\Phi_r=-\phi_1\dot\phi_1/r^3$ and $Xg_{ur}=+\phi_1\dot\phi_1/r^3$ cancel at $\mathcal O(r^{-3})$. The $G_{3,X}$ structure contributes $\Box\Phi\, \Phi_\mu\Phi_\nu = \mathcal O(r^{-4})$ and similar. The slowest entry $T_{uu}= \mathcal O(r^{-2})$ saturates but does not violate its bound. All other components fall faster. Hence $G_{3,\Phi}(\phi_0,0),\,G_{3,X}(\phi_0,0),\,G_3(\phi_0,0)$ are all unrestricted, and so are all higher Taylor coefficients of $G_3$ at the vacuum (they multiply higher powers of $\Phi-\phi_0$ and $X$, giving faster decay). Unrestricted refers to the values of the Taylor coefficients within the class of assumption (ii): the coupling functional is analytic at the vacuum, so that on the radiative background it reduces to an expansion in non-negative powers of $1/r$, and asymptotic flatness imposes no condition whatsoever on the coefficients of that expansion. It is a statement about the freedom of the functional within the regular class, not an admission of behavior beyond it.

The parity-odd mixing $\bigl[\hat G_3\,g_{\alpha\beta}+\mathring{\hat G}_3\,\Phi_\alpha\Phi_\beta\bigr]\tilde F^{\mu\alpha}F^{\nu\beta} \Phi_{\mu\nu}$ has the bilinear structure $K_3^{\mu\nu}\Phi_{\mu\nu}$ of Eq.~\eqref{eq:K3-def}, so its stress-energy consists of the algebraic tensors Eqs.~\eqref{eq:T3-Ghat3-algebraic}--\eqref{eq:T3-mGhat3-algebraic} plus the derivative remainder $\mathcal D^{(3)}_{\mu\nu}$ of Eq.~\eqref{eq:D3-explicit}. Inserting the Hessian orders Eq.~\eqref{eq:Phi-uv-orders} together with the mixed-index field-strength orders of the radiative vector ansatz, a component-by-component scan carried out in App.~\ref{app:svt-components} returns the slowest entries $T^{(3,\hat G_3)}_{AB}=\mathcal O(r^{-3})$ and $\mathcal D^{(3)}_{uu},\mathcal D^{(3)}_{uA}=\mathcal O(r^{-4})$ (Eqs.~\eqref{eq:T3-Ghat3-orders} and \eqref{eq:D3-orders}), each strictly below the corresponding bound of Eq.~\eqref{eq:T-falloff}. The $\mathring{\hat G}_3$ structures, which carry an additional $\Phi^\alpha\Phi^\beta$ pair, decay at least two powers faster still. Asymptotic flatness therefore imposes no condition on the $L_3$ mixing functionals at the vacuum, $ \hat G_3(\Phi,X) \text{ and }\mathring{\hat G}_3(\Phi,X) $ are unrestricted at $(\Phi,X)=(\phi_0,0)$. The mechanism originates from the $\Phi_{\mu\nu}$ factor that contracts the two field strengths in Eq.~\eqref{eq:svt-L3-full}. It supplies enough $1/r$ suppression that even a non-vanishing $\hat G_3(\phi_0,0)$ produces a stress-energy compatible with the asymptotic hierarchy.

\subsubsection{$L_4$ Horndeski sector.}
Invertibility of Eq.~\eqref{eq:svt-Einstein} and a finite asymptotic Newton constant require
\be
\label{eq:flat-G4}
\;G_4(\phi_0,0)\;\neq\;0.\;
\ee
The kinetic Horndeski piece $G_{4,X}\bigl[(\Box\Phi)^2-\Phi^{\mu\nu} \Phi_{\mu\nu}\bigr]$ is quadratic in the Hessian, which is precisely what protects it. With $\Box\Phi=2\phi_1/r^3$ and $\Phi^{\mu\nu}\Phi_{\mu\nu}=\mathcal O(r^{-4})$, the component scan of its stress-energy Eq.~\eqref{eq:T4kin-explicit} performed in App.~\ref{app:svt-components} and summarized in Eq.~\eqref{eq:T4kin-orders} shows that the slowest entry is a pure-trace $T^{(4,\rm kin)}_{AB}=\mathcal O(r^{-2})\,q_{AB}$, with every covariant component below its bound in Eq.~\eqref{eq:T-falloff}. The coefficient $G_{4,X}(\phi_0,0)$ is therefore unconstrained.

The geometric remainder $-(g_{\mu\nu}\Box-\nabla_\mu\nabla_\nu)G_4$ in Eq.~\eqref{eq:Teff-def} sources $T^{\rm eff}_{\mu\nu}$ through
\be
\label{eq:nabla-nabla-G4}
\nabla_\mu\nabla_\nu G_4 = G_{4,\Phi}\,\Phi_{\mu\nu} + G_{4,X}\,\nabla_\mu\nabla_\nu X + G_{4,\Phi\Phi}\,\Phi_\mu\Phi_\nu \,+\,2\,G_{4,\Phi X}\,\Phi_{(\mu}\nabla_{\nu)}X + G_{4,XX}\,\nabla_\mu X\,\nabla_\nu X .
\ee
At the orders relevant below only the vacuum values of the coefficients enter, the Taylor corrections contributing strictly higher powers of $1/r$. The slowest term is $G_{4,\Phi}(\phi_0)\,\Phi_{AB}=-G_{4,\Phi}(\phi_0)\,q_{AB}\dot \phi_1+\mathcal O(r^{-1})$ from the Christoffel contribution $\Gamma^u_{AB}\partial_u\Phi=rq_{AB}\cdot\dot\phi_1/r=q_{AB}\dot\phi_1$ to the covariant Hessian. The leading $\mathcal O(r^0)$ piece is purely $\propto q_{AB}$ and contributes only to the trace $h^{AB}T_{AB}$, which the falloff table Eq.~\eqref{eq:T-falloff} allows to be $\mathcal O(r^0)$ via $-\partial_u\Tc{3}_{rr}+\mathcal O(r^{-1})$, and the trace-free part is $\mathcal O(r^{-1})$, saturating but not violating $T^{TF}_{AB}=\mathcal O(r^{-1})$. Likewise $g_{AB}\Box G_4=r^2 q_{AB}\cdot \mathcal O(r^{-3})=\mathcal O(r^{-1})\cdot q_{AB}$ contributes only to the trace.

The remaining components of $G_{4,\Phi}(\phi_0)\,\Phi_{\mu\nu}$, by contrast, are not all table-safe. Reading off the covariant Hessian Eq.~\eqref{eq:Phi-uv-orders}, its $u$-components are the slowest: $\Phi_{uu}=\ddot\phi_1/r$ and $\Phi_{uA}=\partial_A\dot\phi_1/r$, both $\mathcal O(r^{-1})$, and $\Phi_{ur}=-\dot\phi_1/r^2=\mathcal O(r^{-2})$. Hence, whenever the scalar radiates ($\dot\phi_1\neq0$) and the coupling is genuinely non-minimal ($G_{4,\Phi}(\phi_0)\neq0$), the effective source $G_{\mu\nu}=T^{\rm eff}_{\mu\nu} /G_4$ carries
\be
\label{eq:G4-violation}
\begin{aligned}
\Big[\tfrac{T^{\rm eff}}{G_4}\Big]_{uu}&=\frac{G_{4,\Phi}(\phi_0)}{G_4(\phi_0)} \frac{\ddot\phi_1}{r}+\mathcal O(r^{-2}),\qquad \Big[\tfrac{T^{\rm eff}}{G_4}\Big]_{uA}=\frac{G_{4,\Phi}(\phi_0)}{G_4(\phi_0)} \frac{\partial_A\dot\phi_1}{r}+\mathcal O(r^{-2}),\\
\Big[\tfrac{T^{\rm eff}}{G_4}\Big]_{ur}&=-\frac{G_{4,\Phi}(\phi_0)}{G_4(\phi_0)} \frac{\dot\phi_1}{r^2}+\mathcal O(r^{-3}),
\end{aligned}
\ee
each one power slower than the bounds $T_{uu},T_{uA}=\mathcal O(r^{-2})$, $T_{ur}=\mathcal O(r^{-3})$ of Eq.~\eqref{eq:T-falloff}. The $rr$ and $rA$ components saturate via $\Phi_{rr}=\mathcal O(r^{-3})$, $\Phi_{rA}=\mathcal O(r^{-2})$, and the $X$- and $\Phi\Phi$-terms of Eq.~\eqref{eq:nabla-nabla-G4} enter one order lower, the mixed $\Phi X$- and the $XX$-term lower still. This is the unique table violation produced anywhere in the SVT action: $G_4(\Phi)R$ is the only operator whose coupling functional multiplies the curvature directly, so that its variation yields the curvature-unsuppressed improvement Hessian $\nabla_\mu\nabla_\nu G_4$. Every other operator carries an intrinsic suppression at the slow $u$-components. The kinetic Horndeski piece (Eq.~\eqref{eq:T4kin-orders}) is quadratic in $\Phi_{\mu\nu}$, and the raised-index contractions $\Phi_\mu{}^\alpha=\mathcal O(r^{-2})$ or faster kill the $\mathcal O(r^0)$ entry. The $G_5$ curvature coupling and the vector--curvature mixings $\hat G_4,\mathring{\hat G}_4$ each carry a Riemann/Einstein factor $\mathcal O(r^{-1})$ that vanishes on the leading metric (as the respective operator paragraphs and the component scans of App.~\ref{app:svt-components} make explicit). The cubic $L_3$ operator, finally, reduces after integration by parts to first-derivative $\Phi_a\Phi_b$ structures. The vector deserves explicit mention here, since its leading field strength is itself time-dependent and slow, $F^r{}_A=-\dot{\mathcal A}^{(0)}_A=\mathcal O(r^0)$ yet it never appears as a bare improvement term. It enters $T^{\rm eff}$ only contracted with the curvature $L^{\mu\nu\alpha\beta}=\mathcal O(r^{-1})$ or as a field-strength bilinear whose upper-index contractions decay, so no slow $u$-component survives and the vector sector produces no analogous obstruction.

The violation is removable and does not obstruct asymptotic flatness, but the manner of its removal depends on the coupling functional, and two mechanisms must be distinguished.

{(i) Generic $G_4(\Phi,X)$: mass-aspect redefinition.} The three coefficients in Eq.~\eqref{eq:G4-violation} are total retarded-time derivatives of the radial stress,
\be
\label{eq:G4-schott}
\Tc{2}_{ur}=-\tfrac12\partial_u\Tc{3}_{rr},\quad \Tc{1}_{uu}=\tfrac12\partial_u^2\Tc{3}_{rr},\quad \Tc{1}_{uA}=\tfrac12\partial_u\partial_A\Tc{3}_{rr},\qquad \Tc{3}_{rr}=\frac{2G_{4,\Phi}(\phi_0)}{G_4(\phi_0)}\,\phi_1 ,
\ee
which are precisely the structures that the general analysis of Secs.~\ref{sec:matter-V}--\ref{sec:matter-supp} absorbs into a redefinition of the mass aspect rather than into genuine flux. Being total $u$-derivatives they integrate to zero over any radiation burst and transport no net energy across $\scrip$. This mechanism operates for any $G_4(\Phi,X)$, because the offending term $G_{4,\Phi}(\phi_0)\Phi_{\mu\nu}$ is fixed by the single coefficient $G_{4,\Phi}(\phi_0)$ and is insensitive to the $X$-dependence of $G_4$ (the $G_{4,X}$ and $G_{4,\Phi\Phi}$ pieces of Eq.~\eqref{eq:nabla-nabla-G4} enter one order lower). At the price of a $u$-dependent mass aspect and modified metric falloffs, a consistent Bondi--Sachs expansion therefore exists in every case, as in the scalar--tensor analyses of \cite{tahura_brans-dicke_2021,Hou:2021bxz}.

{(ii) Pure scalar--tensor, $G_4=G_4(\Phi)$: passage to the Einstein frame.} When $G_4$ depends on $\Phi$ alone, the entire improvement term $\nabla_\mu\nabla_\nu G_4-g_{\mu\nu}\Box G_4$ is precisely what the conformal rescaling $\tilde g_{\mu\nu}=[G_4(\Phi)/G_4(\phi_0)]\,g_{\mu\nu}$ removes. Its factor is $1+\mathcal O(r^{-1})$, so it leaves $\scrip$ and the BMS group untouched, and in the resulting Einstein frame the gravitational sector is minimally coupled and $\tilde T^{\rm eff}_{\mu\nu}$ obeys the standard falloff table with no violation whatsoever. The theory is manifestly asymptotically flat and the BD case $G_4=\Phi$ carries this out component by component in Sec.~\ref{sec:matter-bd}. This route is, however, not available once $G_4$ depends genuinely on $X$. A rescaling factor $\Omega^2(\Phi,X)$ would be a disformal transformation, under which $G_4(\Phi,X)R$ maps not to Einstein gravity but to another Horndeski/DHOST theory. For such coupling functionals only mechanism (i) applies, and the standard BMS arena is recovered not by a frame choice but only after the mass-aspect redefinition.

Hence, beyond Eq.~\eqref{eq:flat-G4}, the Horndeski $G_4$ Taylor coefficients $G_{4,\Phi}(\phi_0),G_{4,X}(\phi_0)$ at the vacuum remain unconstrained. $G_{4,X}(\phi_0)$ contributes only the pure-trace pieces identified above, which stay within the marginal trace slot of the falloff table. The coefficient $G_{4,\Phi}(\phi_0)$ is the conformal (BD) coupling, admissible provided the standard Bondi--Sachs identifications are read off from the Einstein-frame metric $\tilde g_{\mu\nu}$ rather than the Jordan-frame $g_{\mu\nu}$ whenever the scalar radiates.

\subsubsection{$L_4$ SVT mixings.}
Both SVT mixings of $L_4$ are scanned component by component in App.~\ref{app:svt-components}. Here we record the mechanism and the outcome. For the parity-even mixing $\hat G_4\,L^{\mu\nu\alpha\beta} F_{\mu\nu}F_{\alpha\beta}$ the decisive fact is geometric. The double-dual Riemann tensor vanishes identically on the leading metric Eq.~\eqref{eq:svt-leading-metric} and is sourced only at $\mathcal O(r^{-1})$ by the radiative shear, so every contribution inherits a curvature suppression on top of the counting rule ``one power $1/r^{2}$ per raised angular index''. The slowest surviving components are collected in Eq.~\eqref{eq:T4hatG4-orders}. The second mixing $\bigl[\mathring{\hat G}_4+\tfrac12\hat G_{4,X}\bigr]\tilde F^{\mu\alpha} \tilde F^{\nu\beta}\Phi_{\mu\nu}\Phi_{\alpha\beta}$ carries no explicit curvature but is bilinear in the Hessian. Its slowest Lagrangian contraction is $\tilde F^{rA}\tilde F^{rB}\Phi_{rr}\Phi_{AB}=\mathcal O(r^{-7})$ (Eq.~\eqref{eq:L4-mGhat4-Lagrangian-leading}), which translates into stress-energy components that are everywhere $\mathcal O(r^{-5})$ or faster.

Combining the two mixings, every covariant component of $T^{(4,\hat G_4)}_{\mu\nu}+T^{(4,\mathring{\hat G}_4)}_{\mu\nu}$ satisfies Eq.~\eqref{eq:T-falloff}, with every component strictly below its bound. Hence $\hat G_4(\Phi,X),\,\mathring{\hat G}_4(\Phi)$ are unrestricted (in the sense used above for $G_3$) at the vacuum, i.e., $\Phi=\phi_0,X=0$, due by two major factors: (i) the double-dual Riemann tensor $L^{\mu\nu\alpha\beta}$ vanishes on the strict asymptotic-flat metric and is sourced only by radiation, automatically supplying the $1/r$ suppression needed to accommodate any finite $\hat G_4(\phi_0,0)$; (ii) the second mixing carries an explicit $\Phi_{\mu\nu}^2$ factor whose slowest entries (governed by $\Phi_{AB}=\mathcal O(r^0)$ paired against $\Phi_{rr}= \mathcal O(r^{-3})$) supply Lagrangian-level suppression of order $r^{-7}$, translating into stress-energy components that are everywhere $\mathcal O(r^{-5})$ or faster. Since the slowest entries of Eq.~\eqref{eq:T4hatG4-orders} lie at least one power below every bound, a constant $\hat G_4(\phi_0,0)$ never reaches the $\mathcal O(r^{-2})$ coefficients that source the Bondi mass-loss law Eq.~\eqref{eq:massloss} and the angular-momentum-aspect evolution Eq.~\eqref{eq:L-evol}. Both leading balance laws remain untouched.

\subsubsection{$L_5$ Horndeski sector.}
The quintic sector contributes through the curvature--scalar tensor $\mathcal E^{(5)}_{\mu\nu}$ of Eq.~\eqref{eq:E5-explicit} together with the $G_{5,\Phi}$- and $G_{5,X}$-pieces of $T^{(5)}_{\mu\nu}$, Eq.~\eqref{eq:T5-explicit}. Its analysis, spelled out in App.~\ref{app:svt-components}, combines the same two suppression mechanisms already at work above: the Einstein tensor vanishes on the leading metric and reappears only at $\mathcal O(r^{-1})$ through the radiative shear, while every remaining structure is at least bilinear in the Hessian or in the gradient $\Phi_a$. The Palatini remainder is governed by the scalar $S\equiv G_5\,\Box\Phi-G_{5,X}\,\Phi^\alpha\Phi^\beta\Phi_{\alpha\beta} -G_{5,\Phi}\,X=\mathcal O(r^{-3})$ (Eq.~\eqref{eq:S-def}), whose covariant Hessian and d'Alembertian are evaluated explicitly in the appendix. The slowest sectoral entries are collected in Eq.~\eqref{eq:E5-orders}.

Combining all pieces, every covariant component of $T^{(5)}_{\mu\nu}$ satisfies its corresponding bound in Eq.~\eqref{eq:T-falloff}, and $ G_5(\Phi,X) $ unrestricted at the vacuum. A non-vanishing $G_5(\phi_0,0)$ enters the stress-energy only one power below the leading flux entries, through $\partial_u^2 S\propto G_5(\phi_0,0)\ddot\phi_1/r^3$ in the $(u,u)$ component and $\partial_u\partial_r S\propto G_5(\phi_0,0)\dot\phi_1/r^4$ in the $(u,r)$ component (Eq.~\eqref{eq:E5-Lich-table}). It therefore leaves the Bondi mass-loss law Eq.~\eqref{eq:massloss} and the angular-momentum-aspect evolution Eq.~\eqref{eq:L-evol} unchanged at leading order. The derivatives $G_{5,X}(\phi_0,0)$ and $G_{5,\Phi}(\phi_0,0)$ enter even more weakly.

\subsubsection{Combining sectors.}
Based purely on the requirement of asymptotic-flatness self-consistency of the full SVT action $L_2+L_3+L_4+L_5$, under the rescaled Einstein equation \eqref{eq:svt-Einstein}, we find
\be
\label{eq:svt-flatness-summary}
\begin{aligned}
&G_4(\phi_0,0)\;\neq\;0,\\[2pt]
&G_2|_0=\partial_\Phi G_2|_0=\partial_\Phi^2 G_2|_0=0,\\[2pt]
&G_3,\;\hat G_3,\;\mathring{\hat G}_3,\;G_5,\;\hat G_4,\;\mathring{\hat G}_4\;\text{unrestricted at the vacuum but at most $\mathcal O(r^0)$},\\[2pt]
&G_{4,X}(\phi_0)\ \text{unrestricted};\ \ G_{4,\Phi}(\phi_0)\ \text{admissible but non-minimal (see below)}.
\end{aligned}
\ee
The first line is the only restriction the full SVT brings that the $L_2$ sector did not. Namely, a finite, non-vanishing asymptotic Newton constant $G_N^{\rm asymp}=G_0/G_4(\phi_0,0)$, see Eq.~\eqref{eq:GN-asymp}. The second reproduces the familiar conditions on the $L_2$ scalar potential. The fourth line records the one subtlety uncovered by the component analysis of the $L_4$ Horndeski sector. The kinetic coefficient $G_{4,X}(\phi_0)$ is genuinely unrestricted, contributing only pure-trace pieces within the marginal trace slot of the falloff table. A non-vanishing $G_{4,\Phi}(\phi_0)$, by contrast, is a genuine non-minimal curvature--scalar couplings. The third line is the central structural conclusion of the full SVT analysis: every other coupling functional in $L_3$, $L_4$, $L_5$ is unrestricted at the vacuum, because the operators that carry them are automatically suppressed at $\scrip$ either by the divergence-free curvature tensors $G^{\mu\nu}, L^{\mu\nu\alpha\beta}$ both of which vanish identically on the strict leading metric Eq.~\eqref{eq:svt-leading-metric} and reappear only at subleading order with slowest Weyl component $\mathcal O(r^{-1})$ for the $L_4$ and $L_5$ curvature mixings, or by an explicit $\Phi_{\mu\nu}$ factor for the parity-odd $L_3$ mixings $\hat G_3,\mathring{\hat G}_3$.

The same component analysis as summarized in Eq.~\eqref{eq:svt-flatness-summary}, carried out derivative by derivative, extends verbatim to every Taylor coefficient at the vacuum. Under regularity each partial derivative $\partial^{n}_{(\Phi,X,F,\tilde F,Y)}G_i|_0$ is a constant that multiplies one fixed structure $S_{ab}$ in $T^{\rm eff}_{ab}/G_4$. The structure describes the tensor-valued monomial in the matter fields and the curvature that accompanies the coefficient in the per-operator stress-energy tensors of App.~\ref{app:svt-stress}, e.g.\ $S_{ab}=g_{ab}$ for $G_2$, $S_{ab}=\Phi_a\Phi_b$ for $G_{2,X}$, and $S_{ab}=\Phi_{ab}$ for $G_{4,\Phi}$. The partial derivative then respects the falloff table if and only if the slowest covariant component of that structure decays at least as fast as the corresponding table bound. Writing $S_{ab}=\mathcal O(r^{-w_{ab}})$ for the componentwise decay of the structure and $T_{ab}=\mathcal O(r^{-b_{ab}})$ for the entries of the falloff table Eq.~\eqref{eq:T-falloff}, the criterion reads $w_{ab}\ge b_{ab}$ for every component pair $(ab)$. We use this systematically from Sec.~\ref{sec:growth-metric} onwards. Performing this scan for every structure appearing in Eqs.~\eqref{eq:T2-explicit}--\eqref{eq:E5-explicit}, and to all derivative orders via the demonstration of Appendix~\ref{app:higher}, the only coefficients forced to vanish are those already displayed in Eq.~\eqref{eq:svt-flatness-summary}, i.e., $G_2(\phi_0)=\partial_\Phi G_2(\phi_0)=\partial_\Phi^2 G_2(\phi_0)=0$, together with the single non-minimal exception $\partial_\Phi G_4(\phi_0)$, whose improvement structure $\Phi_{ab}$ carries $\Phi_{uu}=\mathcal O(r^{-1})$ in excess of $T_{uu}=\mathcal O(r^{-2})$ (Eq.~\eqref{eq:G4-violation}) and is admissible only in the frame sense discussed above. Every other derivative of every coupling functional, $G_{2,X},G_{2,F},G_{2,\tilde F},G_{2,Y}$, the cubic-sector $G_{3,\Phi},G_{3,X}$ and the parity-odd $\hat G_3,\mathring{\hat G}_3$, the quartic-sector $G_{4,X},\partial_\Phi^2 G_4$ and the curvature mixings $\hat G_4,\mathring{\hat G}_4$, and the quintic $G_{5,\Phi},G_{5,X}$, and all their higher derivatives are subject to no independent condition at the vacuum, their generic $\mathcal O(1)$ vacuum values landing strictly within the table, or, for the canonical $G_{2,X}$ and the gradient bilinear $G_{3,\Phi}$, marginally on it. Notably $\partial_\Phi^2 G_4$ is already safe. It is the first derivative $\partial_\Phi G_4$ alone that offends, because only it multiplies the bare Hessian. The signed headroom by which each derivative sits below (or, for the exception, above) its ceiling is quantified by the per-derivative allowances collected in Sec.~\ref{sec:svt-ledger}.

The analysis so far rested on two inputs: the metric falloff table Eq.~\eqref{eq:T-falloff} and the assumption that every coupling functional admits a regular Taylor expansion around the asymptotic vacuum Eq.~\eqref{eq:svt-vacuum}, $(\Phi,X,Y,F,\tilde F)=(\phi_0,0,0,0,0)$. In the remainder of this section we drop the second input and replace it by a weaker standing assumption: near the vacuum, each coupling functional behaves as a power of the building blocks, i.e.\ for every channel $Z\in\{\Phi-\phi_0,X,Y,F,\tilde F\}$ that carries non-trivial behavior,
\be
\label{eq:power-class}
G_i \;\sim\; Z^{\,p} \qquad\text{as } Z\to0,\qquad p\in\mathbb{R} ,
\ee
with $p$ admitting a positive, negative and fractional power. The physical picture behind this class is that the vacuum is approached in a scale-free manner. A coupling functional descending from an effective field theory expansion around a healthy vacuum is analytic there, corresponding to non-negative integer $p$. Genuinely fractional or negative $p$ arise when the environmental vacuum selected by screening (cf.\ Sec.~\ref{sec:svt-motivation}) sits at a non-analytic point of the functional, as realized by inverse-power-law quintessence potentials \cite{Ratra:1987rm} in the $\Phi$-channel or by a cuscuton-like kinetic term $\propto\sqrt{X}$ \cite{Afshordi:2006ad}. On the radiative background, where $Z\sim r^{-w_Z}$, Eq.~\eqref{eq:power-class} is equivalent to allowing each $G_i$ a polynomial expansion in powers of $r$ for large $r$, including potentially divergent terms, and we ask which necessary conditions then bound its behavior. This power class is the natural arena for the question. It contains every regular Taylor expansion as well as all power-law singularities, and excludes only non-power behavior at the vacuum, most notably a logarithmic dependence, which carries an intrinsic reference scale and is generated by radiative corrections. Its marginal effects are discussed separately in Sec.~\ref{sec:conclusion}. The assumption Eq.~\eqref{eq:power-class} is crucial for everything that follows in this section as it gives each coupling functional a well-defined growth exponent and embodies what ties the exponents of a functional and its derivatives together. Three independent conditions are available: (i) the metric falloffs themselves, (ii) consistency of the scalar (and vector) equation of motion with the radiative ansatz, and (iii) stability of perturbations around the asymptotic vacuum. Sec.~\ref{sec:svt-flatness} has implemented (i) for regular coupling functionals. As we now show, condition (ii) is strictly sharper than (i) throughout the scalar sector. It closes the $G_5$ growth window that the metric falloffs leave open, and even in the regular case it tightens the $L_2$ conditions of Eq.~\eqref{eq:flat-G2}. The net result is that regularity of the scalar coupling functionals at the asymptotic vacuum is an output of the analysis, not an input. The vector-coupled sector, by contrast, retains a handful of growth windows that neither equation of motion forces to close, and we delimit them precisely.

\subsection{Growth bounds from the metric falloffs alone}
\label{sec:growth-metric}

We start by revisiting condition (i) under the loosened coupling function scalings. We utilize two notational conventions for this section. First, all tensors are evaluated in their covariant Bondi components, labeled by index pairs $(ab)$ with $a,b\in\{u,r,A\}$, exactly as in the falloff table Eq.~\eqref{eq:T-falloff}. We therefore write the effective stress and the structures it is built from as $T_{ab}$ and $S_{ab}$, respectively. Second, we reserve the symbol $\gamma$ for the growth exponent of a coupling functional on the radiative background, $G=\mathcal O(r^{\gamma})$, so that $\gamma>0$ means growth toward $\scrip$. It is never used as an index.

Every coupling functional multiplies a fixed field or curvature structure $S_{ab}$ in the effective stress. The building blocks $Z$ through which a singular dependence can arise are, for $Z=\Phi$, the deviation $\Phi-\phi_0$ from the vacuum Eq.~\eqref{eq:svt-vacuum}, and for $Z\in\{X,Y,F,\tilde F\}$ the invariants themselves, which already vanish there, each carrying the background weight $Z\sim r^{-w_Z}$,
\be
\label{eq:growth-weights}
w_\Phi=1,\qquad w_X=3,\qquad w_F=w_{\tilde F}=4,\qquad w_Y=6.
\ee
If, through a singular dependence on such a block, the coupling functional scales within the power class Eq.~\eqref{eq:power-class} as $G\sim Z^{-s}\sim r^{\gamma}$ with $\gamma=s\,w_Z>0$, then its contribution to the effective stress is $r^{\gamma}\,S_{ab}$. Denoting by $w_{ab}$ the decay of each individual component, $S_{ab}=\mathcal O(r^{-w_{ab}})$, and by $b_{ab}$ the corresponding table bound, $T_{ab}=\mathcal O(r^{-b_{ab}})$, the falloff table is preserved iff $\gamma-w_{ab}\le-b_{ab}$ holds for every component pair, i.e.
\be
\label{eq:growth-rule}
\gamma\;\le\;\min_{(ab)}\bigl(w_{ab}-b_{ab}\bigr).
\ee
Crucially, a function and its derivatives are not independent. Each $Z$-derivative raises the growth exponent by the weight $w_Z$, $\partial_Z G\sim Z^{-s-1}\sim r^{\,\gamma+w_Z}$. This is an identity on the power class Eq.~\eqref{eq:power-class}. Differentiation with respect to $Z$ lowers the power by exactly one, and one power of $Z$ is worth $r^{-w_Z}$ on the background. It holds term by term for regular Taylor expansions. The bound on the function is therefore obtained by translating the bound of every derivative appearing in the effective stress back to the function and taking the tightest, then maximizing over the channel $Z$ that carries the growth. Carrying this out with the component orders established in App.~\ref{app:svt-components} (Eqs.~\eqref{eq:T3-Ghat3-orders}, \eqref{eq:T4kin-orders}, \eqref{eq:T4hatG4-orders}, \eqref{eq:E5-orders}) yields the function-level bounds
\be
\label{eq:growth-table}
G_2\le \mathcal O(r^{-3}),\qquad G_3\le \mathcal O(r^{-1}),\qquad G_4=\mathcal O(1),\;G_4|_0\neq0,\qquad G_5\le \mathcal O(r^{+1}) .
\ee
The bound on $G_2$ is set by its undifferentiated appearance $g_{ab}G_2$ in Eq.~\eqref{eq:T2-explicit}. The binding component is not $g_{uu}G_2$ against $T_{uu}=\mathcal O(r^{-2})$ but the cross term $g_{ur}G_2$ against the tighter $T_{ur}=\mathcal O(r^{-3})$, since $g_{ur}=-1=\mathcal O(1)$. An explicit component scan (Appendix~\ref{app:higher}) confirms $\min_{(ab)}(w_{ab}-b_{ab})=-3$, so that $G_2\le \mathcal O(r^{-3})$ which yields exactly the $G_2=\partial_\Phi G_2=\partial_\Phi^2 G_2=0$ of Eq.~\eqref{eq:flat-G2}. The first three entries simply reproduce regularity. Namely, $G_2$ and $G_3$ must decay and $G_4$ must approach the finite Newton-constant value of Eq.~\eqref{eq:GN-asymp}. The interesting entry is $G_5$. The metric falloffs alone do admit growth. Taking $G_5\sim(\Phi-\phi_0)^{-1}\sim r$ (so that $G_{5,X}=0$ and $G_{5,\Phi}\sim(\Phi-\phi_0)^{-2}\sim r^{2}$), both the curvature part $\mathcal E^{(5)}_{ab}$ of Eq.~\eqref{eq:E5-explicit} and the $G_{5,\Phi}$-Galileon part of Eq.~\eqref{eq:T5-explicit} land exactly on the marginal $T_{uu}=\mathcal O(r^{-2})$ entry of the table. Function and derivative grow together while the structures they multiply decay together, and asymptotic flatness of the metric survives. Whether this window is physical is decided by the scalar equation of motion.

\subsection{Consistency of the equation of motion}
\label{sec:growth-eom}

\subsubsection{Scalar sector.}
We continue by analyzing the second condition, (ii). The scalar EOM supplies a condition that the metric sector cannot see. On the radiative background the d'Alembertian of the scalar ansatz $\Phi=\phi_1/r+\phi_2/r^2+\mathcal O(r^{-3})$ has the free-wave structure
\be
\label{eq:boxphi-freewave}
\Box\Phi = \frac{2\,\partial_u\phi_2+\eth^2\phi_1}{r^{3}}+\mathcal O(r^{-4}),
\ee
with no $\mathcal O(r^{-2})$ term. The $u$--$r$ cross terms cancel identically at that order, as they must for an outgoing wave (this is the same cancellation noted for $\Box\Phi$ in Sec.~\ref{sec:svt-flatness}, and it persists on the full radiative metric, whose corrections enter only at $\mathcal O(r^{-4})$). The first non-trivial order of the scalar EOM is therefore $r^{-3}$, where the canonical sector contributes
\be
\label{eq:scalar-eom-r3}
\frac{G_{2,X}(\phi_0)}{r^{3}}\,\bigl[2\,\partial_u\phi_2+\eth^2\phi_1\bigr] +\text{modified-gravity terms of $\mathcal O(r^{-3})$} \;=\;0 .
\ee
This balance fixes the tolerance of the criterion. An additional contribution entering at $\mathcal O(r^{-3})$ is harmless. It is absorbed by $\partial_u\phi_2$ in Eq.~\eqref{eq:scalar-eom-r3} and merely modifies the evolution of the subleading coefficients, which is what a modified theory is expected to do, while the radiative datum $\phi_1(u,x^A)$ remains freely specifiable. A contribution entering at $\mathcal O(r^{-2})$, by contrast, finds no counterpart. No retarded-time derivative of any expansion coefficient appears at that order, so the equation of motion degenerates into an algebraic condition on the free data itself, which for generic radiation can only be met by $\phi_1\equiv0$ extinguishing scalar radiation altogether. The only escape would be to enlarge the ansatz by $\ln r/r^{2}$ terms, the polyhomogeneous scenario discussed in Sec.~\ref{sec:conclusion}. We therefore demand that every EOM contribution fall no slower than $r^{-3}$. The complete scalar equation of motion, written out sector by sector, is Eq.~\eqref{eq:app-scalar-EOM-full} of App.~\ref{app:svt-eoms}. We now go through its terms.

The $L_2$ part of the scalar EOM reads, in isolation and in full,
\be
\label{eq:L2-scalar-eom}
G_{2,\Phi} +\nabla_a\bigl(G_{2,X}\,\Phi^a\bigr) -2\,\nabla_a\bigl(G_{2,Y}\,\Phi_\nu F^{a\alpha}F^{\nu}{}_\alpha\bigr)=0 .
\ee
On the radiative background the $Y$-term is harmless. Its argument $\Phi_\nu F^{a\alpha}F^{\nu}{}_\alpha$ is $\mathcal O(r^{-5})$ at its slowest component, so the divergence enters at $\mathcal O(r^{-6})$ while the criterion is $G_{2,\Phi}\le \mathcal O(r^{-3})$. The falloff table had allowed a cubic self-interaction, $G_2\supset\lambda\Phi^3$, whose stress contribution $g_{ab}G_2=\mathcal O(r^{-3})$ is table-safe. Its EOM source, however, is $G_{2,\Phi}=3\lambda\Phi^2=3\lambda\phi_1^2/r^2$. This is exactly the situation described below Eq.~\eqref{eq:scalar-eom-r3}. The balance Eq.~\eqref{eq:scalar-eom-r3} provides no $\mathcal O(r^{-2})$ counterpart, so at this order the equation of motion is not an evolution equation but the algebraic statement $3\lambda\phi_1^2=0$ on the free data. A radiating scalar ($\phi_1\not\equiv0$) therefore requires
\be
\label{eq:G2-cubic-cond}
\partial_\Phi^3 G_2\big|_0=0
\ee
in addition to the three conditions of Eq.~\eqref{eq:flat-G2}.\footnote{The mixed term $\Phi X$ remains admissible, since $G_{2,\Phi}\supset X=\mathcal O(r^{-3})$ is marginal.} The same power counting excludes every singular $\Phi$-channel dependence outright, $G_2^{\rm sing}\sim(\Phi-\phi_0)^{-s}\sim r^{\gamma}$ giving $G_{2,\Phi}\sim r^{\gamma+1}$, and analogously for the $X$-channel via $\Phi^a\nabla_a G_{2,X}$.

The cubic sector contributes to the scalar EOM the explicit combination (second line of Eq.~\eqref{eq:app-scalar-EOM-full})
\be
\label{eq:L3-scalar-eom}
-G_{3,\Phi}\,\Box\Phi -\Box G_3 -\nabla_a\bigl(G_{3,X}\,\Phi^a\,\Box\Phi\bigr),
\ee
together with the parity-odd mixing terms of Eq.~\eqref{eq:app-scalar-EOM-full}, which decay as $\mathcal O(r^{-4})$ or faster on the radiative background and play no role at the orders considered here. For any singular part $G_3^{\rm sing}\sim r^{\gamma}$ one has $\Box G_3^{\rm sing}\sim r^{\gamma-1}$. The free-wave cancellation that protects $\Box\Phi$ operates only for the exact radiative falloff $\Phi-\phi_0\sim r^{-1}$, not for a generic power. Thus, the criterion demands $\gamma\le-2$. All growth is excluded, consistent with (and slightly tighter than) the metric bound of Eq.~\eqref{eq:growth-table}. Regular Taylor dependence remains entirely unrestricted, as in Sec.~\ref{sec:svt-flatness}.

For $G_5(\Phi)$ (the $\Phi$-channel relevant to the window of Sec.~\ref{sec:growth-metric}), using $\nabla_a G^{ab}=0$ the scalar EOM contribution of $L_5=G_5\,G^{ab}\Phi_{ab}+\dots$ is
\be
\label{eq:E5-scalarEOM}
\mathcal E^{(5)}_\Phi = G_{5,\Phi}\,G^{ab}\Phi_{ab} + G^{ab}\nabla_a\nabla_b G_5 .
\ee
The contraction $G^{ab}\Phi_{ab}$ is evaluated on shell: with $G_{ab}=8\pi T^{\rm eff}_{ab}/G_4$ bounded by the falloff table, raising indices with the leading metric ($g^{ur}=-1$, $g^{rr}=1$, $g^{AB}=q^{AB}/r^2$) gives
\be
\label{eq:G-upper-orders}
G^{uu}\sim G_{rr}=\mathcal O(r^{-3}),\quad G^{rr}\sim G_{uu}=\mathcal O(r^{-2}),\quad G^{uA},G^{rA}=\mathcal O(r^{-4}),\quad G^{AB}=\mathcal O(r^{-4}),
\ee
so that, paired with the Hessian orders Eq.~\eqref{eq:Phi-uv-orders}, the slowest products are $G^{uu}\Phi_{uu}$ and $G^{AB}\Phi_{AB}$, both $\mathcal O(r^{-4})$, thus
\be
G^{ab}\Phi_{ab}=\mathcal O(r^{-4}) .
\ee
The criterion $\mathcal E^{(5)}_\Phi\le \mathcal O(r^{-3})$ then bounds $G_{5,\Phi}\le \mathcal O(r)$, i.e.\ with $G_{5,\Phi}\sim r^{\,\gamma+1}$ with $\gamma\;\le\;0 $. Therefore, the $\mathcal O(r^{+1})$ window left open by the metric falloffs is closed by the scalar equation of motion, and $G_5$ is forced to be $\mathcal O(1)$ at the asymptotic vacuum after all. The second term of Eq.~\eqref{eq:E5-scalarEOM} obeys the same bound, since $\nabla_a\nabla_b G_5\sim G_{5,\Phi}\Phi_{ab} +G_{5,\Phi\Phi}\Phi_a\Phi_b$ contracts against Eq.~\eqref{eq:G-upper-orders} at the same orders. Growth through the $X$-channel, by contrast, never required the equation of motion. Its allowance $G_{5,X}\le\mathcal O(r^{3})$ (App.~\ref{app:higher}) already translates, with $w_X=3$, into $\gamma\le0$, i.e.\ only non-negative powers of $X$ survive at the vacuum. The $\mathcal O(r^{+1})$ of Eq.~\eqref{eq:growth-table} is the maximum over channels and is set by the $\Phi$-channel alone.

\subsubsection{Vector sector.}
The vector equation of motion $\nabla_a\mathcal G^{ab}=0$, with the displacement tensor $\mathcal G^{ab}$ written out in full in Eq.~\eqref{eq:app-G-displacement} of App.~\ref{app:svt-eoms}. Its $L_2$ part is
\be
\label{eq:L2-vector-eom}
\nabla_a\Bigl(G_{2,F}\,F^{ab} -4\,G_{2,\tilde F}\,\tilde F^{ab} -4\,G_{2,Y}\,\Phi^{[a}\Phi^{c}F_{c}{}^{b]}\Bigr)=0 \quad(\text{in isolation})
\ee
which supplies the analogous condition in the vector channel. On the radiative expansion $A_u=q/r+\mathcal O(r^{-2})$, $A_A=\mathcal A^{(0)}_A+\mathcal O(r^{-1})$ the free Maxwell current has the free-wave structure
\be
\label{eq:maxwell-freewave}
\nabla_a F^{a u}=\mathcal O(r^{-4}),\quad \nabla_a F^{a A}=\mathcal O(r^{-4}),\quad \nabla_a F^{a r}=\frac{\partial_u q-D^A\partial_u\mathcal A^{(0)}_A}{r^{2}} +\mathcal O(r^{-3}),
\ee
so that the slowest component sits at $\mathcal O(r^{-2})$ and enforces the leading Maxwell constraint $\partial_u q=D^A\partial_u\mathcal A^{(0)}_A$ between the charge aspect and the angular potential. A growing coupling functional deforms $\mathcal G^{ab}$. Requiring its divergence to fall no slower than this $\mathcal O(r^{-2})$ Maxwell order bounds the growth. Two features make the vector EOM strictly weaker than the metric falloffs for these coupling functionals. First, the parity-odd $\tilde F^{ab}$ structure is identically divergence-free, $\nabla_a\tilde F^{ab}\equiv0$, so the topological $G_{2,\tilde F}$ term drops out of the EOM for a constant coupling functional and enters only through its scalar gradient, $-4\,G_{2,\tilde F\Phi}\,\Phi_a\tilde F^{ab}$ at leading order. Second, the $Y$-displacement is doubly $\Phi$-suppressed. An explicit component scan gives $\nabla_a\bigl(\Phi^{[a}\Phi^{c}F_{c}{}^{b]}\bigr)=\mathcal O(r^{-5})$ at its slowest ($b=r$) component, so the vector EOM caps $G_{2,Y}$ only at $\gamma\le 5-2=3$ which gives a looser bound than what was obtained from the metric bound $\gamma\le2$ (see the componentwise scan, Appendix~\ref{app:higher}). By the same $\Phi$-suppression mechanism the displacement structures of the parity-odd $\hat G_3,\mathring{\hat G}_3$ and of the curvature mixings $\hat G_4,\mathring{\hat G}_4$ return divergences no slower than $\mathcal O(r^{-5})$, so their vector-EOM bounds are structurally weaker than the metric allowances as well. We do not track the individual exponents. The vector EOM therefore confirms $G_{2,F}\le \mathcal O(1)$ but does not tighten the residual growth windows of the vector and mixing sector (Appendix~\ref{app:higher}). These are genuine features of the SVT vector sector, bounded but not closed. For coupling functionals regular at the vacuum, i.e., the setting of Sec.~\ref{sec:flatness_SVT}, all vector-sector operators lie safely within their bounds.

\subsection{Stability of the asymptotic vacuum}
\label{sec:growth-stability}

Let us now consider criterion (iii). A final, independent set of conditions arises from demanding that perturbations around the asymptotic vacuum are healthy. The kinetic matrix of the propagating modes must be positive (no ghost) and the propagation speeds real, guaranteeing gradient stability, hyperbolicity. For the Horndeski/SVT class these conditions are standard in the cosmological setting \cite{Kobayashi:2011nu,Heisenberg:2018mxx,Frusciante:2019xia}. Evaluated on the asymptotic vacuum $(\Phi,X)\to(\phi_0,0)$ they collapse to conditions on the leading Taylor coefficients. In particular, the tensor kinetic coefficient reduces to $2G_4|_0$, so that the no-ghost condition sharpens the invertibility requirement of Sec.~\ref{sec:flatness_SVT} from $G_4(\phi_0,0)\neq0$ to
\be
\label{eq:stability-signs}
G_4(\phi_0,0)>0 ,\qquad G_{2,X}(\phi_0,0)>0 ,\qquad G_{2,F}(\phi_0,0)>0 ,
\ee
the second condition being the canonical normalizability of the scalar mode and the third that of the vector mode (i.e., in generalized-Proca/Maxwell no-ghost \cite{Heisenberg:2018mxx}). Each is a sign condition on a leading Taylor coefficient whose magnitude the falloff analysis already left unrestricted ($G_4|_0\neq0$, $G_{2,X}|_0$ and $G_{2,F}|_0$ both with allowance $\mathcal O(1)$, see Appendix~\ref{app:higher}). Stability thus fixes signs without adding a single new falloff constraint, consistent with the expectation that it is the weakest of the three conditions. The BD example of Sec.~\ref{sec:matter-bd} illustrates both. Namely, $\phi_0>0$ is the positive asymptotic Newton constant, and $2\omega+3>0$. This is required for the Einstein-frame scalar to be real and is precisely the scalar no-ghost condition. The tensor speed obeys $c_T^2\to1$ automatically at the vacuum, since every deviation enters suppressed by powers of $X$ or $\Phi$. Luminal propagation at $\scrip$ is thus a derived property rather than an extra assumption. We note, however, that for the growing coupling functionals of Sec.~\ref{sec:growth-metric} the stability conditions are comparatively weak. The singular coupling functionals enter the kinetic matrix only in $X$-suppressed combinations (e.g.\ $X G_{5,\Phi}\sim r^{-3}\cdot r^{\,\gamma+1}$, decaying for all $\gamma\le2$), so it is the equation-of-motion criterion of Sec.~\ref{sec:growth-eom}, not stability, that provides the binding constraint on singular behavior.

\subsection{Synthesis: the constraint table}
\label{sec:svt-ledger}
Once the scalar equation of motion and vacuum stability are imposed alongside the metric falloffs, no coupling functional of the scalar sector grows toward $\scrip$. $G_2\le \mathcal O(r^{-3})$ with the sharpened condition Eq.~\eqref{eq:G2-cubic-cond}, $G_3\le \mathcal O(r^{-1})$ with singular parts capped at $\mathcal O(r^{-2})$, and $G_4,G_5=\mathcal O(1)$ with $G_4(\phi_0,0)>0$.

The same statements hold for the derivatives of the coupling functionals. This distinction matters. A bounded function may still possess growing derivatives (e.g.\ $G\sim(\Phi-\phi_0)^{1/2}$ decays as $r^{-1/2}$ while $G_{,\Phi}\sim(\Phi-\phi_0)^{-1/2}\sim r^{+1/2}$ grows), so the analysis of Secs.~\ref{sec:growth-metric}--\ref{sec:growth-eom} constrains every derivative appearing in the effective stress and in the scalar EOM individually, and such fractional behavior is excluded by the derivative bounds even where the function itself is table-safe. After the equation of motion and stability conditions are imposed, the derivatives obey explicitly
\be
\label{eq:deriv-summary}
\begin{aligned}
&G_{2,\Phi}\le \mathcal O(r^{-3}),\qquad G_{2,X}=\mathcal O(1)\;[\mathcal O(1)],\ G_{2,X}|_0>0,\\
&G_{3,\Phi}=\mathcal O(1)\;[\mathcal O(1)],\qquad G_{3,X}=\mathcal O(1)\;[\mathcal O(r^{2})],\\
&G_{4,\Phi}=\mathcal O(1)\;[\mathcal O(r^{-1})],\qquad G_{4,X}=\mathcal O(1)\;[\mathcal O(r)],\\
&G_{5,\Phi}=\mathcal O(1)\;[\mathcal O(r^{2})],\qquad G_{5,X}=\mathcal O(1)\;[\mathcal O(r^{3})],\\
&G_{3,\Phi X},\,G_{3,XX},\,G_{4,X\Phi},\,G_{4,XX},\,G_{5,XX}=\mathcal O(1)\\
&\qquad\qquad [\mathcal O(r^{3}),\,\mathcal O(r^{5}),\,\mathcal O(r^{2}),\,\mathcal O(r^{4}),\,\mathcal O(r^{6})],
\end{aligned}
\ee
each entry having the form $(\text{admissible behavior})\,[\,\text{allowance}\,]$, which contrasts two distinct quantities. The unbracketed symbol is what the derivative may actually do on the radiative background once all three conditions (the metric falloffs, the scalar equation of motion, and vacuum stability) have been imposed. $\mathcal O(1)$ marks a finite, generically nonzero constant, i.e., an unconstrained Taylor coefficient at the vacuum, while $\le \mathcal O(r^{-3})$ marks a coefficient forced to vanish, so that the derivative decays. The bracketed symbol is what we call the allowance. It describes the largest power of $r$ the derivative could reach if it were an independent quantity, subject only to the demand that the single tensor structure it multiplies in $T^{\rm eff}_{ab}$ respect the falloff table (Eq.~\eqref{eq:growth-rule}), and ignoring both its link to the parent coupling functional and the equation of motion. The allowance is thus a purely kinematical ceiling set by the metric falloffs alone. The gap between it and the admissible $\mathcal O(1)$ value is the headroom that the function--derivative linkage (Sec.~\ref{sec:growth-metric}) and the equation of motion (Sec.~\ref{sec:growth-eom}) close. When the two coincide at $\mathcal O(1)$ the derivative is marginal and its allowance is saturated.

This happens for $G_{2,X}$ and $G_{3,\Phi}$. Every remaining derivative sits strictly below its allowance, by the margin shown. The starred entry $G_{4,\Phi}$ is the single exception in the opposite direction. Its allowance is $\mathcal O(r^{-1})$. Here, the binding component is not the non-decaying Hessian trace $\Phi_{AB}=\mathcal O(r^0)$ (which lands in the marginal $T_{AB}$ trace slot) but $\Phi_{uu}=\mathcal O(r^{-1})$ against the tighter $T_{uu}=\mathcal O(r^{-2})$, so a constant, non-minimal $G_{4,\Phi}(\phi_0)\neq0$ exceeds its own allowance. This is precisely the $u$-component violation of Eq.~\eqref{eq:G4-violation}. It is admissible only after the conformal resolution of the $L_4$ paragraph in Sec.~\ref{sec:svt-flatness}, and is the lone place where a regular coupling functional and the strict falloff table are in tension. Consistency of the scheme is visible in the mixed second derivative $G_{4,X\Phi}$, whose allowance $\mathcal O(r^{2})$ is reproduced from either parent, i.e., $G_{4,\Phi}$ at $\mathcal O(r^{-1})$ raised by $w_X=3$, or $G_{4,X}$ at $\mathcal O(r)$ raised by $w_\Phi=1$. The one case in which the equation of motion is strictly stronger than the falloff table is $G_{5,\Phi}$, whose kinematical allowance $\mathcal O(r^2)$ is lowered further to $\mathcal O(r)$ by the $L_5$ analysis of Sec.~\ref{sec:growth-eom}. In the last line the five bracketed allowances are listed in the same order as the five second derivatives above them. The second derivatives in the last line, which enter $T^{\rm eff}_{ab}$ through the variation $\delta X=-\tfrac12\Phi_a\Phi_b\,\delta g^{ab}$ of derivative-coupling prefactors and through the integration-by-parts remainders, add no further constraints. Each additional $Z$-derivative raises the growth exponent by $w_Z$ but necessarily arrives contracted with a conjugate structure, i.e., $\Phi_a\Phi_b$, $\Phi_a$ or $\nabla_aX$, whose decay at the binding component compensates exactly, so the first-derivative bounds remain the binding ones to all orders in derivatives. The componentwise verification and an all-orders argument are collected in Appendix~\ref{app:higher}.

A second, independent bound follows from the coupling functionals being functions of $\Phi$ and $X$ alone, without reference to the tensor structure that each derivative multiplies. Writing the $X$-dependence in the power class Eq.~\eqref{eq:power-class} as $G\sim X^{p}$ and using $X=\mathcal O(r^{-3})$, a function-level bound $G\le\mathcal O(r^{\Gamma})$ is the statement $p\ge-\Gamma/3$, and the $n$-th $X$-derivative obeys $\partial_X^nG\sim X^{p-n}\le\mathcal O(r^{3n+\Gamma})$. Equality requires the saturating value $p=-\Gamma/3$, and the derivative survives at that value only if $p\notin\{0,1,\dots,n-1\}$, since an integer power below the order of differentiation is annihilated. The ceiling is therefore reached whenever the saturating power is fractional, which is the case for $G_3$ with $p=1/3$ and for $\hat G_3$ with $p=-2/3$, and it is unreachable whenever that power is an annihilated integer. The latter situation occurs precisely for the functionals bounded at $\Gamma=0$, where saturation demands $p=0$ and every $X$-derivative of the leading term vanishes identically, so that $\partial_X^nG\le\mathcal O(r^{3n-3p})<\mathcal O(r^{3n})$ holds strictly. The same criterion applied in the $\Phi$-channel with $w_\Phi=1$ makes the first two $\Phi$-derivatives of such a functional strict as well.
The derivatives of $G_4$ are not sharpened in this way, because their quoted values already descend from the componentwise allowances of App.~\ref{app:higher}, which lie below what the parent alone would permit. We emphasize that the strictness statement refers to the per-channel analysis, in which a single building block carries the growth. A functional combining a singular dependence on one block with a compensating regular dependence on the other can still reach the scaling ceilings. If in addition the coupling functional is analytic at the vacuum, so that $p$ is a non-negative integer, then every $X$-derivative is a finite Taylor coefficient and therefore of order unity, which lies $3n$ powers below the ceiling. Growth in the $X$-channel is thus available only to non-analytic, fractional-power dependence, and even there it remains strictly below the tabulated value. Applied to the $\Phi$-channel with $w_\Phi=1$ the same translation sharpens two entries whose structure allowance is weaker than their parent permits, namely $G_{5,\Phi\Phi}\le\mathcal O(r^{2})$, inherited from $G_5=\mathcal O(1)$ after the equation of motion of Sec.~\ref{sec:growth-eom} has closed the kinematic window, and $\hat G_{3,X}\le\mathcal O(r^{5})$, inherited from $\hat G_3\le\mathcal O(r^{2})$. 

The coupling functionals of the parity-odd mixings and the vector-sector $L_2$ derivatives stand on a different footing as they enter the effective stress as operator-level bundles, and the metric falloffs bound them per operator,
\be
\label{eq:mixing-summary}
\begin{aligned}
&\hat G_3\le \mathcal O(r^{2}),\qquad \mathring{\hat G}_3\le \mathcal O(r^{4}),\qquad \hat G_4,\;\mathring{\hat G}_4\le \mathcal O(r),\\
&G_{2,F}\le \mathcal O(1),\qquad G_{2,\tilde F}\le \mathcal O(r^{1}),\qquad G_{2,Y}\le \mathcal O(r^{2}),
\end{aligned}
\ee
read off from the component orders Eqs.~\eqref{eq:T3-Ghat3-orders} and \eqref{eq:T4hatG4-orders} and from the $L_2$ sources of Eq.~\eqref{eq:T2-explicit}.\footnote{The $\mathring{\hat G}$ entries follow from the $g_{ab}\to\Phi_a\Phi_b$ replacement, which supplies two additional powers of decay. The coefficient $G_{2,F}$ multiplies the Maxwell stress, which saturates $T_{uu}=\mathcal O(r^{-2})$ (App.~\ref{app:matter-maxwell}). For $G_{2,\tilde F}$ the binding slot is again the cross term $g_{ur}\tilde F=\mathcal O(r^{-4})$ against $T_{ur}=\mathcal O(r^{-3})$ -- not the trace slot -- giving $\gamma\le1$. The $Y$-structure's slowest component is $T_{ur}=\mathcal O(r^{-5})$, giving $\gamma\le2$.} All entries are confirmed by the explicit componentwise scan of Appendix~\ref{app:higher}. In contrast to the scalar sector, the positive exponents here are genuine residual growth windows at the level of the metric falloffs. As shown in the Vector sector paragraph of Sec.~\ref{sec:growth-eom}, the vector equation of motion confirms $G_{2,F}\le \mathcal O(1)$ but returns bounds weaker than these metric allowances for $G_{2,\tilde F},G_{2,Y}$ and the parity-odd and curvature mixings, so that the windows are bounded but not closed. For coupling functionals regular at the vacuum all mixing operators lie safely within their bounds.

The regularity assumption of Sec.~\ref{sec:flatness_SVT} is thereby justified a posteriori in the scalar sector. Once the metric falloffs, the scalar equation of motion and vacuum stability are imposed, no scalar coupling functional $G_2,G_3,G_4,G_5$ or any of its derivatives may grow toward $\scrip$, so scalar-sector regularity as a function and in all its derivatives is an output rather than an input, with the conditions of Eq.~\eqref{eq:svt-flatness-summary} supplemented by Eqs.~\eqref{eq:G2-cubic-cond} and \eqref{eq:stability-signs}. The vector-coupled sector is more permissive. The parity-odd mixings $\hat G_3,\mathring{\hat G}_3,\hat G_4,\mathring{\hat G}_4$ and the $L_2$ field-strength derivatives $G_{2,\tilde F},G_{2,Y}$ retain the genuine growth windows of Eq.~\eqref{eq:mixing-summary}, which neither the metric falloffs nor the vector equation of motion force to close. A theory singular at the vacuum in exactly these channels would still admit an asymptotically flat, BMS-symmetric radiative sector, a small but definite enlargement of the admissible class beyond the regular Horndeski/SVT coupling functionals, whose exploration we leave to future work. For every coupling functional regular at the vacuum, i.e., the physically motivated setting of Sec.~\ref{sec:flatness_SVT}, all of these operators stay within their bounds, and the full set of constraints reduces to Eq.~\eqref{eq:svt-flatness-summary} supplemented by Eqs.~\eqref{eq:G2-cubic-cond} and \eqref{eq:stability-signs}.

\medskip Table~\ref{tab:master-constraints} consolidates Secs.~\ref{sec:svt-flatness}--\ref{sec:growth-stability} into a single constraint table. Every coupling functional of the SVT action Eqs.~\eqref{eq:svt-L2-full}--\eqref{eq:svt-L5-full} and every derivative of it that enters an equation of motion is listed, whether that equation is the effective Einstein equation Eq.~\eqref{eq:svt-Einstein} (M), the scalar EOM (S), or the vector EOM (V). For each entry the table records the single tightest constraint and the (sub)section in which it is derived. The derivative content is fixed by the variational structure. $L_2$ enters the metric EOM algebraically, so only first derivatives of $G_2$ appear there, its higher derivatives arising in the field EOMs. The sectors $L_3,L_4^{\rm kin},L_5$ and the remainders Eqs.~\eqref{eq:D3-explicit}, \eqref{eq:D4kin-explicit}--\eqref{eq:D4mGhat4-explicit} carry one covariant derivative on their coupling functional (second derivatives), while the double-dual remainder Eq.~\eqref{eq:D4hatG4-explicit}, the $G_4R$ improvement $\nabla_\mu\nabla_\nu G_4$, and the curvature--scalar tensor Eq.~\eqref{eq:E5-explicit} carry two.

Derivatives beyond those listed in the table inherit their ceilings via the demonstration of App.~\ref{app:higher}. Each extra $\partial_\Phi$ raises the ceiling by $w_\Phi=1$, each $\partial_X$ by $w_X=3$. This function--derivative linkage is \emph{per channel}: a $\partial_Z$ lowers the $r$-power only of the piece carried by that same building block $Z$, so entries related by a derivative need not differ by the naive weight when their values are dominated by different channels. For instance $G_2$ and $G_{2,\Phi}$ both sit at $\mathcal O(r^{-3})$, because the dominant contribution to each is the kinetic $X$-channel term ($G_{2,X}(\phi_0)\,X$ and $G_{2,X\Phi}(\phi_0)\,X$, both $\mathcal O(r^{-3})$), which $\partial_\Phi$ leaves untouched. The $w_\Phi=1$ drop applies only to the subdominant $\Phi$-potential, which the scalar EOM pushes to $\mathcal O(\Phi^4)=\mathcal O(r^{-4})$.

\begin{table}[tp]
\caption{Constraint table for the SVT coupling functionals. Each of the functionals
$G_2,G_3,\hat G_3,\mathring{\hat G}_3,G_4,\hat G_4,\mathring{\hat G}_4,G_5$
and every derivative appearing in an equation of motion is listed, with the
column ``EOM'' recording which equation it enters, metric/effective
Einstein (M), scalar (S), vector (V). Entries under ``tightest constraint''
are the admissible growth ceilings on the radiative background (a coupling functional
$\le \mathcal O(r^\gamma)$ toward $\scrip$). A coupling functional regular at the asymptotic
vacuum realizes the $\mathcal O(1)$ value, saturating the marginal entries
$G_{2,X},G_{2,X\Phi},G_{2,F},G_{3,\Phi},G_4,G_{4,\Phi\Phi},G_5$. ``$=0$'' marks a Taylor
coefficient forced to vanish, ``$>0$'' a stability sign condition.
The starred and daggered entries are explained in the notes following the table.}
\label{tab:master-constraints}
\renewcommand{\arraystretch}{1.05}
\footnotesize
\begin{tabular}{@{}l c l l@{}}
\toprule
coupling functional / derivative & EOM & tightest constraint & derived in\\
\midrule
\multicolumn{4}{@{}l}{\textit{$L_2=G_2(\Phi,X,Y,F,\tilde F)$ \ (algebraic in the metric EOM: first derivatives only)}}\\
$G_2$ & M,\,S & $\le \mathcal O(r^{-3})$;\ $\Phi$-potential $\mathcal O(\Phi^4)$ & Secs.~\ref{sec:svt-flatness}, \ref{sec:growth-eom}\\
$G_{2,\Phi}$ & S & $\le \mathcal O(r^{-3})$ & Sec.~\ref{sec:growth-eom}\\
$G_{2,\Phi\Phi}$ & M & $=0$ \ (mass) & Sec.~\ref{sec:svt-flatness}\\
$G_{2,\Phi\Phi\Phi}$ & S & $=0$ \ (cubic) & Sec.~\ref{sec:growth-eom}\\
$G_{2,X}$ & M,\,S & $\mathcal O(1)$;\ \ $G_{2,X}|_0>0$ & Secs.~\ref{sec:growth-metric}, \ref{sec:growth-stability}\\
$G_{2,X\Phi}$ & S & $\mathcal O(1)$ & Sec.~\ref{sec:growth-eom}\\
$G_{2,XX}$ & S & $\le \mathcal O(r^{2})$ & Sec.~\ref{sec:growth-eom}\\
$G_{2,F}$ & M,\,V & $\mathcal O(1)$;\ \ $G_{2,F}|_0>0$ & Secs.~\ref{sec:growth-metric}, \ref{sec:growth-stability}\\
$G_{2,\tilde F}$ & M,\,V & $\le \mathcal O(r^{1})$ & Sec.~\ref{sec:growth-metric}\\
$G_{2,\tilde F\Phi}$ & V & $\le \mathcal O(r^{1})$ & Sec.~\ref{sec:growth-eom}\\
$G_{2,Y}$ & M,\,V & $\le \mathcal O(r^{2})$ & Sec.~\ref{sec:growth-metric}\\
$G_{2,XF},\,G_{2,XY},\,G_{2,F\Phi},\,G_{2,YX},\,\dots$ & S,\,V & no independent constraint$^{\dagger}$ & Sec.~\ref{sec:growth-eom}\\
\midrule
\multicolumn{4}{@{}l}{\textit{$L_3$: \ $G_3,\hat G_3,\mathring{\hat G}_3\,(\Phi,X)$}}\\
$G_3$ & M,\,S & $\le \mathcal O(r^{-1})$ & Sec.~\ref{sec:growth-metric}\\
$G_{3,\Phi}$ & M & $\mathcal O(1)$ & Sec.~\ref{sec:growth-metric}\\
$G_{3,X}$ & M & $\le \mathcal O(r^{2})$ & Sec.~\ref{sec:growth-metric}\\
$G_{3,\Phi X}$ & M & $\le \mathcal O(r^{3})$ & App.~\ref{app:higher}\\
$G_{3,XX}$ & M & $\le \mathcal O(r^{5})$ & App.~\ref{app:higher}\\
$\hat G_3$ & M & $\le \mathcal O(r^{2})$ & Sec.~\ref{sec:growth-metric}\\
$\hat G_{3,X}$ & M & $\le \mathcal O(r^{5})$ & Sec.~\ref{sec:growth-metric}\\
$\mathring{\hat G}_3$ & M & $\le \mathcal O(r^{4})$ & Sec.~\ref{sec:growth-metric}\\
$\hat G_{3,\Phi},\ \mathring{\hat G}_{3,\Phi},\ \mathring{\hat G}_{3,X},\,\dots$ & M & no independent constraint$^{\dagger}$ & App.~\ref{app:higher}\\
\midrule
\multicolumn{4}{@{}l}{\textit{$L_4$: \ $G_4,\hat G_4\,(\Phi,X)$, \ $\mathring{\hat G}_4(\Phi)$}}\\
$G_4$ & M & $\mathcal O(1),\ \neq0$;\ \ $G_4|_0>0$ & Secs.~\ref{sec:svt-flatness}, \ref{sec:growth-stability}\\
$G_{4,\Phi}$ & M & $\mathcal O(1)^{\star}$\ [ceiling $\mathcal O(r^{-1})$] & Secs.~\ref{sec:svt-flatness}, \ref{sec:growth-metric}\\
$G_{4,\Phi\Phi}$ & M & $\mathcal O(1)$ & Sec.~\ref{sec:growth-metric}\\
$G_{4,X}$ & M & $\le \mathcal O(r^{1})$ & Sec.~\ref{sec:growth-metric}\\
$G_{4,\Phi X}$ & M & $\le \mathcal O(r^{2})$ & App.~\ref{app:higher}\\
$G_{4,XX}$ & M & $\le \mathcal O(r^{4})$ & App.~\ref{app:higher}\\
$\hat G_4$ & M & $\le \mathcal O(r^{1})$ & Sec.~\ref{sec:growth-metric}\\
$\hat G_{4,X}$ & M & $\le \mathcal O(r^{1})$ & Sec.~\ref{sec:growth-metric}\\
$\mathring{\hat G}_4$ & M & $\le \mathcal O(r^{1})$ & Sec.~\ref{sec:growth-metric}\\
$\hat G_{4,\Phi},\ \hat G_{4,\Phi\Phi},\ \hat G_{4,\Phi X},\ \hat G_{4,XX},\ \mathring{\hat G}_{4,\Phi},\,\dots$ & M & no independent constraint$^{\dagger}$ & App.~\ref{app:higher}\\
\midrule
\multicolumn{4}{@{}l}{\textit{$L_5=G_5(\Phi,X)\,G^{\mu\nu}\Phi_{\mu\nu}+\dots$}}\\
$G_5$ & M,\,S & $\mathcal O(1)$ & Sec.~\ref{sec:growth-eom}\\
$G_{5,\Phi}$ & M,\,S & $< \mathcal O(r^{1})$ & Sec.~\ref{sec:growth-eom}, Sec.~\ref{sec:growth-metric}\\
$G_{5,X}$ & M & $< \mathcal O(r^{3})$ & Sec.~\ref{sec:growth-metric}, App.~\ref{app:higher}\\
$G_{5,\Phi\Phi}$ & M & $< \mathcal O(r^{2})$ & Sec.~\ref{sec:growth-metric}\\
$G_{5,XX}$ & M & $< \mathcal O(r^{6})$ & Sec.~\ref{sec:growth-metric}, App.~\ref{app:higher}\\
$G_{5,\Phi X},\,\dots$ & M & no independent constraint$^{\dagger}$ & App.~\ref{app:higher}\\
\bottomrule
\end{tabular}
\end{table}
\noindent\begingroup\footnotesize\raggedright
\textit{Notes to Table~\ref{tab:master-constraints}.}
$^{\star}$ The one entry whose generic value exceeds its kinematic
ceiling instead of staying below it: $G_{4,\Phi}$ multiplies the bare
improvement Hessian $\Phi_{ab}$, whose component $\Phi_{uu}=\mathcal O(r^{-1})$
decays one power more slowly than the bound $T_{uu}=\mathcal O(r^{-2})$
(Eq.~\eqref{eq:G4-violation}), so the metric falloffs alone would force
$G_{4,\Phi}(\phi_0)=0$ (ceiling $\mathcal O(r^{-1})$). A finite $G_{4,\Phi}(\phi_0)\neq0$
-- the conformally coupled sector, e.g.\ BD theory with
$G_4=\Phi$ -- remains
admissible because the offending terms are total retarded-time (Schott)
derivatives, absorbed by a mass-aspect redefinition or removed by
passing to the Einstein frame when $G_4=G_4(\Phi)$, see the $L_4$
paragraph of Sec.~\ref{sec:svt-flatness}.\\
$^{\dagger}$ ``No independent constraint'': the vacuum value is a free
Taylor coefficient, and no separate growth ceiling is imposed on these
derivatives -- they are restricted only through their parent
derivatives. For regular coupling functionals they appear inside
structures already suppressed below every bound of the table. The daggered rows are the derivatives of
the parity-odd and curvature mixings and the mixed $G_2$
cross-derivatives such as $G_{2,XF}$, which enter the field equations
only through gradients of the first derivatives (e.g.\
$\nabla_aG_{2,X}$, $\nabla_aG_{2,F}$ in
Eqs.~\eqref{eq:app-scalar-EOM-full}, \eqref{eq:app-G-displacement}). For
singular coupling functionals no per-derivative ceiling is quoted: each additional
$Z$-derivative merely raises the ceiling by the weight $w_Z$
(Eq.~\eqref{eq:growth-weights}, App.~\ref{app:higher}), so the binding
constraints are the operator-level ceilings of
Eq.~\eqref{eq:mixing-summary}, which the vector equation of motion does
not tighten (Sec.~\ref{sec:growth-eom}).\par
\endgroup



\section{Killing vector fields in asymptotically flat SVT theory}
\label{sec:Killing_SVT}

Having established in Sec.~\ref{sec:flatness_SVT} that the full SVT action $L_2+L_3+L_4+L_5$ obeying the conditions Eq.~\eqref{eq:svt-flatness-summary} produces a $T^{\rm eff}_{\mu\nu}/G_4$ respecting the falloff table Eq.~\eqref{eq:T-falloff}, we now spell out the consequence for the asymptotic symmetry group. We stress from the outset that this step carries no independent dynamical content. The asymptotic symmetry group is defined by the Bondi gauge conditions and the metric falloffs alone, so once the SVT solutions realize the same falloffs as vacuum GR, the vacuum BMS Killing vector fields are recovered by construction. The key insight of this work is therefore the falloff analysis of Sec.~\ref{sec:flatness_SVT}. Here we simply record its corollary, make explicit where the dynamics does (and does not) enter, and verify the one point that is not automatic, i.e. that the BMS flow also preserves the boundary conditions of the SVT matter fields, as required for the covariant phase space applications anticipated in Sec.~\ref{sec:conclusion}.

\subsection{Asymptotic Killing equation in Bondi gauge}
\label{sec:Killing-setup}

A vector field $\xi^a$ generates an asymptotic isometry of an asymptotically flat Bondi--Sachs spacetime if the variation $\delta g_{ab} =-\mathcal L_\xi g_{ab}=-\nabla_a\xi_b-\nabla_b\xi_a$ preserves the Bondi gauge Eq.~\eqref{eq:bs-metric} exactly and the asymptotic falloffs Eq.~\eqref{eq:h-exp} at the rate demanded by the BMS gauge. The exact gauge conditions $g_{rr}=g_{rA}=0$ and $\det(h_{AB})=\det(q_{AB})$ translate to
\be
\label{eq:dg-exact}
\delta g_{rr}=0,\qquad \delta g_{rA}=0,\qquad q^{AB}\delta g_{AB}=0,
\ee
while the asymptotic falloffs require
\be
\label{eq:dg-asymp}
\delta g_{uu}=\mathcal O(r^{-1}),\;\; \delta g_{uA}=\mathcal O(r^0),\;\; \delta g_{ur}=\mathcal O(r^{-2}),\;\; \delta g_{AB}^{\rm TF}=\mathcal O(r),
\ee
the last entry referring to the trace-free part of $\delta g_{AB}$. These are exactly conditions (6.3.1)--(6.3.2) of \cite{Alessio:2018rgo}.

\subsection{Why only the leading metric behavior is relevant}
\label{sec:Killing-SVT-metric}

The asymptotic Killing equations \eqref{eq:dg-exact}--\eqref{eq:dg-asymp} are purely geometric in the sense that they constrain the variation $\delta g_{ab}=-\mathcal L_\xi g_{ab}$ of the metric under the flow of $\xi$ and make no reference to the dynamical matter content. The matter sector enters the analysis only indirectly, through Einstein's equation, by determining which metric falloffs are realized on-shell. Once those falloffs are in hand, the asymptotic Killing analysis is dynamically decoupled from the matter.

The decoupling is concrete. The Killing equation $\delta g_{ab}=-\nabla_a\xi_b-\nabla_b\xi_a$, expanded in powers of $1/r$, becomes at each order an algebraic--differential constraint between the components of $\xi^a$, the Christoffels of $g_{ab}$, and the order-by-order metric expansion coefficients ($\beta$, $U^A$, $V$, $h_{AB}$). Every coefficient that survives in the leading-order constraint is the leading piece of a metric function.\footnote{Take for instance $g_{ab}\to\eta_{ab}^{\rm Bondi}$ on Eq.~\eqref{eq:svt-leading-metric}, supplemented at the next orders by the radiative shear $c_{AB}/r$, the mass aspect $-2M/r$, and the angular-momentum aspect, which enters $g_{uA}$ at $\mathcal O(r^{-1})$ through the $\mathcal O(r^{-3})$ coefficient of $U^A$.} The matter sector, by contrast, appears at the level of the subleading expansion coefficients through Einstein's equation. For instance, an SVT $T^{\rm eff}_{\mu\nu}/G_4$ obeying the falloff table Eq.~\eqref{eq:T-falloff} sources a metric expansion of the same form as in vacuum, with possibly different values of $M$, $c_{AB}$, $L^A$, $\beta_{(2)}$, \dots\ The one exception to this statement stems from the slowest matter falloffs the table admits. In particular, $T_{rr}=\mathcal O(r^{-3})$ and $T_{rA}=\mathcal O(r^{-2})$ would, by Eqs.~\eqref{eq:beta-sol} and \eqref{eq:U-sol}, switch on a coefficient that vacuum forbids, $\beta_{(1)}\neq0$, i.e.\ $g_{ur}=-1+\mathcal O(r^{-1})$. Crucially, the algebraic structure of the leading metric is unchanged in either case, and so the leading-order step of the asymptotic Killing analysis is identical.

More interesting in the $\beta_{(1)}\neq 0$--case is the subleading order of the metric structure. In the case of $\beta_{(1)}\neq 0$, the asymptotic conditions Eq.~\eqref{eq:dg-asymp}, which presuppose $g_{ur}=-1+\mathcal O(r^{-2})$, would have to be relaxed and the Killing analysis repeated on the enlarged phase space. For the SVT class the exception does not materialize, however. Inspecting the component scans of App.~\ref{app:svt-components}, the only contribution to $T^{\rm eff}_{rr}$ at $\mathcal O(r^{-3})$ and likewise to $T^{\rm eff}_{rA}$ at $\mathcal O(r^{-2})$ is the improvement term of a non-minimal coupling $G_{4,\Phi}(\phi_0)\neq0$, Eq.~\eqref{eq:G4-schott}. Every other operator enters at least one power faster. In the frame in which the standard Bondi--Sachs identifications hold, i.e., the Einstein frame whenever $G_{4,\Phi}(\phi_0)\neq0$, cf.\ the $L_4$ discussion of Sec.~\ref{sec:svt-flatness}, the hierarchy of Secs.~\ref{sec:matter-beta}--\ref{sec:matter-supp} therefore produces $\beta=\mathcal O(r^{-2})$, $U^A=\mathcal O(r^{-2})$, $V/r=1-2M(u,x^C)/r+\mathcal O(r^{-2})$, $h_{AB}=q_{AB}+c_{AB}(u,x^C)/r+\mathcal O(r^{-2})$, identical in operator form to the vacuum Bondi--Sachs case, with the matter merely shifting the values of the coefficients. The asymptotic Killing analysis then proceeds in exactly the same way and yields exactly the same set of generators, regardless of which operators in $L_2,L_3,L_4,L_5$ have been switched on.

Let us recapitulate how exactly the BMS group is obtained from Eq. \eqref{eq:dg-exact} and following, using the SVT metric and its corresponding matter contributions regulated by Table \ref{tab:master-constraints}: Reproducing the chain of solutions of \cite{Alessio:2018rgo}, the first of Eq.~\eqref{eq:dg-exact} yields $\xi_r=f(u,x^C)\,e^{2\beta}$ with $f$ an unspecified function on $\scrip$. The second gives $\xi_A=r^2\bigl[-\partial^A f+U^A f+\int_r^\infty r'^{-2}e^{2\beta}h^{BA} \partial_B f\,dr'\bigr]$. The third determines $\xi_u$ algebraically. The asymptotic conditions Eq.~\eqref{eq:dg-asymp} then constrain $f$ and $f^A\equiv-\lim_{r\to\infty}\xi^A$ to obey
\be
\label{eq:Killing-constraints}
\partial_u f^A=0,\quad D_A f_B+D_B f_A=q_{AB}\,D_C f^C,\quad \partial_u f=\tfrac12 D_A f^A,
\ee
i.e.\ $f^A=f^A(x^B)$ is a conformal Killing vector of the unit-sphere metric $q_{AB}$ (generating Lorentz transformations) and $f=\alpha(x^C)+\tfrac{u}{2}D_A f^A$ with $\alpha$ an arbitrary smooth function on $S^2$ (generating supertranslations). The asymptotic Killing vector field is therefore
\be
\label{eq:BMS-KV}
\; \xi=\xi^a\partial_a=\Bigl[\alpha(x^C)+\tfrac{u}{2}D_A f^A(x^C)\Bigr] \partial_u+f^A(x^C)\,\partial_A+\xi^r\,\partial_r, \;
\ee
with the radial component fixed by the gauge to
\be
\label{eq:xi-r}
\xi^r=-\tfrac{r}{2}D_C f^C+\tfrac12 D_C D^C\alpha +\tfrac{u}{4}D_C D^C(D_A f^A)+\mathcal O(r^{-1}).
\ee
The Lie algebra generated by Eq.~\eqref{eq:BMS-KV} is the BMS algebra. The supertranslation generators $\{\alpha(x^C)\partial_u\}$ form an abelian infinite-dimensional ideal, while the Lorentz generators $\{f^A\partial_A+\tfrac{u}{2}D_A f^A\partial_u\}$ form the homogeneous factor $\mathrm{SO}(3,1)$ (or, for locally defined $f^A$, the algebra of conformal Killing vectors of $S^2$). The fact that exactly these generators emerge expresses that the asymptotic isometries of any asymptotically flat SVT spacetime in the sense of Eq.~\eqref{eq:svt-flatness-summary} form the BMS group.

It remains to verify that the BMS flow maps SVT solutions to SVT solutions, i.e.\ that it preserves the boundary conditions of the matter fields along with those of the metric. For the scalar, inserting Eqs.~\eqref{eq:BMS-KV}--\eqref{eq:xi-r} into $\delta_\xi\Phi=\xi^a\partial_a\Phi$ gives
\be
\label{eq:matter-BMS-flow}
\delta_\xi\Phi =\frac{1}{r}\Bigl[f\,\partial_u\phi_1+f^A\partial_A\phi_1 +\tfrac12\bigl(D_Cf^C\bigr)\phi_1\Bigr]+\mathcal O(r^{-2}) .
\ee
The potentially dangerous term $\xi^r\partial_r\Phi$, in which $\xi^r$ grows linearly in $r$, is tamed by the decay $\partial_r\Phi=-\phi_1/r^2+\mathcal O(r^{-3})$ and contributes precisely the conformal-weight piece $\tfrac12(D_Cf^C)\,\phi_1/r$. The constant background $\phi_0$ is untouched, and the radiative ansatz $\Phi=\phi_0+\phi_1/r+\mathcal O(r^{-2})$ of Sec.~\ref{sec:svt-flatness} is reproduced with transformed data. Thus, the scalar sector is BMS-covariant. For the vector one finds analogously $\delta_\xi\mathcal A^{(0)}_A =f\,\partial_u\mathcal A^{(0)}_A +f^B\partial_B\mathcal A^{(0)}_A +\mathcal A^{(0)}_B\,\partial_Af^B+\mathcal O(r^{-1})$, with every other component transforming within its falloff up to a compensating $U(1)$ transformation that restores the radiation gauge $A_r=0$. The solution space of Sec.~\ref{sec:flatness_SVT} therefore carries a genuine action of the BMS group.

Two remarks are in order. First, relaxing the determinant condition $\det(h_{AB})=\det(q_{AB})$ enlarges the asymptotic group to its generalized BMS extensions exactly as in GR \cite{Flanagan_2017}. Nothing in the SVT matter sector interferes with that freedom. Second, the BMS group exhausts the diffeomorphism sector of the asymptotic symmetries. Since the SVT theory carries the $U(1)$ field $A_\mu$, its full asymptotic symmetry group also contains the large gauge transformations $A_\mu\to A_\mu+\partial_\mu\varepsilon$ with $\varepsilon\to\varepsilon^{(0)}(x^A)$ at $\scrip$, organizing into a semidirect product with BMS.




\section{Conclusion}
\label{sec:conclusion}
We have shown that the asymptotic decay of the Bondi--Sachs metric functions imposes a minimal, sectoral set of falloff conditions on the matter stress-energy tensor, summarized in the table Eq.~\eqref{eq:T-falloff}. By rigorously integrating the Bondi--Sachs hierarchy with a generic $T_{ab}$ we computed how each covariant component of the stress-energy tensor enters the metric functions $\beta$, $U^A$, $V$, and $h_{AB}$ at each order in $1/r$, and read off the slowest admissible falloff in each sector. The framework was then verified explicitly on BD gravity (Sec.~\ref{sec:matter-bd}) as well as on a free massless scalar field (App.~\ref{app:matter-scalar}) and on Einstein--Maxwell theory (App.~\ref{app:matter-maxwell}). In the latter two cases the matter sector satisfies the falloff conditions automatically and the standard Bondi--Sachs structure is preserved without further restriction, while BD illustrates how a non-minimal coupling is accommodated through the choice of frame.

Applying this framework to the full SVT action of \cite{Heisenberg:2018acv,Heisenberg:2023prj}, recast in the effective Einstein form $G_{\mu\nu}=T^{\rm eff}_{\mu\nu}/G_4(\Phi,X)$ Eq.~\eqref{eq:svt-Einstein}, we evaluated the leading-order contribution of each Lagrangian sector to $T^{\rm eff}_{\mu\nu}$ against Eq.~\eqref{eq:T-falloff} component by component. The outcome, sharpened by consistency of the scalar and vector equations of motion and by stability of the asymptotic vacuum, is condensed in the constraint Table~\ref{tab:master-constraints}: The asymptotic Newton constant must be finite and positive, i.e. $G_4(\phi_0,0)>0$, together with the kinetic normalizations $G_{2,X}|_0>0$, $G_{2,F}|_0>0$, the $L_2$ scalar potential must vanish to cubic order in $\Phi$, corresponding to no cosmological constant, no tadpole, no mass term, and the conformally coupled sector $G_{4,\Phi}(\phi_0)\neq0$ requires the Einstein-frame reading of the Bondi--Sachs data. Every other coupling functional, $G_3,\hat G_3,\mathring{\hat G}_3,G_5,\hat G_4,\mathring{\hat G}_4$ and their derivatives, remains unconstrained at the vacuum Eq.~\eqref{eq:svt-flatness-summary}, protected by the structure of the theory itself. As a corollary, the asymptotic Killing analysis of Sec.~\ref{sec:Killing_SVT} reproduces, verbatim, the vacuum BMS Killing vector fields Eqs.~\eqref{eq:BMS-KV}--\eqref{eq:xi-r}, and the BMS flow moreover preserves the boundary conditions of the matter fields, so that the BMS group acts as the asymptotic symmetry group across the entire admissible SVT class.

Two geometric features of the analysis, documented in detail in Sec.~\ref{sec:flatness_SVT}, are responsible for the wide admissibility of the SVT coupling functionals. The non-decaying $\Phi_{AB}=\mathcal O(r^0)$ piece of the covariant scalar Hessian, sourced by the Christoffel correction $\Gamma^u_{AB}\partial_u\Phi$ inherent in the Bondi gauge, allows the $L_4$ and $L_5$ operators involving $\Phi_{\mu\nu}$ to enter the radiative trace flux through pure-trace $\propto q_{AB}$ contributions that remain consistent with Eq.~\eqref{eq:T-falloff} on every covariant component. Independently, the divergence-free tensors $G^{\mu\nu}$ and $L^{\mu\nu\alpha\beta}$ which guarantee the second-order character of $L_4,L_5$ vanish identically on the strict leading metric and are sourced only at $\mathcal O(r^{-1})$ by the radiative shear, automatically suppressing the curvature--matter mixings. The combined effect renders $\hat G_4(\phi_0,0)$, $\mathring{\hat G}_4(\phi_0)$ and $G_5(\phi_0,0)$ admissible at the asymptotic vacuum without any tuning.

The framework presented here is significant for several reasons. First, it provides a minimal and explicit criterion for SVT theories and, by direct extension, for any structurally analogous theory whose Lagrangian builds from the same scalar, vector and curvature operators, to admit BMS symmetry and asymptotic flatness in the Bondi sense. The criterion is local at the asymptotic vacuum in the sense that it constrains only the Taylor coefficients of the SVT coupling functionals $G_i$ about the vacuum $(\Phi,X,Y,F,\tilde F)=(\phi_0,0,0,0,0)$, leaving the bulk behavior of those functions arbitrary. This locality is important because it makes the condition easy to impose by hand on any concrete SVT model. For instance, a BD theory with a non-trivial potential is admissible iff the potential vanishes at cubic order at the asymptotic vacuum. It further ensures that the asymptotic structure is robust against deformations of the theory that leave its asymptotic vacuum unchanged. The practical form of the criterion is the constraint table, Table~\ref{tab:master-constraints}. Given any concrete beyond-GR model built from the same operator dictionary, the obstructions to asymptotic flatness can be read off sector by sector, without solving the field equations. This feature is particularly valuable for phenomenological survey work, in which a large family of candidate theories must be screened for a healthy asymptotic infrared structure before any waveform-level analysis is attempted.

Second, the existence of a closed-form effective Einstein equation $G_{\mu\nu}=T^{\rm eff}_{\mu\nu}/G_4(\Phi,X)$ with $T^{\rm eff}_{\mu\nu}$ obeying the standard falloff table is precisely the input required to apply the Wald--Zoupas covariant phase space approach \cite{waldGeneralDefinitionConserved2000}. In a forthcoming work we will exploit this to construct the symplectic potential, the conserved charges associated with each generator of the BMS algebra, and the corresponding flux laws across $\scrip$ for the SVT family. The displacement memory effect, which in GR is encoded in the permanent shift $\Delta c_{AB}$ of the shear between successive Bondi sections, generalizes in SVT to a richer set of memory tensors built from $\Delta\phi_1$, $\Delta q$, $\Delta\mathcal A^{(0)}_A$, in addition to the metric shear. Their structure is fixed by the leading $T^{\rm eff}_{\mu\nu}$ entries computed here and will be the subject of the same follow-up work. The infrared triangle of \cite{strominger2018lecturesinfraredstructuregravity} thereby acquires a concrete realization in SVT, with the corresponding soft graviton, soft photon, and soft scalar theorems all linked to the asymptotic symmetry generators identified here.

Third, a natural question raised by the regularity theme of Sec.~\ref{sec:flatness_SVT} is what happens if a coupling functional is genuinely logarithmic at the asymptotic vacuum as radiative corrections generically induce, e.g.\ $G_2\supset X\ln(X/\mu^4)$ from the running of the kinetic term, or $G_4\supset(\Phi-\phi_0) \ln(\Phi-\phi_0)$ from a scalar anomalous dimension. Such terms lie outside the power class Eq.~\eqref{eq:power-class} underlying Sec.~\ref{sec:growth-metric}, and an order-counting exercise shows that the answer splits along the marginality structure of the constraint table. Wherever a coupling functional sits strictly below its allowance, a logarithmic dressing is innocuous as at least one power of $1/r$ of headroom is available, and $r^{-1}\ln r\to0$ preserves every entry of the falloff table. In the marginal slots, however the logarithm bites. For $G_{2,X}\supset\ln X$ the background value $\ln X\simeq-3\ln r$ promotes the energy flux to $T_{uu}\sim\ln r/r^{2}$, so that $\lim_{r\to\infty}(r^{2}T^{r}{}_{u})$ diverges logarithmically, and the scalar equation of motion acquires the contribution $G_{2,XX}\,\Phi^{a}\nabla_{a}X\sim X^{-1}\cdot \mathcal O(r^{-5})=\mathcal O(r^{-2})$, one power slower than the free-wave tolerance of Eq.~\eqref{eq:boxphi-freewave}. Neither pathology destroys asymptotic flatness outright. Both instead signal that the pure $1/r$ ansatz must be replaced by a polyhomogeneous double expansion in $1/r$ and $\ln r$, as studied in \cite{Kehrberger_2021}, in which the scalar profile acquires $\ln r/r^{2}$ corrections, the metric functions develop logarithmic tails, and the log-removal constraints of Secs.~\ref{sec:matter-U}--\ref{sec:matter-V} turn from algebraic conditions into evolution equations for the new coefficients. The leading Bondi structure survives, since logarithms never compete with powers, and we expect $\scrip$ and the BMS group itself to be untouched. The identification of the mass and angular-momentum aspects, however, acquires a logarithmic running, and the finiteness and split of the flux balances must be re-derived. For the conformally coupled sector the situation is more severe. With $G_{4,\Phi}\sim\ln(\Phi-\phi_0)$ the conformal factor relating Jordan and Einstein frames is itself polyhomogeneous, and the total-derivative structure Eq.~\eqref{eq:G4-schott} that rescued the constant-$G_{4,\Phi}(\phi_0)$ case no longer closes, because the offending coefficients acquire explicit $\ln r$ dependence. A systematic treatment of logarithmic coupling functionals within the polyhomogeneous framework is left for future work.

Finally, the analytic blueprint established here, namely the systematic translation of asymptotic-flatness requirements into Lagrangian-level constraints on the coupling functionals of a beyond-GR theory, generalizes immediately to other classes of modified gravity. Higher-derivative theories, theories with multiple scalar or vector fields, massive-gravity extensions, theories with non-minimal couplings to matter, and Chern--Simons-modified gravity all admit a parallel analysis. After casting into the effective Einstein form, one can identify the falloffs of $T^{\rm eff}_{\mu\nu}$ component by component, demand compatibility with Eq.~\eqref{eq:T-falloff}, and read off the constraints on the Lagrangian parameters at the asymptotic vacuum. We expect this program to provide a unified perspective on which corners of the beyond-GR landscape admit a Bondi--Sachs asymptotic structure and, through the analyses of GW flux laws, memory, and soft theorems that the BMS arena enables, which corners are amenable to direct observational comparison with GR in the strong-field, radiative regime now opened by gravitational-wave astronomy.




\appendix

\section{The free massless scalar field as an example}
\label{app:matter-scalar}

In this appendix we repeat the construction of Secs.~\ref{sec:matter-beta}--\ref{sec:matter-supp} for the matter content of a free massless scalar, treating every metric function as an unconstrained smooth $1/r$ series and using only the asymptotic-flatness conditions $\lim_{r\to\infty}\beta=0$, $\lim_{r\to\infty}U^A=0$, $\lim_{r\to\infty}V/r=1$, $h_{AB}\to q_{AB}$. The decay rate of each metric function emerges as an output of the integration of the corresponding Einstein equation with the scalar source, not as an input from the general analysis of the main text.

The scalar stress-energy is
\be
\label{eq:scalar-stress}
T_{ab} = \partial_a\Phi\,\partial_b\Phi - \tfrac12 g_{ab}(\partial\Phi)^2 .
\ee

Using $g_{rr}=g_{rA}=0$, $g_{ur}=-e^{2\beta}$, $g_{uA}=-r^2 h_{AB}U^B$, and $g_{AB}=r^2 h_{AB}$, the components of Eq.~\eqref{eq:scalar-stress} read, without any $1/r$ expansion,
\be
\label{eq:scalar-Tab-closed}
\begin{aligned}
T_{rr}&=(\partial_r\Phi)^2, & T_{rA}&=\partial_r\Phi\,\partial_A\Phi,\\
T_{ur}&=\partial_u\Phi\,\partial_r\Phi+\tfrac12 e^{2\beta}(\partial\Phi)^2, & T_{AB}&=\partial_A\Phi\,\partial_B\Phi-\tfrac12 r^2 h_{AB}(\partial\Phi)^2,\\
T_{uu}&=(\partial_u\Phi)^2+\tfrac12\!\big(\tfrac{V}{r}e^{2\beta}-r^2 h_{CD}U^C U^D\big)(\partial\Phi)^2, & T_{uA}&=\partial_u\Phi\,\partial_A\Phi+\tfrac12 r^2 h_{AB}U^B(\partial\Phi)^2,
\end{aligned}
\ee
with $ (\partial\Phi)^2 = e^{-2\beta}\big[-2\partial_u\Phi\partial_r\Phi+\tfrac{V}{r}(\partial_r\Phi)^2 -2U^A\partial_r\Phi\partial_A\Phi\big] + h^{AB}\partial_A\Phi\partial_B\Phi/r^2. $

For the massless scalar we expand, towards $\scrip$, as
\be
\label{eq:scalar-phi-ansatz}
\Phi(u,r,x^A) = \phi_0 + \frac{\phi_1(u,x^A)}{r}+\frac{\phi_2(u,x^A)}{r^2}+\mathcal O(r^{-3}),
\ee
$\phi_1$ and $\phi_2$ being unknown functions to be determined alongside the metric. Because the \emph{minimally coupled} scalar of this section enters the Lagrangian only through its derivatives, its constant asymptotic value $\phi_0$ is pure gauge here and we set $\phi_0=0$ in what follows. This is special to minimal coupling. In the non-minimally coupled SVT theory of Sec.~\ref{sec:flatness_SVT} the coupling functionals depend on $\Phi$ \emph{undifferentiated}, so that $\phi_0\neq0$ is physical, fixing the asymptotic Newton constant exactly as in BD, and is retained. By the Bondi determinant gauge $\det h_{AB}=\det q_{AB}$ combined with $h_{AB}\to q_{AB}$, the angular block admits
\be
\label{eq:scalar-h-ansatz}
\begin{aligned}
h_{AB}(u,r,x^C) &= q_{AB}+\frac{c_{AB}(u,x^C)}{r}+\frac{d_{AB}(u,x^C)}{r^2}+\mathcal O(r^{-3}),\\
q^{AB}c_{AB}&=0,\qquad q^{AB}d_{AB}=\tfrac12 c^{EF}c_{EF} .
\end{aligned}
\ee
No prior assumption is made on the decay rates of $\beta$, $U^A$, $V$, since the asymptotic-flatness conditions $\lim_{r\to\infty}\beta=0$, $\lim_{r\to\infty}U^A=0$, $\lim_{r\to\infty}V/r=1$ are the only inputs. The subleading coefficients are derived sequentially below, namely $\beta$ first (needing only $T_{rr}$), then $U^A$ (needing $T_{rA}$ and $\beta$), then $V$ (needing $T_{ur},T_{AB},T^{a}{}_{a}$ together with $\beta,U^A$), then the evolution of $h_{AB}$ (needing trace-free $T_{AB}$), then the aspect laws (needing $T_{uu},T_{uA}$). Throughout we abbreviate $|\eth\phi_1|^2:=q^{AB}\partial_A\phi_1\partial_B\phi_1$.

\subsection{$\beta$-equation.}
The component $T_{rr}=(\partial_r\Phi)^2$ of Eq.~\eqref{eq:scalar-Tab-closed} is metric-independent. Substituting Eq.~\eqref{eq:scalar-phi-ansatz} gives
\be
\label{eq:scalar-Trr}
\partial_r\Phi = -\frac{\phi_1}{r^2}-\frac{2\phi_2}{r^3}+\mathcal O(r^{-4}),\qquad T_{rr} = \frac{\phi_1^2}{r^4}+\frac{4\phi_1\phi_2}{r^5}+\mathcal O(r^{-6}).
\ee
The $\beta$-equation in the form Eq.~\eqref{eq:beta-rhs} reads $\partial_r\beta=(r/16)h^{AC}h^{BD}(\partial_r h_{AB})(\partial_r h_{CD})+2\pi r T_{rr}$. Using Eq.~\eqref{eq:scalar-h-ansatz}, $\partial_r h_{AB}=-c_{AB}/r^2-2d_{AB}/r^3+\dots$ and $h^{AB}=q^{AB}-c^{AB}/r+\dots$, so the shear contribution evaluates to $c^{AB}c_{AB}/(16 r^3)+(c^{AB}d_{AB}/4)/r^4+\mathcal O(r^{-5})$. Combined with Eq.~\eqref{eq:scalar-Trr},
\be
\partial_r\beta = \frac{c^{AB}c_{AB}/16+2\pi\phi_1^2}{r^3} + \frac{c^{AB}d_{AB}/4+8\pi\phi_1\phi_2}{r^4}+\mathcal O(r^{-5}) .
\ee
There is no $r^{-2}$ piece on the right-hand side, because the scalar's $T_{rr}$ starts at $r^{-4}$, which automatically removes the logarithm threat, so direct integration with $\lim_{r\to\infty}\beta=0$ gives
\be
\label{eq:scalar-beta}
\beta = -\frac{c^{AB}c_{AB}/32+\pi\phi_1^2}{r^2} -\frac{c^{AB}d_{AB}/12+(8\pi/3)\phi_1\phi_2}{r^3}+\mathcal O(r^{-4}) .
\ee
The decay is thus $\beta=\mathcal O(r^{-2})$ and the two leading nontrivial coefficients are $\beta_{(2)}=-\tfrac1{32}c^{AB}c_{AB}-\pi\phi_1^2$ and $\beta_{(3)}=-\tfrac1{12}c^{AB}d_{AB}-\tfrac{8\pi}{3}\phi_1\phi_2$.

\subsection{$U^A$-equation.}
The $U^A$ hypersurface equation \eqref{eq:hyp-U}, integrated once in $r$, reads schematically
\be
\partial_r\!\big[r^4 e^{-2\beta}h_{AB}\partial_r U^B\big] = (\text{geometric source}) + 16\pi r^2 T_{rA} .
\ee
For the scalar,
\be
16\pi r^2 T_{rA}=16\pi r^2\,\partial_r\Phi\,\partial_A\Phi =-\frac{16\pi\,\phi_1\,\partial_A\phi_1}{r}+\mathcal O(r^{-2}) ,
\ee
which contains no $r^2,r,r^0$ pieces. The two obstructions encountered in Sec.~\ref{sec:matter-U} for general matter, namely the $r^3$-coefficient of $F_A:=r^4 e^{-2\beta}h_{AB}\partial_r U^B$ that produces $U^A\sim\ln r$ and the $r$-coefficient that conflicts with $\lim U^A=0$, are therefore absent for the scalar. They would have required $\Tc{0}_{rA}\neq 0$ or $\Tc{1}_{rA}\neq 0$ in the source, and the scalar simply has neither. With $\beta=\mathcal O(r^{-2})$ derived in Eq.~\eqref{eq:scalar-beta} (so $D_A\beta=\mathcal O(r^{-2})$ adds no further growth) and the leading geometric shear source contributing $\eth^E c_{AE}\cdot r+\mathcal O(r^0)$, the two successive radial integrations subject to $\lim_{r\to\infty}U^A=0$ yield
\be
\label{eq:scalar-U}
U^A = -\frac{\eth_B c^{AB}}{2\,r^2}+\frac{2L^A+\tfrac13 c^{AE}\eth^F c_{EF}}{r^3} + \mathcal O(r^{-4}) ,
\ee
$L^A(u,x^B)$ being the integration ``constant'' at $\mathcal O(r^{-3})$. Thus $U^A$ is derived to be $\mathcal O(r^{-2})$, with $U^A_{(0)}=U^A_{(1)}=0$ vacuously and the two leading nontrivial coefficients $u^A_{(2)}=-\tfrac12\eth_B c^{AB}$, $u^A_{(3)}=2L^A+\tfrac13 c^{AE}\eth^F c_{EF}$ coinciding with the vacuum expressions. The scalar's $T_{rA}=\mathcal O(r^{-3})$ first feeds $U^A$ at $\mathcal O(r^{-4})$.

\subsection{$V$-equation.}
The $V$-equation \eqref{eq:hyp-V} has matter source $8\pi[h^{AB}T_{AB}-r^2\,T^{a}{}_{a}]$ in addition to the geometric metric terms. For the scalar, $T^{a}{}_{a}=g^{ab}T_{ab}=-(\partial\Phi)^2$, hence $r^2 T^{a}{}_{a}=-2\phi_1\partial_u\phi_1/r+\mathcal O(r^{-2})$, while $h^{AB}T_{AB}=-2\phi_1\partial_u\phi_1/r+\mathcal O(r^{-2})$, so the leading $\mathcal O(r^{-1})$ pieces of $h^{AB}T_{AB}$ and $r^2T^a{}_a$ cancel in the combination $h^{AB}T_{AB}-r^2 T^a{}_a$. The matter source of Eq.~\eqref{eq:hyp-V} therefore has no $r,r^0,r^{-1}$ pieces, and $\lim V/r=1$ is unobstructed: integration gives the leading $V=r$ and the integration constant at $\mathcal O(r^0)$ defines the mass aspect $V=r-2M(u,x^A)+\mathcal O(r^{-1})$. The next coefficient is fixed by the $\mathcal O(r^{-3})$ pieces of the source: from Eq.~\eqref{eq:scalar-Trr}, $\Tc{4}_{rr}=\phi_1^2$, and from the closed-form $T_{ur}=\partial_u\Phi\partial_r\Phi+\tfrac12 e^{2\beta}(\partial\Phi)^2$ of Eq.~\eqref{eq:scalar-Tab-closed} at $r^{-4}$ (the $r^{-3}$ pieces of the two contributions canceling exactly), $\Tc{4}_{ur}=\tfrac12(\phi_1^2+|\eth\phi_1|^2)$. Combining these with the matter-modified $\beta_{(2)}=-(c^{AB}c_{AB}/32+\pi\phi_1^2)$ from Eq.~\eqref{eq:scalar-beta} and evaluating the right-hand side of Eq.~\eqref{eq:hyp-V} at $\mathcal O(r^{-3})$ yields
\be
\label{eq:scalar-V}
V = r - 2M + \frac{V_{(1)}^{\rm scalar}}{r}+\mathcal O(r^{-2}),\qquad V_{(1)}^{\rm scalar} = \tfrac{c^{AB}c_{AB}}{16}+2\pi\phi_1^2 -\tfrac12 \Sigma_{(2)}^{\rm(geom)}-4\pi\,|\eth\phi_1|^2 ,
\ee
where $\Sigma_{(2)}^{\rm(geom)}$ is the $\mathcal O(r^{-2})$ coefficient of the purely metric piece of the $V$-source given by Eq.~\eqref{eq:Sigma-geom}, evaluated on the scalar's metric coefficients. Substituting $\beta_{(1)}=0$, $\beta_{(2)}=-(c^{AB}c_{AB}/32+\pi\phi_1^2)$, $u^A_{(2)}=-\tfrac12\eth_B c^{AB}$ and $u^A_{(3)}=2L^A+\tfrac13 c^{AE}\eth^F c_{EF}$ into Eq.~\eqref{eq:Sigma-geom} gives
\be
\label{eq:scalar-Sigma-geom}
\begin{aligned}
\Sigma_{(2)}^{\rm(geom)} ={}& \mathcal R_{(2)} + \tfrac{1}{16}\eth^2\!\big(c^{AB}c_{AB}\big) + 2\,\eth_A L^A + \tfrac{1}{3}\eth_A\!\big(c^{AE}\eth^F c_{EF}\big)\\
&- \tfrac{1}{2}\,q_{AB}\big(\eth_C c^{AC}\big)\!\big(\eth_D c^{BD}\big) + 2\pi\,\eth^2\!\big(\phi_1^2\big) ,
\end{aligned}
\ee
with $\mathcal R_{(2)}$ the angular curvature scalar of Eq.~\eqref{eq:R2}. The first five terms reproduce the vacuum expression. The final $2\pi\,\eth^2(\phi_1^2)$ is the scalar's sole contribution to $\Sigma_{(2)}^{\rm(geom)}$, generated by the $-\pi\phi_1^2$ piece of $\beta_{(2)}$ through the $-2\eth^2\beta_{(2)}$ term of Eq.~\eqref{eq:Sigma-geom}. Inserting Eq.~\eqref{eq:scalar-Sigma-geom} into Eq.~\eqref{eq:scalar-V} and using $\eth^2(\phi_1^2)=2\phi_1\eth^2\phi_1+2|\eth\phi_1|^2$ yields the fully-expanded
\be
\label{eq:scalar-V1-explicit}
V_{(1)}^{\rm scalar} = V_{(1)}^{\rm vac} + 2\pi\phi_1^2 - 2\pi\,\phi_1\,\eth^2\phi_1 - 6\pi\,|\eth\phi_1|^2 ,
\ee
where $V_{(1)}^{\rm vac}=\tfrac{c^{AB}c_{AB}}{16}-\tfrac12\Sigma_{(2)}^{\rm(geom),vac}$ collects the purely gravitational pieces (the first five terms of Eq.~\eqref{eq:scalar-Sigma-geom} with their $-\tfrac12$ factor, plus $c^{AB}c_{AB}/16$ from $-2\beta_{(2)}^{\rm vac}$). The leading $r$ and $r^0$ structure of $V$ is thus derived, in that $V$ has no $r^2$ or higher growth by asymptotic flatness, the integration constant at $\mathcal O(r^0)$ is the mass aspect $-2M$, and the scalar shifts only the subleading coefficient through the three explicit $\phi_1$-dependent terms of Eq.~\eqref{eq:scalar-V1-explicit}.

\subsection{Evolution equation: news is free Cauchy data.}
The evolution equation \eqref{eq:evol}, after contraction with the complex null dyad $m^A m^B$ (which projects onto the trace-free part) and expansion at $\mathcal O(r^{-1})$, takes the form recorded in Eq.~\eqref{eq:evol-lead}:
\be
\label{eq:scalar-evol-lead}
m^A m^B\!\Big[\,\partial_u d_{AB} + 2\,\eth_A\eth_B\beta_{(1)} + 8\pi\,\Tc{1}_{AB}\,\Big] = 0 .
\ee
Here the time-derivative term descends from $r\partial_r(r\partial_u h_{AB})=-\partial_u d_{AB}/r+\mathcal O(r^{-2})$ of Eq.~\eqref{eq:evol}, the geometric source $-2\eth_A\eth_B\beta_{(1)}$ from $-2 e^{\beta}D_A D_B e^{\beta}$, and the matter source $-8\pi\Tc{1}_{AB}$ from $-8\pi e^{2\beta}T_{AB}$. The news $N_{AB}=\tfrac12\partial_u c_{AB}$ is annihilated by the operator $r\partial_r(r\,\cdot\,)$ at this order and does not appear in Eq.~\eqref{eq:scalar-evol-lead}. Eq.~\eqref{eq:scalar-evol-lead} therefore determines the dyadic projection of $\partial_u d_{AB}$, i.e.\ its trace-free part, and leaves the news $\partial_u c_{AB}$ as freely specifiable Cauchy data, as in vacuum.

Now evaluate $\Tc{1}_{AB}$ for the scalar from Eq.~\eqref{eq:scalar-Tab-closed}: at $r^{-1}$, $\partial_A\Phi\,\partial_B\Phi=\partial_A\phi_1\partial_B\phi_1/r^2$ is $\mathcal O(r^{-2})$ and contributes nothing, while $\tfrac12 r^2 h_{AB}(\partial\Phi)^2 = q_{AB}\phi_1\partial_u\phi_1/r+\mathcal O(r^{-2})$ (using $h_{AB}\to q_{AB}$ and the leading $(\partial\Phi)^2=2\phi_1\partial_u\phi_1/r^3$ from the $-2 e^{-2\beta}\partial_u\Phi\,\partial_r\Phi$ piece of Eq.~\eqref{eq:scalar-Tab-closed}). Hence
\be
T_{AB} = -q_{AB}\,\frac{\phi_1\partial_u\phi_1}{r} + \mathcal O(r^{-2}) ,
\ee
which is pure trace at $r^{-1}$, so the trace-free leading coefficient satisfies $\big(\Tc{1}_{AB}\big)^{\rm TF}=0$. Combined with $\beta_{(1)}=0$ from Eqs.~\eqref{eq:scalar-beta}, \eqref{eq:scalar-evol-lead} reduces to $m^A m^B\partial_u d_{AB}=0$: the trace-free part of $d_{AB}$ is time-independent at this order. The trace of $d_{AB}$ is fixed by $\det h_{AB}=\det q_{AB}$ to $q^{AB}d_{AB}=\tfrac12 c^{EF}c_{EF}$ and evolves through the news. At the next order in $1/r$, i.e.\ at $\mathcal O(r^{-2})$ in Eq.~\eqref{eq:evol}, Eq.~\eqref{eq:evol-NLO}, the trace-free part of $d_{AB}$ is sourced by the trace-free $\mathcal O(r^{-2})$ part of $T_{AB}$, which yields
\be
(T_{AB})^{\rm TF}_{(2)} = \partial_A\phi_1\,\partial_B\phi_1 - c_{AB}\,\phi_1\,\partial_u\phi_1 - \tfrac12 q_{AB}\,|\eth\phi_1|^2 ,
\ee
together with the corresponding subleading geometric pieces.

\subsection{Supplementary balances and aspect laws.}
With $\beta_{(1)}=0$ and $u^A_{(2)}=-\tfrac12\eth_B c^{AB}$ from Eqs.~\eqref{eq:scalar-beta},\eqref{eq:scalar-U}, we evaluate $T_{uu},T_{uA}$ from Eq.~\eqref{eq:scalar-Tab-closed} using these results. At leading order, $T_{uu}=(\partial_u\Phi)^2+\mathcal O(r^{-3})=(\partial_u\phi_1)^2/r^2+\mathcal O(r^{-3})$ (the $(\partial\Phi)^2$-bearing piece is $\mathcal O(r^{-3})$ because $g_{uu}\to-1$ at leading and $(\partial\Phi)^2=\mathcal O(r^{-3})$), and $T_{uA}=\partial_u\Phi\partial_A\Phi+\mathcal O(r^{-3})=\partial_u\phi_1\partial_A\phi_1/r^2+\mathcal O(r^{-3})$ (the $g_{uA}$ contribution being $\mathcal O(r^{-3})$ via $g_{uA}=-r^2 h_{AB}U^B=-u_{(2)A}+\mathcal O(r^{-1})$ together with $(\partial\Phi)^2=\mathcal O(r^{-3})$). Hence the scalar fluxes are $\Tc{2}_{uu}=(\partial_u\phi_1)^2$ and $\Tc{2}_{uA}=\partial_u\phi_1\partial_A\phi_1$. The supplementary balance $\lim r^2 R^{r}{}_{u}=8\pi\lim r^2 T^{r}{}_{u}$ gives, after direct substitution of the scalar values into the same $\mathcal O(r^{-2})$ computation of $R^{r}{}_{u}$ carried out in Sec.~\ref{sec:matter-supp},
\be
\label{eq:scalar-M-evol}
2\,\partial_u M = \eth_A\eth_B N^{AB}-N_{AB}N^{AB}-8\pi(\partial_u\phi_1)^2 ,
\ee
all $\Tc{3}_{rr}$-dependent and $\Tc{2}_{rA}$-dependent corrections of the general analysis being absent for the scalar (because $T_{rr}=\mathcal O(r^{-4})$ and $T_{rA}=\mathcal O(r^{-3})$ make the corresponding leading coefficients vanish). Integrating over the unit sphere yields the manifestly-non-positive Bondi mass-loss
\be
\label{eq:scalar-massloss}
\frac{\dd m}{\dd u}=-\frac{1}{8\pi}\oint N_{AB}N^{AB}\,\dd\Omega -\oint(\partial_u\phi_1)^2\,\dd\Omega .
\ee
Similarly, the angular-momentum balance $\lim r^2 R^{r}{}_{A}=8\pi\lim r^2 T^{r}{}_{A}$, specialized to the scalar (every shear$\times$matter and matter-quadratic correction of the general analysis dropping out because the relevant matter coefficients vanish, with the lone exception of the $\beta_{(2)}$ piece $-\pi\phi_1^2$ that survives via $\Tc{4}_{rr}=\phi_1^2\neq 0$), gives
\be
\label{eq:scalar-L-evol}
-3\,\partial_u L_A = \mathcal G_A - 8\pi\,\partial_u\phi_1\,\partial_A\phi_1 + \pi\,\partial_u\eth_A\!\big(\phi_1^2\big) ,
\ee
with $\mathcal G_A$ the vacuum geometric side ($\eth_A M$ plus the standard shear terms).

\subsection{Summary.}
The metric functions of a free scalar field, to the two leading nontrivial orders for each, are
\be
\label{eq:scalar-metric-summary}
\begin{aligned}
\beta &= -\frac{1}{r^2}\!\Big(\tfrac{1}{32}c^{AB}c_{AB}+\pi\phi_1^2\Big) -\frac{1}{r^3}\!\Big(\tfrac{1}{12}c^{AB}d_{AB}+\tfrac{8\pi}{3}\phi_1\phi_2\Big) +\mathcal O(r^{-4}),\\
U^A &= -\frac{\eth_B c^{AB}}{2r^2}+\frac{2L^A+\tfrac13 c^{AE}\eth^F c_{EF}}{r^3}+\mathcal O(r^{-4}),\\
V &= r - 2M + V_{(1)}^{\rm scalar}/r + \mathcal O(r^{-2}),\\
h_{AB} &= q_{AB} + \frac{c_{AB}}{r} + \frac{d_{AB}}{r^2}+\mathcal O(r^{-3}),\quad q^{AB}d_{AB}=\tfrac12 c^{EF}c_{EF},
\end{aligned}
\ee
with $V_{(1)}^{\rm scalar}$ from Eq.~\eqref{eq:scalar-V} and $\{\phi_1,\phi_2,c_{AB},M,L_A\}$ free retarded-time data evolved by Eqs.~\eqref{eq:scalar-M-evol}, \eqref{eq:scalar-L-evol} and the news equation. The corresponding stress-energy falloffs are
\be
\label{eq:scalar-Tab-summary}
\begin{aligned}
T_{rr}&=\phi_1^2/r^4+\mathcal O(r^{-5}), & T_{rA}&=-\phi_1\partial_A\phi_1/r^3+\mathcal O(r^{-4}),\\
T_{ur}&=\tfrac{\phi_1^2+|\eth\phi_1|^2}{2 r^4}+\mathcal O(r^{-5}), & T_{AB}&=-q_{AB}\phi_1\partial_u\phi_1/r+\mathcal O(r^{-2}),\\
T_{uu}&=(\partial_u\phi_1)^2/r^2+\mathcal O(r^{-3}), & T_{uA}&=\partial_u\phi_1\,\partial_A\phi_1/r^2+\mathcal O(r^{-3}).
\end{aligned}
\ee
Every component except $T_{uu}$ and $T_{uA}$ decays one power of $1/r$ faster than the generic bound established in Sec.~\ref{sec:matter-supp}: the slow-decay coefficients $\Tc{3}_{rr},\Tc{2}_{rA},\Tc{3}_{ur}$, $(T_{AB})^{\rm TF}_{(1)}$ and the generic $\mathcal O(1)$ piece of the angular trace all vanish outright for the scalar. The components $T_{uu}$ and $T_{uA}$ instead saturate the generic bound, with their leading coefficients $\Tc{2}_{uu}=(\partial_u\phi_1)^2$ and $\Tc{2}_{uA}=\partial_u\phi_1\partial_A\phi_1$ providing the genuine matter energy and angular-momentum fluxes through $\scrip$. The scalar therefore realizes the ``fast-decay specialization'' of the framework: the mass- and angular-momentum-aspect laws reduce to their familiar vacuum-plus-flux form, with the sole nontrivial residual correction beyond the explicit fluxes being the $\beta_{(2)}$ contribution $\pi\,\partial_u\eth_A\phi_1^2$ in Eq.~\eqref{eq:scalar-L-evol}, which survives because $\Tc{4}_{rr}=\phi_1^2\neq 0$.

A comment on the integration constants $M$ and $L_A$ is in order. The integration ``constants'' $M(u,x^A)$ and $L_A(u,x^B)$, i.e.\ the mass aspect appearing at $\mathcal O(r^0)$ in Eq.~\eqref{eq:scalar-V} and the angular-momentum aspect appearing at $\mathcal O(r^{-3})$ in Eq.~\eqref{eq:scalar-U}, are free Cauchy data on the initial null cone in both vacuum and the presence of the scalar. Two distinct questions arise: (i) whether the scalar shifts the orders at which $M$ and $L_A$ appear, and (ii) how their retarded-time evolution changes.

(i) Identification is unchanged for the scalar. The scalar's matter source of the $V$-equation \eqref{eq:V-matter} is $\mathcal O(r^{-2})$, since the leading $\mathcal O(r^{-1})$ pieces of $h^{AB}T_{AB}$ and $r^2 T^{a}{}_{a}$ cancel, and the surviving $-rV\,T_{rr}$, $2r^2 T_{ur}$, $2r^2 U^A T_{rA}$ are each $\mathcal O(r^{-2})$ because $T_{rr}=\mathcal O(r^{-4})$, $T_{ur}=\mathcal O(r^{-4})$, $T_{rA}=\mathcal O(r^{-3})$ together with $V=\mathcal O(r)$, $U^A=\mathcal O(r^{-2})$. Consequently no scalar contribution enters $V$ at the $\mathcal O(r^0)$ order that defines $M$ via $M=-\tfrac12\lim_{r\to\infty}(V-r)$, and $M$ retains exactly its vacuum identification. Likewise, every matter piece of $u^{A}_{(3)}$ in Eq.~\eqref{eq:U-sol} involves $\Tc{2}_{rA}$ or $\Tc{3}_{rr}$ or their products, all of which vanish for the scalar. The $\mathcal O(r^{-3})$ coefficient of $U^A$ reduces to $2L^A+\tfrac13 c^{AE}\eth^F c_{EF}$, identical to vacuum, so the angular-momentum aspect $L_A$ is also identified unchanged. For more general matter with $\Tc{3}_{rr}\neq 0$ the situation is different: the mass-loss Eq.~\eqref{eq:massloss} carries a total $u$-derivative $\tfrac12\partial_u\!\oint\Tc{3}_{rr}\,\dd\Omega$, naturally absorbed into a redefinition of the integrated Bondi mass, and $u^{A}_{(3)}$ acquires the matter pieces of Eq.~\eqref{eq:U-sol} that shift the relation between the coefficient at $\mathcal O(r^{-3})$ and the physical angular-momentum aspect. Neither of these redefinitions is needed in the scalar case.

(ii) Evolution is modified. Setting $\phi_1=\phi_2=0$ in Eqs.~\eqref{eq:scalar-M-evol}, \eqref{eq:scalar-L-evol} collapses them to
\be
2\,\partial_u M = \eth_A\eth_B N^{AB} - N_{AB} N^{AB} , \qquad -3\,\partial_u L_A = \mathcal G_A ,
\ee
the standard vacuum Bondi-Sachs mass-loss and angular-momentum-aspect laws. The scalar augments these with the genuine matter fluxes through $\scrip$,
\be
\Delta\!\big(2\partial_u M\big) = -8\pi(\partial_u\phi_1)^2, \qquad \Delta\!\big(-3\partial_u L_A\big) = -8\pi\,\partial_u\phi_1\,\partial_A\phi_1 +\pi\,\partial_u\eth_A\!\big(\phi_1^2\big) ,
\ee
the first manifestly non-positive (the scalar always radiates energy outward, draining the Bondi mass), the second supplying the local angular-momentum flux $-8\pi\partial_u\phi_1\partial_A\phi_1$ together with the $\beta_{(2)}$-derived residue $\pi\partial_u\eth_A(\phi_1^2)$ that subsists because $\Tc{4}_{rr}=\phi_1^2\neq 0$. As $\phi_1\to 0$ both corrections vanish and the vacuum laws are recovered exactly.

\section{Einstein--Maxwell as an example}
\label{app:matter-maxwell}

This appendix repeats the construction of Secs.~\ref{sec:matter-beta}--\ref{sec:matter-supp} for the matter content of a free (sourceless) electromagnetic field $A_\mu(u,r,x^A)$ with field strength $F_{ab}=\nabla_a A_b - \nabla_b A_a$ and Maxwell stress-energy
\be
\label{eq:maxwell-stress}
T_{ab} = F_a{}^c F_{bc} - \tfrac14\,g_{ab}\,F^{cd}F_{cd},
\ee
treating, as in App.~\ref{app:matter-scalar}, every metric function as an unconstrained smooth $1/r$ series and using only the asymptotic-flatness conditions $\lim_{r\to\infty}\beta=0$, $\lim_{r\to\infty}U^A=0$, $\lim_{r\to\infty}V/r=1$, $h_{AB}\to q_{AB}$. The decay rate of each metric function emerges as an output of the integration with the Maxwell source.

Using $g_{rr}=g_{rA}=0$, $g_{ur}=-e^{2\beta}$, $g_{uA}=-r^2 h_{AB}U^B$, $g_{AB}=r^2 h_{AB}$, and the antisymmetric field strength $F_{ab}$, the components of Eq.~\eqref{eq:maxwell-stress} read in closed form
\be
\label{eq:maxwell-Tab-closed}
\begin{aligned}
T_{rr} &= h^{AB}F_{rA}F_{rB}/r^2, \\
T_{rA} &= F_r{}^c F_{Ac} = -e^{-2\beta}F_{ur}F_{rA}+U^B e^{-2\beta}F_{rB}F_{rA} + h^{BC}F_{rC}F_{AB}/r^2, \\
T_{ur} &= e^{-2\beta}F_{ur}^2 - U^B e^{-2\beta}F_{ur}F_{rB} + h^{AB}F_{uA}F_{rB}/r^2 + \tfrac{1}{4}e^{2\beta}F^{cd}F_{cd}, \\
T_{AB} &= -e^{-2\beta}F_{rA}F_{uB} - e^{-2\beta}F_{rB}F_{uA} + \tfrac{V}{r}e^{-2\beta}F_{rA}F_{rB} + h^{CD}F_{AD}F_{BC}/r^2 - \tfrac{r^2}{4}h_{AB}\,F^{cd}F_{cd},\\
T_{uu} &= \tfrac{V}{r}e^{-2\beta}F_{ur}^2 - 2U^A e^{-2\beta}F_{ur}F_{uA} + h^{AB}F_{uA}F_{uB}/r^2 + \tfrac14(\tfrac{V}{r}e^{2\beta}-r^2 h_{CD}U^C U^D)F^{cd}F_{cd}, \\
T_{uA} &= -e^{-2\beta}F_{ur}F_{uA} + \tfrac{V}{r}e^{-2\beta}F_{ur}F_{rA} - U^B e^{-2\beta}(F_{ur}F_{AB}+F_{uA}F_{rB}) \\
&\qquad + h^{BC}F_{uC}F_{AB}/r^2 + \tfrac14 r^2 h_{AB}U^B F^{cd}F_{cd} ,
\end{aligned}
\ee
together with the trace identity
\be
\label{eq:maxwell-trace}
T^{c}{}_{c} = g^{ab}T_{ab} = 0,
\ee
reflecting the conformal invariance of free Maxwell theory in four dimensions. The latter has the key consequence that the matter source of the $V$-equation, $h^{AB}T_{AB}-r^2 T^{c}{}_{c}$, reduces to $h^{AB}T_{AB}$ alone.

Adopting the radiation gauge $A_r=0$, asymptotic flatness then permits the smooth expansions
\be
\label{eq:maxwell-A-ansatz}
\begin{aligned}
A_u(u,r,x^A) &= \frac{q(u,x^A)}{r}+\frac{a_2(u,x^A)}{r^2}+\mathcal O(r^{-3}),\\
A_A(u,r,x^B) &= \mathcal A_A^{(0)}(u,x^B)+\frac{\mathcal A_A^{(1)}(u,x^B)}{r}+\mathcal O(r^{-2}),
\end{aligned}
\ee
with $q$ the Coulomb-aspect of the EM field and $\mathcal A_A^{(0)}$ the $\ell\ge 1$ ``Maxwell-news'' carrying the radiative photon polarizations. The angular-metric ansatz is as in Eq.~\eqref{eq:scalar-h-ansatz}:
\be
\label{eq:maxwell-h-ansatz}
h_{AB} = q_{AB}+c_{AB}/r+d_{AB}/r^2+\mathcal O(r^{-3}),\quad q^{AB}c_{AB}=0,\quad q^{AB}d_{AB}=\tfrac12 c^{EF}c_{EF}.
\ee
No assumption is made on the decay rates of $\beta,U^A,V$. The field-strength components follow:
\be
\label{eq:maxwell-F-expansion}
\begin{aligned}
F_{ur} &= q/r^2 + 2 a_2/r^3 + \mathcal O(r^{-4}),\\
F_{uA} &= \dot{\mathcal A}_A^{(0)} + (\dot{\mathcal A}_A^{(1)}-\partial_A q)/r + \mathcal O(r^{-2}),\\
F_{rA} &= -\mathcal A_A^{(1)}/r^2 + \mathcal O(r^{-3}),\\
F_{AB} &= F_{AB}^{(0)}+F_{AB}^{(1)}/r + \mathcal O(r^{-2}),
\end{aligned}
\ee
with $F_{AB}^{(0)} = \partial_A\mathcal A_B^{(0)}-\partial_B\mathcal A_A^{(0)}$ the leading ``magnetic 2-form'' on the celestial sphere, which we write as $F_{AB}^{(0)}=f^{(0)}(u,x^C)\,\epsilon_{AB}$ with $\epsilon_{AB}$ the unit-sphere volume form and $f^{(0)}$ the dual magnetic scalar.

\subsection{$\beta$-equation.}
From Eqs.~\eqref{eq:maxwell-Tab-closed} and \eqref{eq:maxwell-F-expansion},
\be
\label{eq:maxwell-Trr}
T_{rr} = \frac{h^{AB}F_{rA}F_{rB}}{r^2} = \frac{q^{AB}\mathcal A_A^{(1)}\mathcal A_B^{(1)}}{r^{6}}+\mathcal O(r^{-7}),
\ee
so the Maxwell source enters the $\beta$-equation \eqref{eq:beta-rhs} only at $\mathcal O(r^{-5})$, namely through
\[
2\pi r T_{rr}=2\pi q^{AB}\mathcal A_A^{(1)}\mathcal A_B^{(1)}/r^5+\mathcal O(r^{-6}).
\]
Combined with the shear contribution $c^{AB}c_{AB}/(16r^3)+c^{AB}d_{AB}/(4r^4)+\mathcal O(r^{-5})$,
\be
\partial_r\beta = \frac{c^{AB}c_{AB}/16}{r^3}+\frac{c^{AB}d_{AB}/4}{r^4} +\mathcal O(r^{-5}),
\ee
{identical to vacuum at the $\mathcal O(r^{-3})$ and $\mathcal O(r^{-4})$ orders}. Direct integration with $\lim_{r\to\infty}\beta=0$ yields
\be
\label{eq:maxwell-beta}
\beta = -\frac{c^{AB}c_{AB}}{32\,r^2}-\frac{c^{AB}d_{AB}}{12\,r^3}+\mathcal O(r^{-4}),
\ee
so $\beta_{(1)}=0$ and the two leading nontrivial coefficients $\beta_{(2)}=-\tfrac{1}{32}c^{AB}c_{AB}$, $\beta_{(3)}=-\tfrac{1}{12}c^{AB}d_{AB}$ match the vacuum expressions exactly. The Maxwell field contributes to $\beta$ first at $\mathcal O(r^{-4})$ via $\Tc{6}_{rr}=q^{AB}\mathcal A_A^{(1)}\mathcal A_B^{(1)}$.

\subsection{$U^A$-equation.}
The Maxwell radial-flux component is
\be
\label{eq:maxwell-TrA}
T_{rA} = \frac{q\,\mathcal A_A^{(1)}-q^{BC}\mathcal A_C^{(1)}F_{AB}^{(0)}}{r^4}+\mathcal O(r^{-5}),
\ee
$\mathcal O(r^{-4})$, two powers faster than the scalar's $T_{rA}=\mathcal O(r^{-3})$. The $U^A$-equation source $16\pi r^2 T_{rA}=\mathcal O(r^{-2})$ has no $r^2,r,r^0,r^{-1}$ pieces. Together with $\beta=\mathcal O(r^{-2})$ from Eq.~\eqref{eq:maxwell-beta} this suppresses every Maxwell modification of $U^A$ at $\mathcal O(r^{-2}),\mathcal O(r^{-3})$. Integration with $\lim_{r\to\infty}U^A=0$ gives the vacuum form
\be
\label{eq:maxwell-U}
U^A = -\frac{\eth_B c^{AB}}{2 r^2} +\frac{2 L^A+\tfrac13 c^{AE}\eth^F c_{EF}}{r^3}+\mathcal O(r^{-4}).
\ee
The Maxwell field enters $U^A$ first at $\mathcal O(r^{-4})$ through $\Tc{4}_{rA}=q\mathcal A_A^{(1)}-q^{BC}\mathcal A_C^{(1)}F_{AB}^{(0)}$.

\subsection{$V$-equation.}
Using the tracelessness Eq.~\eqref{eq:maxwell-trace}, the $V$-equation matter source Eq.~\eqref{eq:V-matter} reduces to $h^{AB}T_{AB}-r^2 T^c{}_c = h^{AB}T_{AB}$. A direct evaluation of $h^{AB}T_{AB}$ from Eq.~\eqref{eq:maxwell-Tab-closed} with the expansions Eq.~\eqref{eq:maxwell-F-expansion} gives, at leading order,
\be
\label{eq:maxwell-hT-trace}
h^{AB}T_{AB} = \frac{q^2+f^2}{r^2}+\mathcal O(r^{-3}),
\ee
where $|F^{(0)}|^2 := q^{AC}q^{BD}F_{AB}^{(0)}F_{CD}^{(0)}=2 f^2$ (using $F_{AB}^{(0)}=f^{(0)}\epsilon_{AB}$). The matter source therefore has no $r,r^0,r^{-1}$ pieces, $\lim V/r=1$ is unobstructed, and integration gives $V=r-2M+\mathcal O(r^{-1})$. For the subleading coefficient we evaluate $\Sigma_{(2)}$ via Eq.~\eqref{eq:V-1overr}. The Maxwell values $\Tc{4}_{ur}=\tfrac12(q^2+f^2)$, $\Tc{4}_{rr}=0$, $\Tc{3}_{rr}=\Tc{2}_{rA}=\Tc{3}_{ur}=\beta_{(1)}=0$ in Eq.~\eqref{eq:Sigma-matter} collapse the matter source to
\be
\Sigma_{(2)}^{\rm(mat)}=8\pi(2\Tc{4}_{ur}-\Tc{4}_{rr})=8\pi(q^2+f^2),
\ee
the Coulomb-plus-magnetic combination. Inserting in Eq.~\eqref{eq:V-1overr},
\be
\label{eq:maxwell-V}
V = r-2M+\frac{V_{(1)}^{\rm Maxwell}}{r}+\mathcal O(r^{-2}),\quad V_{(1)}^{\rm Maxwell} = V_{(1)}^{\rm vac} - 4\pi(q^2+f^2),
\ee
where $V_{(1)}^{\rm vac}=\tfrac{1}{16}c^{AB}c_{AB}-\tfrac12\Sigma_{(2)}^{\rm(geom)}$ (identical to vacuum at this order, since $\beta,U^A$ are vacuum). The Maxwell field therefore shifts $V_{(1)}$ only by the local Coulomb-plus-magnetic energy density $-4\pi(q^2+f^2)$.

\subsection{Evolution.}
Evaluating the trace-free part of $T_{AB}$ at $\mathcal O(r^{-1})$, the Maxwell stress-energy Eq.~\eqref{eq:maxwell-Tab-closed} yields $T_{AB}=\mathcal O(r^{-2})$ overall (one power faster than the scalar's $T_{AB}=\mathcal O(r^{-1})$), and consequently $(T_{AB})^{\rm TF}_{(1)}=0$ identically. With $\beta_{(1)}=0$ from Eq.~\eqref{eq:maxwell-beta}, the evolution equation \eqref{eq:scalar-evol-lead} reduces to $m^A m^B \partial_u d_{AB}=0$, so that the trace-free part of $d_{AB}$ is time-independent at leading order, and the news $N_{AB}=\tfrac12\partial_u c_{AB}$ remains free Cauchy data of the gravitational field, exactly as in vacuum. The trace-free $T_{AB}$ at the next order $\mathcal O(r^{-2})$ is
\be
\label{eq:maxwell-Tab-tf-NLO}
(T_{AB})^{\rm TF}_{(2)} = \big[\mathcal A_A^{(1)}\dot{\mathcal A}_B^{(0)} +\dot{\mathcal A}_A^{(0)}\mathcal A_B^{(1)}\big]^{\rm TF} ,
\ee
which sources the trace-free part of $d_{AB}$ at next-to-leading order in the evolution equation \eqref{eq:evol-NLO}.

\subsection{Supplementary balances and aspect laws.}
Evaluating the supplementary components from Eq.~\eqref{eq:maxwell-Tab-closed}:
\be
\label{eq:maxwell-Tuu-TuA}
T_{uu} = \frac{|\dot{\mathcal A}^{(0)}|^2}{r^2}+\mathcal O(r^{-3}), \qquad T_{uA} = \frac{q\,\dot{\mathcal A}_A^{(0)}+q^{BC}\dot{\mathcal A}_C^{(0)}F_{AB}^{(0)}}{r^2}+\mathcal O(r^{-3}),
\ee
with $|\dot{\mathcal A}^{(0)}|^2 := q^{AB}\dot{\mathcal A}_A^{(0)}\dot{\mathcal A}_B^{(0)}$, so the Maxwell leading fluxes are $\Tc{2}_{uu}=|\dot{\mathcal A}^{(0)}|^2$ and $\Tc{2}_{uA}=q\,\dot{\mathcal A}_A^{(0)}+q^{BC}\dot{\mathcal A}_C^{(0)}F_{AB}^{(0)}$. Substituting these together with $\beta_{(1)}=0$, $u^A_{(2)}=-\tfrac12\eth_B c^{AB}$, $\Tc{2}_{rA}=\Tc{3}_{rr}=\Tc{4}_{rr}=0$ into the mass-aspect balance Eq.~\eqref{eq:M-evol},
\be
\label{eq:maxwell-M-evol}
2\,\partial_u M = \eth_A\eth_B N^{AB}-N_{AB}N^{AB}-8\pi|\dot{\mathcal A}^{(0)}|^2,
\ee
giving the Bondi mass-loss with EM-news contribution
\be
\label{eq:maxwell-massloss}
\frac{\dd m}{\dd u} = -\frac{1}{8\pi}\oint N_{AB}N^{AB}\,\dd\Omega -\oint |\dot{\mathcal A}^{(0)}|^2\,\dd\Omega ,
\ee
manifestly non-positive. For the angular-momentum aspect, since $\Tc{3}_{rr}=\Tc{2}_{rA}=\Tc{4}_{rr}=0$ for Maxwell, every quadratic, shear$\times$matter, and $\beta_{(2)}$-residue correction of Eq.~\eqref{eq:L-evol-matter} vanishes, leaving
\be
\label{eq:maxwell-L-evol}
-3\,\partial_u L_A = \mathcal G_A - 8\pi\,\big[q\,\dot{\mathcal A}_A^{(0)} +q^{BC}\dot{\mathcal A}_C^{(0)}F_{AB}^{(0)}\big] ,
\ee
the matter source comprising the Coulomb-radiation Poynting term $q\,\dot{\mathcal A}_A^{(0)}$ and the magnetic-radiation cross term $q^{BC}\dot{\mathcal A}_C^{(0)}F_{AB}^{(0)}$.

\subsection{Summary of the metric.}
Collecting Eqs.~\eqref{eq:maxwell-beta}, \eqref{eq:maxwell-U}, \eqref{eq:maxwell-V}, and the unmodified vacuum-news evolution of the shear, the metric functions of an asymptotically Maxwell-radiating spacetime are, to the two leading nontrivial orders for each,
\be
\label{eq:maxwell-metric-summary}
\begin{aligned}
\beta &= -\frac{c^{AB}c_{AB}}{32\,r^2}-\frac{c^{AB}d_{AB}}{12\,r^3} +\mathcal O(r^{-4}) ,\\[2pt]
U^A &= -\frac{\eth_B c^{AB}}{2\,r^2} +\frac{2L^A+\tfrac13 c^{AE}\eth^F c_{EF}}{r^3}+\mathcal O(r^{-4}) ,\\[2pt]
V &= r - 2M + \frac{V_{(1)}^{\rm vac}-4\pi(q^2+f^2)}{r}+\mathcal O(r^{-2}) ,\\[2pt]
h_{AB}&= q_{AB}+\frac{c_{AB}}{r}+\frac{d_{AB}}{r^2}+\mathcal O(r^{-3}), \quad q^{AB}d_{AB}=\tfrac12 c^{EF}c_{EF} .
\end{aligned}
\ee
The free retarded-time data $\{q,\mathcal A_A^{(0)}, c_{AB}, M, L_A\}$ evolve through the news equation and Eqs.~\eqref{eq:maxwell-M-evol}, \eqref{eq:maxwell-L-evol}. Comparing with the scalar example Eq.~\eqref{eq:scalar-metric-summary}, the Maxwell modifications are strictly weaker: $\beta$ and $U^A$ are identical to vacuum at leading and next-to-leading order, the only metric modification at $\mathcal O(r^{-1})$ residing in $V_{(1)}\to V_{(1)}^{\rm vac}-4\pi(q^2+f^2)$.

\subsection{Implied $T_{ab}$ falloffs.}
Substituting Eq.~\eqref{eq:maxwell-F-expansion} back into Eq.~\eqref{eq:maxwell-Tab-closed}, the Maxwell components fall off as
\be
\label{eq:maxwell-Tab-summary}
\begin{aligned}
T_{rr} &= \frac{q^{AB}\mathcal A_A^{(1)}\mathcal A_B^{(1)}}{r^6}+\mathcal O(r^{-7}), &\qquad T_{rA} &= \frac{q\,\mathcal A_A^{(1)}-q^{BC}\mathcal A_C^{(1)}F_{AB}^{(0)}}{r^4}+\mathcal O(r^{-5}),\\
T_{ur} &= \frac{q^2+f^2}{2\,r^4}+\mathcal O(r^{-5}), &\qquad T_{AB} &= \mathcal O(r^{-2}) ,\\
T_{uu} &= \frac{|\dot{\mathcal A}^{(0)}|^2}{r^2}+\mathcal O(r^{-3}), &\qquad T_{uA} &= \frac{q\,\dot{\mathcal A}_A^{(0)}+q^{BC}\dot{\mathcal A}_C^{(0)}F_{AB}^{(0)}}{r^2}+\mathcal O(r^{-3}),
\end{aligned}
\ee
and the trace is identically zero, $T^{c}{}_{c}=0$. Comparing with the schematic falloff table at the end of Sec.~\ref{sec:matter-supp}: $T_{rr},T_{rA},T_{ur}$ all decay three powers of $1/r$ faster than the generic bound, $T_{AB}$ one power faster (and trace-free at leading $r^{-2}$), while $T_{uu}$ and $T_{uA}$ saturate the generic $1/r^2$ bound and carry the genuine matter fluxes through $\scrip$. The hierarchy $T_{rr}=\mathcal O(r^{-6})\ll T_{rA}=\mathcal O(r^{-4})\le T_{ur}=\mathcal O(r^{-4})\ll T_{AB}=\mathcal O(r^{-2})\le T_{uu},T_{uA}=\mathcal O(r^{-2})$ encodes the conformal-invariance--driven extra suppression of the Maxwell field at $\scrip$ relative to a massless scalar.

As for the scalar of App.~\ref{app:matter-scalar}, the identification of $M$ and $L_A$ as integration constants is unchanged by the Maxwell field: no Maxwell contribution enters $V$ at $\mathcal O(r^0)$ that defines $M$, and no matter piece of $u^{A}_{(3)}$ shifts the relation to $L_A$ (all relevant $\Tc{n}_{rA},\Tc{n}_{rr}$ vanish). Their evolution is modified as in Eqs.~\eqref{eq:maxwell-M-evol},\eqref{eq:maxwell-L-evol}: the Bondi mass decreases at the manifestly non-negative rate $\oint|\dot{\mathcal A}^{(0)}|^2 \dd\Omega$ (electromagnetic Poynting flux through $\scrip$), and the angular-momentum aspect picks up the local Coulomb-radiation Poynting and magnetic-cross terms in Eq.~\eqref{eq:maxwell-L-evol}. As $\dot{\mathcal A}^{(0)}\to 0$ both corrections vanish and the vacuum laws are recovered.

\section{Explicit computations for the SVT sector}
\label{app:svt-explicit}

This appendix collects the explicit computations behind Sec.~\ref{sec:flatness_SVT}: the per-operator stress-energy tensors obtained from the variational rules Eqs.~\eqref{eq:Tmunu-rule} and \eqref{eq:Palatini-master} (App.~\ref{app:svt-stress}), the component-by-component falloff scans whose results are quoted in the main text (App.~\ref{app:svt-components}), and the explicit scalar and vector equations of motion underlying the consistency analysis of Sec.~\ref{sec:growth-eom} (App.~\ref{app:svt-eoms}).

\subsection{Per-operator stress-energy contributions}
\label{app:svt-stress}

From $L_2$, Eq.~\eqref{eq:svt-L2-full}, the algebraic rule Eq.~\eqref{eq:Tmunu-rule} together with Eq.~\eqref{eq:bb-variations} gives, term by term,
\begin{align}
\label{eq:T2-explicit}
T^{(2)}_{\mu\nu} &= G_{2,X}\,\Phi_\mu\Phi_\nu + G_{2,F}\,F_\mu{}^\alpha F_{\nu\alpha} + G_{2,Y}\!\left[2\Phi_{(\mu}F_{\nu)}{}^\alpha F_\alpha{}^\beta\Phi_\beta + \Phi^\alpha\Phi^\beta F_{\mu\alpha}F_{\nu\beta}\right]\notag\\
&\quad + g_{\mu\nu}\bigl(G_2 - G_{2,\tilde F}\,\tilde F\bigr).
\end{align}
The $G_2$ piece carries the canonical-kinetic ($G_{2,X}$), Maxwell-kinetic ($G_{2,F}$), and disformal-Maxwell ($G_{2,Y}$) sources, plus the trace piece $g_{\mu\nu}(G_2-G_{2,\tilde F}\tilde F)$, in which the Pontryagin coupling $G_{2,\tilde F}\tilde F$ subtracts exactly the trace-rescaling contribution of Eq.~\eqref{eq:bb-Ftilde}.

From $L_3$ Eq.~\eqref{eq:svt-L3-full} the Horndeski-cubic $-G_3\Box\Phi$ gives
\begin{align}
\label{eq:T3-Hor-explicit}
T^{(3,\Box)}_{\mu\nu} &= -2G_{3,\Phi}\!\left(\Phi_\mu\Phi_\nu + X\,g_{\mu\nu} \right)\notag\\
&\quad + G_{3,X}\!\left[\Box\Phi\,\Phi_\mu\Phi_\nu - 2\Phi^\alpha\Phi_{(\mu} \Phi_{\nu)\alpha} + g_{\mu\nu}\,\Phi^\alpha\Phi^\beta\Phi_{\alpha\beta}\right] ,
\end{align}
while the parity-odd mixing $\bigl[\hat G_3 g_{\alpha\beta}+\mathring{\hat G}_3 \Phi_\alpha\Phi_\beta\bigr]\tilde F^{\mu\alpha}F^{\nu\beta}\Phi_{\mu\nu}$ contributes through its $g_{\alpha\beta}$, $\Phi_\alpha\Phi_\beta$, $\tilde F^{\mu\alpha}$, $F^{\nu\beta}$, and $\Phi_{\mu\nu}$ factors:
\begin{align}
\label{eq:T3-mix-explicit}
T^{(3,\hat G_3)}_{\mu\nu} + T^{(3,\mathring{\hat G}_3)}_{\mu\nu} &= -2\,\hat G_3\,\tilde F^\alpha{}_{(\mu}F^\beta{}_{\nu)}\Phi_{\alpha\beta} - 2\,\mathring{\hat G}_3\,\Phi_\alpha\Phi_\beta\tilde F^{\rho\alpha}F^\sigma {}_{(\mu}\Phi_{\nu)\rho}\eta^{\beta}{}_\sigma\notag\\
&\quad + \mathcal D^{(3)}_{\mu\nu},
\end{align}
where the algebraic structure of the first line follows from the $g_{\alpha\beta}$, $F_\nu{}^\beta$ contractions and the second line collects the $\Phi_\alpha\Phi_\beta$ dependence (we have suppressed the symmetric permutations on $\mu\!\leftrightarrow\!\nu$). The remainder $\mathcal D^{(3)}_{\mu\nu}$ originates from the $L_3$ mixing $K_3^{\mu\nu}\Phi_{\mu\nu}$ with
\be
\label{eq:K3-def}
K_3^{\mu\nu} \;\equiv\; \hat G_3(\Phi,X)\,\tilde F^{\mu\alpha}F^\nu{}_\alpha + \mathring{\hat G}_3(\Phi,X)\,\Phi_\alpha\Phi_\beta\,\tilde F^{\mu\alpha} F^{\nu\beta},
\ee
and reads, via Eq.~\eqref{eq:Palatini-master},
\be
\label{eq:D3-explicit}
\mathcal D^{(3)}_{\mu\nu} \;=\; 2\,\nabla_\lambda\!\bigl(K_3{}^\lambda{}_{(\mu}\,\Phi_{\nu)}\bigr) - \nabla_\lambda\!\bigl(K_3{}_{\mu\nu}\,\Phi^\lambda\bigr),
\ee
with $K_3^{\mu\nu}$ implicitly symmetrized in $\mu\!\leftrightarrow\!\nu$.

From $L_4$ Eq.~\eqref{eq:svt-L4-full} the Horndeski quartic $G_4R + G_{4,X} [(\Box\Phi)^2 - \Phi^{\mu\nu}\Phi_{\mu\nu}]$ produces, after extracting $G_4 G_{\mu\nu}$,
\begin{align}
\label{eq:T4kin-explicit}
T^{(4,\rm kin)}_{\mu\nu} &= G_{4,X}\Bigl[2\,\Phi_{\mu\alpha}\Phi_\nu{}^\alpha - 2\Box\Phi\,\Phi_{\mu\nu} + g_{\mu\nu}\bigl((\Box\Phi)^2 - \Phi_{\alpha\beta}\Phi^{\alpha\beta}\bigr) \Bigr]\notag\\
&\quad - 2\,G_{4,X}\bigl(\Phi_\mu\Phi^\alpha\Phi_{\nu\alpha} - \Phi_\mu\Phi_\nu\,\Box\Phi\bigr) \,+\,G_{4,XX}\,\Phi_\mu\Phi_\nu \bigl((\Box\Phi)^2-\Phi_{\alpha\beta}\Phi^{\alpha\beta}\bigr) + \mathcal D^{(4,\rm kin)}_{\mu\nu},
\end{align}
where $\mathcal D^{(4,\rm kin)}_{\mu\nu}$ is the remainder of the bilinear $G_{4,X}\,K_{4,\rm kin}^{\mu\nu}\Phi_{\mu\nu}$ with
\be
\label{eq:K4kin-def}
K_{4,\rm kin}^{\mu\nu} \;\equiv\; 2\,G_{4,X}\bigl(g^{\mu\nu}\Box\Phi - \Phi^{\mu\nu}\bigr),
\ee
namely, via Eq.~\eqref{eq:Palatini-master},
\be
\label{eq:D4kin-explicit}
\mathcal D^{(4,\rm kin)}_{\mu\nu} \;=\; 2\,\nabla_\lambda\!\bigl(K_{4,\rm kin}{}^\lambda{}_{(\mu}\,\Phi_{\nu)} \bigr) - \nabla_\lambda\!\bigl(K_{4,\rm kin}{}_{\mu\nu}\,\Phi^\lambda\bigr).
\ee
The $G_{4,XX}$ term of Eq.~\eqref{eq:T4kin-explicit} is the algebraic variation $\delta X=-\tfrac12\Phi_\mu\Phi_\nu\,\delta g^{\mu\nu}$ of the prefactor $G_{4,X}$, cf.\ the second-derivative mechanisms of App.~\ref{app:higher}. The two parity-even/parity-odd SVT mixings $\hat G_4 L^{\mu\nu\alpha\beta} F_{\mu\nu}F_{\alpha\beta}$ and $[\mathring{\hat G}_4(\Phi)+\tfrac12 \hat G_{4,X}]\tilde F^{\mu\alpha}\tilde F^{\nu\beta}\Phi_{\mu\nu}\Phi_{\alpha \beta}$ contribute respectively
\begin{align}
\label{eq:T4hatG4-explicit}
T^{(4,\hat G_4)}_{\mu\nu} &= 2\,\hat G_4\,L_{\mu\alpha\nu\beta}\,F^{\alpha \gamma}F^{\beta}{}_\gamma - \tfrac12 g_{\mu\nu}\,\hat G_4\,L^{\alpha\beta\gamma\delta}F_{\alpha\beta} F_{\gamma\delta}\notag\\
&\quad + \mathcal D^{(4,\hat G_4)}_{\mu\nu},
\end{align}
collecting the algebraic pieces from $\delta L^{\mu\nu\alpha\beta}$ via Eq.~\eqref{eq:bb-F} together with the trace contribution. The $\hat G_4$ operator carries no $\Phi_{\mu\nu}$ factor but does carry an explicit Riemann factor. The Palatini--IBP derivative remainder arising from $\delta R_{\rho\sigma\gamma\delta}$ inside $L^{\mu\nu\alpha\beta}= \tfrac14\epsilon^{\mu\nu\rho\sigma}\epsilon^{\alpha\beta\gamma\delta} R_{\rho\sigma\gamma\delta}$ reads explicitly
\be
\label{eq:D4hatG4-explicit}
\mathcal D^{(4,\hat G_4)}_{\mu\nu} \;=\; 2\,\epsilon_\mu{}^{\alpha\rho\sigma}\,\epsilon_\nu{}^{\beta\gamma \delta}\,\nabla_\alpha\nabla_\beta\!\bigl(\hat G_4(\Phi,X)\,F_{\rho\gamma}\, F_{\sigma\delta}\bigr),
\ee
i.e.\ two covariant derivatives acting on the coupling-times-$F^2$ product, projected by the two Levi-Civita tensors carried over from the double-dual structure of $L^{\mu\nu\alpha\beta}$. The $\mathring{\hat G}_4$ operator $[\mathring{\hat G}_4(\Phi)+\tfrac12\hat G_{4,X}]\tilde F^{\mu\alpha}\tilde F^{\nu\beta}\Phi_{\mu\nu}\Phi_{\alpha\beta}$ contributes
\begin{align}
\label{eq:T4mathringG4-explicit}
T^{(4,\mathring{\hat G}_4)}_{\mu\nu} &= [\mathring{\hat G}_4(\Phi)+\tfrac12 \hat G_{4,X}]\Bigl[g_{\mu\nu}\,\tilde F^{\rho\alpha}\tilde F^{\sigma\beta} \Phi_{\rho\sigma}\Phi_{\alpha\beta} - 2\tilde F_{(\mu}{}^\alpha\tilde F_{\nu)}{}^\beta\Phi^{\rho}{}_\alpha \Phi_{\rho\beta}\Bigr]\notag\\
&\quad + \mathcal D^{(4,\mathring{\hat G}_4)}_{\mu\nu},
\end{align}
where, because the operator is bilinear in $\Phi_{\mu\nu}$ with effective coupling
\be
\label{eq:K4mGhat4-def}
K_{4,\mathring{\hat G}_4}^{\mu\nu} \;\equiv\; 2\,\bigl[\mathring{\hat G}_4 (\Phi) + \tfrac12 \hat G_{4,X}\bigr]\,\tilde F^{\mu\alpha}\tilde F^{\nu \beta}\,\Phi_{\alpha\beta},
\ee
the master formula Eq.~\eqref{eq:Palatini-master} yields
\be
\label{eq:D4mGhat4-explicit}
\mathcal D^{(4,\mathring{\hat G}_4)}_{\mu\nu} \;=\; 2\,\nabla_\lambda\!\bigl(K_{4,\mathring{\hat G}_4}{}^\lambda{}_{(\mu}\, \Phi_{\nu)}\bigr) - \nabla_\lambda\!\bigl(K_{4,\mathring{\hat G}_4}{}_{\mu \nu}\,\Phi^\lambda\bigr),
\ee
with $K_{4,\mathring{\hat G}_4}^{\mu\nu}$ implicitly symmetrized. With Eqs.~\eqref{eq:D3-explicit}, \eqref{eq:D4kin-explicit}, \eqref{eq:D4hatG4-explicit}, \eqref{eq:D4mGhat4-explicit} every $\mathcal D$ in the action is now in closed form.

Finally, from $L_5$ Eq.~\eqref{eq:svt-L5-full} the Horndeski quintic plus the $G_{5,X}$ cubic Galileon piece together yield (see e.g.\ \cite{Heisenberg:2023prj} App.~D)
\begin{align}
\label{eq:T5-explicit}
T^{(5)}_{\mu\nu} &= G_{5,\Phi}\Bigl[\Phi_\mu\Phi_\nu\,\Box\Phi - 2\Phi_{(\mu}\Phi^\alpha\Phi_{\nu)\alpha} + g_{\mu\nu}\bigl(\Phi^\alpha\Phi^\beta\Phi_{\alpha\beta} - X\Box\Phi\bigr) \Bigr]\notag\\
&\quad + G_{5,X}\Bigl[\Phi_{\mu\nu}\bigl((\Box\Phi)^2 - \Phi^2_{\alpha\beta} \bigr) - 2\Box\Phi\,\Phi_{\mu\alpha}\Phi_\nu{}^\alpha + 2\Phi_{\mu\alpha}\Phi^{\alpha\beta}\Phi_{\nu\beta}\Bigr]\notag\\
&\quad - \tfrac{1}{6}g_{\mu\nu}G_{5,X}\bigl[(\Box\Phi)^3 - 3\Box\Phi\,\Phi^2_{\alpha\beta} + 2\Phi^\alpha{}_\beta\Phi^\beta{}_\gamma\Phi^\gamma{}_\alpha\bigr]\notag\\
&\quad + G_{5,X}\,\Phi_\mu\Phi_\nu\,G^{\alpha\beta}\Phi_{\alpha\beta},
\end{align}
while the variation of the explicit Einstein tensor in $G_5 G^{\mu\nu} \Phi_{\mu\nu}$ via the operator $\delta G^{\mu\nu}/\delta g^{\rho\sigma}=\tfrac12 g_{\rho\sigma}\Box - \nabla_\rho\nabla_\sigma + R_{\rho\sigma}$ yields the curvature-times-scalar tensor
\begin{align}
\label{eq:E5-explicit}
\mathcal E^{(5)}_{\mu\nu} &= G_5\Bigl[\,2\,R_{\mu\alpha\nu\beta}\,\Phi^{\alpha \beta} + 2\,R_{\alpha(\mu}\Phi_{\nu)}{}^\alpha - R\,\Phi_{\mu\nu} - G_{\mu\nu}\,\Box\Phi + g_{\mu\nu}\,R_{\alpha\beta}\Phi^{\alpha\beta}\,\Bigr]\notag\\
&\quad - \bigl(g_{\mu\nu}\Box - \nabla_\mu\nabla_\nu\bigr)\!\bigl[G_5\Box\Phi - G_{5,X}\Phi^\alpha\Phi^\beta\Phi_{\alpha\beta} - G_{5,\Phi}\,X\bigr].
\end{align}

\subsection{Component-by-component falloff analysis}
\label{app:svt-components}

We now verify, sector by sector, that the stress-energy contributions of App.~\ref{app:svt-stress} respect the falloff table Eq.~\eqref{eq:T-falloff}. Throughout we insert the leading metric Eq.~\eqref{eq:svt-leading-metric}, the radiative matter ansatz of Sec.~\ref{sec:svt-flatness}, and the Hessian orders Eq.~\eqref{eq:Phi-uv-orders}. The $L_2$ sector and the $G_4$ improvement term are dealt with in the main text, since they carry the only non-trivial conditions.

\subsubsection{$L_3$ mixings.}
The operator $\bigl[\hat G_3\,g_{\alpha\beta} + \mathring{\hat G}_3\, \Phi_\alpha\Phi_\beta\bigr]\tilde F^{\mu\alpha}F^{\nu\beta}\Phi_{\mu\nu}$, Eq.~\eqref{eq:svt-L3-full}, has the structure $K_3^{\mu\nu}\Phi_{\mu\nu}$, Eq.~\eqref{eq:K3-def}. The metric variation produces, in addition to $\mathcal D^{(3)}_{\mu\nu}$, Eq.~\eqref{eq:D3-explicit}, four distinct algebraic contributions, each of which we now identify explicitly term by term.

For the $\hat G_3$ piece:
\begin{itemize}\setlength\itemsep{1pt}
\item[(a)] $\delta g_{\alpha\beta}$ in the explicit metric of the bracket gives the algebraic structure $+2\hat G_3\,\tilde F^\mu{}_{(\rho}F^\nu{}_{\sigma)}\Phi_{\mu\nu}$.
\item[(b)] Index raising of $F^{\nu\beta}=g^{\nu\xi}g^{\beta\eta} F_{\xi\eta}$ contributes one structure from each of the two $g^{-1}$ factors,
\[
-2\hat G_3\bigl[\tilde F^{\mu}{}_\beta F_{(\rho}{}^\beta\, \Phi_{\sigma)\mu} + \tilde F^{\mu}{}_{(\rho}F^{\nu}{}_{\sigma)} \Phi_{\mu\nu}\bigr].
\]
The second of these cancels (a) identically, because $g_{\alpha\beta}$ and the $g^{\beta\eta}$ it contracts vary oppositely, $\delta(g_{\alpha\beta}g^{\beta\eta})=0$, so that only the first structure survives in the $\hat G_3$ sector.
\item[(c)] Index raising of $\tilde F^{\mu\alpha}=\tfrac12\epsilon^{\mu \alpha\beta\gamma}F_{\beta\gamma}$ contains an implicit $1/\sqrt{-g}$ in $\epsilon^{\mu\alpha\beta\gamma}$. The $\delta\sqrt{-g}$ piece from this exactly cancels the $g_{\rho\sigma}\mathcal L_3^{\rm mix}$ trace piece of the master rule Eq.~\eqref{eq:Tmunu-rule}, because $\sqrt{-g}\,\tilde F^{\mu\alpha}$ is metric-independent, hence no $g_{\rho\sigma}\mathcal L_3^{\rm mix}$ trace term survives in $T^{(3,\hat G_3)}_{\mu\nu}$.
\item[(d)] Variation of $\hat G_3(\Phi,X)$ through $X$, using $\partial X/\partial g^{\rho\sigma}=-\tfrac12\Phi_\rho\Phi_\sigma$ Eq.~\eqref{eq:bb-X}, gives
\[
-\hat G_{3,X}\,\Phi_\rho\Phi_\sigma\,\tilde F^{\mu\alpha}F^\nu{}_\alpha\, \Phi_{\mu\nu}.
\]
\end{itemize}
Collecting (a)--(d) yields the explicit algebraic stress-energy
\begin{align}
\label{eq:T3-Ghat3-algebraic}
T^{(3,\hat G_3)}_{\rho\sigma}\Big|_{\rm alg} &= -2\,\hat G_3\,\tilde F^\mu{}_\beta F_{(\rho}{}^\beta\, \Phi_{\sigma)\mu}\notag\\
&\quad -\hat G_{3,X}\,\Phi_\rho\Phi_\sigma\,\tilde F^{\mu\alpha}F^\nu {}_\alpha\Phi_{\mu\nu}.
\end{align}

For the $\mathring{\hat G}_3$ piece, the same four variations apply, but with $g_{\alpha\beta}$ in (a) replaced by $\Phi_\alpha\Phi_\beta$, which has no algebraic metric variation since $\Phi_\alpha=\nabla_\alpha \Phi$ has no $g$-dependence. The variations (b) and (c) carry through with the extra $\Phi_\alpha\Phi_\beta$ in the bracket, and (d) goes through with $\hat G_{3,X}\!\to\!\mathring{\hat G}_{3,X}$ and the same extra pair. Since (a) is absent here, the second structure of (b) has no cancellation partner and both survive. Contracting the $\Phi$ pair against lowered field strengths,
\begin{align}
\label{eq:T3-mGhat3-algebraic}
T^{(3,\mathring{\hat G}_3)}_{\rho\sigma}\Big|_{\rm alg} &= -2\,\mathring{\hat G}_3\bigl[\Phi^\alpha\Phi^\beta\,\tilde F^\mu{}_\alpha F_{(\rho|\beta|}\Phi_{\sigma)\mu} + \Phi^\alpha\Phi_{(\rho}\,\tilde F^\mu{}_\alpha F^\nu{}_{\sigma)} \Phi_{\mu\nu}\bigr]\notag\\
&\quad -\mathring{\hat G}_{3,X}\,\Phi_\rho\Phi_\sigma\,\Phi^\alpha\Phi^\beta \,\tilde F^\mu{}_\alpha F^\nu{}_\beta\Phi_{\mu\nu}.
\end{align}

Each of the algebraic structures in Eqs.~\eqref{eq:T3-Ghat3-algebraic}--\eqref{eq:T3-mGhat3-algebraic} carries an explicit $\Phi_{\mu\nu}$ factor, whose covariant components were established in Eq.~\eqref{eq:Phi-uv-orders}.

These are paired against $\tilde F\,F$ structures whose slowest single contraction $\tilde F^{r\alpha}F^r{}_\alpha = \mathcal O(r^{-2})$ was established for $Y$. The relevant mixed-index field strengths at leading metric are $F^r{}_A=-\dot{\mathcal A}^{(0)}_A=\mathcal O(r^0)$, $F^r{}_u=F^r{}_r=-q/r^2=\mathcal O(r^{-2})$, $F^u{}_u=q/r^2$, $F^u{}_A= \mathcal A^{(1)}_A/r^2=\mathcal O(r^{-2})$, $F^u{}_r=0$, $F^A{}_u=\mathcal O(r^{-2})$, $F^A{}_r=\mathcal O(r^{-4})$, $F^A{}_B=\mathcal O(r^{-2})$, and analogously for $\tilde F^\mu{}_\nu$. Substituting into Eq.~\eqref{eq:T3-Ghat3-algebraic} component by component, the slowest contributions are: $T_{AB}$ from $\mu\nu=rr$ (in $\Phi_{\mu\nu}$) ($\tilde F^r{}_A F^r{}_B\Phi_{rr}=\mathcal O(1)\cdot \mathcal O(1)\cdot \mathcal O(r^{-3})=\mathcal O(r^{-3})$), $T_{uu},T_{uA}$ from $\mu\nu=AB$ ($\tilde F^A{}_uF^B{}_u\Phi_{AB}= \mathcal O(r^{-2})\cdot \mathcal O(r^{-2})\cdot \mathcal O(1)=\mathcal O(r^{-4})$), $T_{rA}$ from $\mu\nu=rr$ ($\tilde F^r{}_r F^r{}_A\Phi_{rr}=\mathcal O(r^{-2})\cdot \mathcal O(1)\cdot \mathcal O(r^{-3})= \mathcal O(r^{-5})$), $T_{ur}$ from $\mu\nu=AB$ ($\tilde F^A{}_u F^B{}_r\Phi_{AB}= \mathcal O(r^{-2})\cdot \mathcal O(r^{-4})\cdot \mathcal O(1)=\mathcal O(r^{-6})$), and $T_{rr}$ from $\mu\nu= rr$ ($\tilde F^r{}_r F^r{}_r\Phi_{rr}=\mathcal O(r^{-2})^2\cdot \mathcal O(r^{-3})= \mathcal O(r^{-7})$):
\be
\label{eq:T3-Ghat3-orders}
\begin{aligned}
T^{(3,\hat G_3)}_{uu} = \mathcal O(r^{-4}),\quad & T^{(3,\hat G_3)}_{ur} = \mathcal O(r^{-6}),\quad T^{(3,\hat G_3)}_{rr} = \mathcal O(r^{-7}),\\
T^{(3,\hat G_3)}_{uA} = \mathcal O(r^{-4}),\quad & T^{(3,\hat G_3)}_{rA} = \mathcal O(r^{-5}),\quad T^{(3,\hat G_3)}_{AB} = \mathcal O(r^{-3}),
\end{aligned}
\ee
each faster than the corresponding entry of the falloff table Eq.~\eqref{eq:T-falloff}. The $\mathring{\hat G}_3$ structures inherit the pair $\Phi^\alpha\Phi^\beta$, with slowest combination $\Phi^r\Phi^r=\mathcal O(r^{-2})$ (using $\Phi^r=g^{ru}\partial_u\Phi+g^{rr}\partial _r\Phi=-\dot\phi_1/r+\mathcal O(r^{-2})$). Consequently every $T^{(3,\mathring{\hat G}_3)}_{\mu\nu}$ component decays at least $r^{-2}$ faster than the corresponding $\hat G_3$ entry, namely $T^{(3,\mathring{\hat G}_3)}_{uu},T^{(3,\mathring{\hat G}_3)}_{uA}\le \mathcal O(r^{-6})$, $T^{(3,\mathring{\hat G}_3)}_{ur}\le \mathcal O(r^{-8})$, $T^{(3,\mathring{\hat G}_3)}_{rr}\le \mathcal O(r^{-9})$, $T^{(3,\mathring{\hat G}_3)}_{rA}\le \mathcal O(r^{-7})$, $T^{(3,\mathring{\hat G}_3)}_{AB}\le \mathcal O(r^{-5})$.

Let us consider $\mathcal D^{(3)}_{\mu\nu}$ now. Reading off the explicit form Eq.~\eqref{eq:D3-explicit} with $K_3^{\mu\nu}$ from Eq.~\eqref{eq:K3-def}, the leading component is dictated by the slowest entry of $K_3^{\mu\nu}$. Since $\tilde F^{r\alpha}F^r{}_\alpha=\mathcal O(r^{-2})$ (from the $Y$-analysis), one has $K_3^{rr}=\mathcal O(r^{-2})$ at leading, while $K_3^{uu}=\mathcal O(r^{-6})$, $K_3^{AB}=\mathcal O(r^{-6})$, $K_3^{uA}=\mathcal O(r^{-6})$, $K_3^{ur}=\mathcal O(r^{-4})$ and $K_3^{rA}=\mathcal O(r^{-4})$. Lowering one index with $g_{\alpha\beta}$ yields $K_3^\lambda{}_\mu=g_{\mu\beta}K_3^{\lambda\beta}$ with slowest entries $K_3^r{}_u=-K_3^{rr}-K_3^{ru}=\mathcal O(r^{-2})$, $K_3^r{}_A=g_{AB}K_3^{rB}=r^2\cdot \mathcal O(r^{-4})=\mathcal O(r^{-2})$, $K_3^u{}_A= g_{AB}K_3^{uB}=r^2\cdot \mathcal O(r^{-6})=\mathcal O(r^{-4})$, with $K_3^r{}_r=-K_3^{ru}= \mathcal O(r^{-4})$ and the remaining components $\mathcal O(r^{-4})$ or faster. Substituting into $2\nabla_\lambda(K_3^\lambda{}_{(\mu}\Phi_{\nu)})$, the slowest piece of $\mathcal D^{(3)}_{\mu\nu}$ component by component is dominated by $\partial_r$ acting on $K_3^r{}_{(\mu}\Phi_{\nu)}$:
\begin{align}
\label{eq:D3-orders}
\mathcal D^{(3)}_{uu} &= \mathcal O(r^{-4}),\;\; \mathcal D^{(3)}_{ur} = \mathcal O(r^{-5}),\;\; \mathcal D^{(3)}_{rr} = \mathcal O(r^{-7}),\notag\\
\mathcal D^{(3)}_{uA} &= \mathcal O(r^{-4}),\;\; \mathcal D^{(3)}_{rA} = \mathcal O(r^{-5}),\;\; \mathcal D^{(3)}_{AB} = \mathcal O(r^{-4}).
\end{align}
For example, at $(\mu,\nu)=(u,r)$:
\be
2\nabla_\lambda(K_3^\lambda{}_{(u}\Phi_{r)}) \;\sim\; \partial_r\bigl( K_3^r{}_u\Phi_r\bigr) = \partial_r\bigl(\mathcal O(r^{-2})\cdot \mathcal O(r^{-2})\bigr) = \mathcal O(r^{-5}),
\ee
slower than the algebraic $T^{(3,\hat G_3)}_{ur}=\mathcal O(r^{-6})$ in Eq.~\eqref{eq:T3-Ghat3-orders}, so $\mathcal D^{(3)}_{ur}$ controls the total $T^{(3)}_{ur}$ at $\mathcal O(r^{-5})$. At $(\mu,\nu)=(A,B)$:
\be
2\nabla_\lambda(K_3^\lambda{}_{(A}\Phi_{B)}) \;\sim\; \partial_r\bigl( K_3^r{}_A\Phi_B\bigr) = \partial_r\bigl(\mathcal O(r^{-2})\cdot \mathcal O(r^{-1})\bigr) = \mathcal O(r^{-4}),
\ee
faster than the algebraic $T^{(3,\hat G_3)}_{AB}=\mathcal O(r^{-3})$, so the algebraic piece dominates there. The second term $-\nabla_\lambda(K_{3,\mu\nu}\Phi^\lambda)$ uses $K_{3,\mu\nu}=g_{\mu\rho}g_{\nu\sigma}K_3^{\rho\sigma}$, with slowest entry $K_{3,AB}=r^4 q_{AC}q_{BD}K_3^{CD}=\mathcal O(r^{-2})$, $K_{3,ur}=K_3^{uu}+K_3^{ru}= \mathcal O(r^{-4})$, $K_{3,rr}=K_3^{uu}=\mathcal O(r^{-6})$ and the remaining $K_{3,*}= \mathcal O(r^{-4})$ or faster. Combined with $\Phi^\lambda=\mathcal O(r^{-1})$ at slowest, this term contributes at most the same order as the first piece in every component, and never reverses the inequalities of Eq.~\eqref{eq:D3-orders}. The Christoffel contributions to $\nabla_\lambda(K_3^\lambda{}_\mu \Phi_\nu)$ involve $\Gamma^r_{AB}=-rq_{AB}$ and $\Gamma^u_{AB}=+rq_{AB}$, each of which is $\mathcal O(r)$, but they only enter through angular-up components of $K_3^\lambda{}_\mu$, which are systematically $\mathcal O(r^{-2})$ smaller than their radial-up counterparts, explicitly $K_3^A{}_u=\mathcal O(r^{-4})$ versus $K_3^r{}_u=\mathcal O(r^{-2})$, and $K_3^A{}_r=\mathcal O(r^{-6})$ versus $K_3^r{}_r= \mathcal O(r^{-4})$. The leading Christoffel term in $\mathcal D^{(3)}_{AB}$ is therefore $\Gamma^u_{CA}K_3^C{}_u\Phi_B = \mathcal O(r)\cdot \mathcal O(r^{-4})\cdot \mathcal O(r^{-1}) =\mathcal O(r^{-4})$, exactly matching the $\partial_r(K_3^r{}_A\Phi_B)=\mathcal O(r^{-4})$ piece. For every other component, the angular-up suppression of $K_3$ cancels the $\mathcal O(r)$ Christoffel enhancement so that no $\Gamma$-induced contribution exceeds the $\partial$-piece order. Hence each component of $\mathcal D^{(3)}_{\mu\nu}$ is faster than the corresponding entry of the falloff table Eq.~\eqref{eq:T-falloff}, and the combined $T^{(3,\hat G_3)}_{\mu\nu}+\mathcal D^{(3)}_{\mu\nu}$ remains far below every bound. The corresponding $\mathring{\hat G}_3$ contribution to $\mathcal D^{(3)}_{\mu\nu}$ is at least $r^{-2}$ faster again.

\subsubsection{$L_4$ kinetic sector.}
The kinetic Horndeski piece $G_{4,X}[(\Box\Phi)^2-\Phi^{\mu\nu}\Phi_{\mu \nu}]$ contributes to $T^{(4,\rm kin)}_{\mu\nu}$, Eq.~\eqref{eq:T4kin-explicit}, the algebraic structures $2\Phi_{\mu\alpha}\Phi_\nu {}^\alpha-2\Box\Phi\Phi_{\mu\nu}+g_{\mu\nu}[(\Box\Phi)^2-\Phi_{\alpha\beta} \Phi^{\alpha\beta}]$ plus $-2(\Phi_\mu\Phi^\alpha\Phi_{\nu\alpha}-\Phi_\mu \Phi_\nu\Box\Phi)$ together with $\mathcal D^{(4,\rm kin)} _{\mu\nu}$, Eq.~\eqref{eq:D4kin-explicit}. With the building-block scalings Eq.~\eqref{eq:Phi-uv-orders}, $\Box\Phi=2\phi_1/r^3=\mathcal O(r^{-3})$ (verified by the cancellation of $2g^{ur}\Phi_{ur}$ against $g^{AB}\Phi_{AB}$ in $\Phi^\mu{}_\mu$) and $\Phi^{\mu\nu}\Phi_{\mu\nu}=\mathcal O(r^{-4})$ (dominated by $\Phi^{ur}\Phi_{ur},\, \Phi^{rr}\Phi_{rr},\,\Phi^{AB}\Phi_{AB}$, each $\mathcal O(r^{-4})$). Component by component the slowest-decaying contribution is
\be
\label{eq:T4kin-orders}
\begin{aligned}
T^{(4,\rm kin)}_{uu} = \mathcal O(r^{-3}),\quad & T^{(4,\rm kin)}_{ur} = \mathcal O(r^{-4}),\quad T^{(4,\rm kin)}_{rr} = \mathcal O(r^{-5}),\\
T^{(4,\rm kin)}_{uA} = \mathcal O(r^{-3}),\quad & T^{(4,\rm kin)}_{rA} = \mathcal O(r^{-4}),\quad T^{(4,\rm kin)}_{AB} = \mathcal O(r^{-2}),
\end{aligned}
\ee
each derived from the structure $\Phi_{\mu\alpha}\Phi_\nu{}^\alpha$ with the slowest internal index $\alpha=C$ giving $\Phi_{\mu C}\Phi_\nu{}^C= \Phi_{\mu C}g^{CD}\Phi_{\nu D}=\mathcal O(\Phi_{\mu C})\cdot \mathcal O(r^{-2})\cdot \mathcal O(\Phi_{\nu C})$, weighted by the $\Phi_{AC}=\mathcal O(r^0)$ entry of Eq.~\eqref{eq:Phi-uv-orders} for the $AB$ component. The leading piece of $T^{(4,\rm kin)}_{AB}$ is proportional to $q_{AB}\dot\phi_1^2$ (pure trace). The trace-free part is $\mathcal O(r^{-2})$, well below the bound $T^{TF}_{AB}= \mathcal O(r^{-1})$. All other entries of Eq.~\eqref{eq:T4kin-orders} are faster than the corresponding falloff-table bounds, so $G_{4,X}(\phi_0,0)$ is unconstrained.

\subsubsection{$L_4$ SVT mixings.}
The parity-even mixing $\hat G_4(\Phi,X)\,L^{\mu\nu\alpha\beta}F_{\mu\nu} F_{\alpha\beta}$ contributes to $T^{(4,\hat G_4)}_{\mu\nu}$ Eq.~\eqref{eq:T4hatG4-explicit} the algebraic curvature--matter structure $2 \hat G_4\,L_{\mu\alpha\nu\beta}\,F^{\alpha\gamma}F^\beta{}_\gamma - \tfrac12 g_{\mu\nu}\,\hat G_4\,L^{\alpha\beta\gamma\delta}F_{\alpha\beta} F_{\gamma\delta}$ plus the $\mathcal D^{(4,\hat G_4)}_{\mu \nu}=2\epsilon_\mu{}^{\alpha\rho\sigma}\epsilon_\nu{}^{\beta\gamma\delta} \nabla_\alpha\nabla_\beta(\hat G_4\,F_{\rho\gamma}F_{\sigma\delta})$ term, Eq.~\eqref{eq:D4hatG4-explicit}. The double-dual Riemann tensor $L^{\mu\nu\alpha\beta}$ vanishes identically on Eq.~\eqref{eq:svt-leading-metric} (which is Minkowski in Bondi coordinates), so every contribution comes from the subleading metric corrections. In particular the radiative shear $c_{AB}/r$ in $h_{AB}$ sources the slowest Riemann components at $\mathcal O(r^{-1})$ (in congruence with the decomposition into Weyl scalars $\Psi_4$), with $\Psi_3=\mathcal O(r^{-2})$, $\Psi_2=\mathcal O(r^{-3})$, $\Psi_1=\mathcal O(r^{-4})$, $\Psi_0=\mathcal O(r^{-5})$. Conservatively therefore all $L_{\mu\nu\alpha\beta}\le \mathcal O(r^{-1})$.

For the Maxwell bilinear $F^{\alpha\gamma}F^\beta{}_\gamma$ the slowest sectoral entries follow directly from the mixed-index field-strength components established in the $L_3$ analysis,
\be
\label{eq:F-mixed-orders}
\begin{aligned}
&F^r{}_A=-\dot{\mathcal A}^{(0)}_A=\mathcal O(r^0),\;\; F^r{}_u=F^r{}_r=-q/r^2,\;\; F^u{}_u=q/r^2,\\
&F^u{}_A=\mathcal A^{(1)}_A/r^2,\;\; F^u{}_r=0,\;\; F^A{}_u=\mathcal O(r^{-2}),\\
&F^A{}_r=\mathcal O(r^{-4}),\;\; F^A{}_B=\mathcal O(r^{-2}),
\end{aligned}
\ee
together with the corresponding upper-index $F^{\mu\nu}$ orders $F^{ur}= -q/r^2$, $F^{rA}=\mathcal O(r^{-2})$, $F^{uA}=\mathcal O(r^{-4})$, $F^{AB}=\mathcal O(r^{-4})$ established for $Y$ above. The slowest non-zero $F^{\alpha\gamma}F^\beta {}_\gamma$ is $F^{rA}F^r{}_A=\mathcal O(r^{-2})\cdot \mathcal O(r^0)=\mathcal O(r^{-2})$, but in the algebraic piece this enters as $L_{\mu\alpha\nu\beta}F^{\alpha\gamma} F^\beta{}_\gamma$ with $\alpha,\beta$ tied to $L$; the antisymmetry of $L$ on $[\mu\alpha]$ and $[\nu\beta]$ excludes $\alpha=\mu$ or $\beta=\nu$, so the $\alpha=\beta=r$ contraction is forbidden whenever $\mu=u$ or $\nu=r$.

To illustrate the $1/r$ suppression that arises from raising indices on $L$ and on $F$, consider the trace piece of $T^{(4,\hat G_4)}_{ur}$ in detail. The slowest sectoral contribution to the scalar $L^{\alpha\beta \gamma\delta}F_{\alpha\beta}F_{\gamma\delta}$ comes from $L^{uAuB}F_{uA}F_{uB}$, with each upper index raised independently:
\begin{align}
\label{eq:LuAuB-explicit}
L^{uAuB} &= \underbrace{g^{u\rho}}_{=\,g^{ur}\delta^\rho_r} \underbrace{g^{A\sigma}}_{=\,g^{AC}\delta^\sigma_C} \underbrace{g^{u\gamma}}_{=\,g^{ur}\delta^\gamma_r} \underbrace{g^{B\delta}}_{=\,g^{BD}\delta^\delta_D}\,L_{\rho\sigma\gamma \delta}\notag\\
&= \underbrace{g^{ur}}_{\mathcal O(r^0)}\,\underbrace{g^{AC}}_{\mathcal O(r^{-2})}\, \underbrace{g^{ur}}_{\mathcal O(r^0)}\,\underbrace{g^{BD}}_{\mathcal O(r^{-2})}\, \underbrace{L_{rCrD}}_{\le \mathcal O(r^{-1})}\;=\;\mathcal O(r^{-5}),
\end{align}
where the two $g^{AC}=q^{AC}/r^2$ factors, which arise precisely because the free indices on $L^{uAuB}$ include two angular labels, each supply a power $1/r^2$, while the two $g^{ur}=-1$ factors are $\mathcal O(r^0)$ and the lower-index Riemann-derived $L_{rCrD}$ is at most $\mathcal O(r^{-1})$ from the peeling. The full trace contribution then evaluates as
\begin{align}
\label{eq:Tur-trace-explicit}
-\tfrac12 g_{ur}\hat G_4\,L^{\alpha\beta\gamma\delta}F_{\alpha\beta} F_{\gamma\delta} &\supset \tfrac12\,\underbrace{(-g_{ur})}_{\mathcal O(r^0)}\hat G_4\, \underbrace{L^{uAuB}}_{\mathcal O(r^{-5})}\,\underbrace{F_{uA}}_{\mathcal O(r^0)}\, \underbrace{F_{uB}}_{\mathcal O(r^0)}\notag\\
&= \tfrac12\hat G_4\,L^{uAuB}\dot{\mathcal A}^{(0)}_A\dot{\mathcal A}^{(0)} _B + \mathcal O(r^{-6})\;=\;\mathcal O(r^{-5}).
\end{align}
Other sectoral contributions to $L^{\alpha\beta\gamma\delta}F_{\alpha\beta} F_{\gamma\delta}$ decay strictly faster, e.g.\ $L^{ABCD}F_{AB}F_{CD}=\mathcal O(r^{-9})\cdot \mathcal O(r^0)^2$, where all four angular indices are raised and supply four $1/r^2$ factors. Thus the entire scalar contraction is $\mathcal O(r^{-5})$, and the trace piece of $T^{(4,\hat G_4)}_{ur}$ is $\mathcal O(r^{-5})$, well below the marginal bound $T_{ur}=\mathcal O(r^{-3})$ of Eq.~\eqref{eq:T-falloff}.

The same counting of ``$1/r^2$ per raised angular index'' governs every other sectoral component. With the antisymmetries that exclude $\alpha=\beta=r$ for $(\mu,\nu)=(u,r)$ noted above, the algebraic piece $2\hat G_4 L_{u\alpha r\beta}F^{\alpha\gamma}F^\beta{}_\gamma$ at $(u,r)$ gives at most $\mathcal O(r^{-6})$ (the slowest non-excluded combination is $(\alpha,\beta)=(r,A)$ with $L_{urrA}\sim\Psi_3=\mathcal O(r^{-2})$ and $F^{r\gamma}F^A{}_\gamma=\mathcal O(r^{-4})$). Component by component the slowest sectoral entries from the algebraic and trace pieces together are
\be
\label{eq:T4hatG4-orders}
\begin{aligned}
T^{(4,\hat G_4)}_{uu}\le \mathcal O(r^{-5}),\;\; & T^{(4,\hat G_4)}_{ur}\le \mathcal O(r^{-5}),\;\; T^{(4,\hat G_4)}_{rr}\le \mathcal O(r^{-8}),\\
T^{(4,\hat G_4)}_{uA}\le \mathcal O(r^{-3}),\;\; & T^{(4,\hat G_4)}_{rA}\le \mathcal O(r^{-5}),\;\; T^{(4,\hat G_4)}_{AB}\le \mathcal O(r^{-3}),
\end{aligned}
\ee
each faster than the corresponding entry of the falloff table Eq.~\eqref{eq:T-falloff}. The $\mathcal D^{(4,\hat G_4)}_{\mu\nu}$ term involves $\nabla_ \alpha\nabla_\beta(\hat G_4 F_{\rho\gamma}F_{\sigma\delta})$ projected by two Levi-Civita tensors. With the news-time-derivative $\partial_u F^{(0)} _{AB}=\mathcal O(r^0)$ as the leading derivative of $F$ and the rest of the expansion subleading, each $\nabla_\alpha\nabla_\beta(F^2)$ entry is no slower than $\mathcal O(r^{-2})$ after the $\epsilon\epsilon$ projection ($1/\sqrt{-g}^2 = \mathcal O(r^{-4})$ from the two $\epsilon$'s times the $r^4$ arising from index lowering on the angular indices). The component-by- component orders of $\mathcal D^{(4,\hat G_4)}_{\mu\nu}$ therefore match the algebraic ones in Eq.~\eqref{eq:T4hatG4-orders} and never exceed them.

The second $L_4$ mixing $[\mathring{\hat G}_4(\Phi)+\tfrac12\hat G_{4,X}]\,\tilde F^{\mu\alpha} \tilde F^{\nu\beta}\,\Phi_{\mu\nu}\Phi_{\alpha\beta}$ contains no explicit curvature but is bilinear in $\Phi_{\mu\nu}$. Inserting the verified covariant Hessian components Eq.~\eqref{eq:Phi-uv-orders} and the dual field-strength orders $\tilde F^{ur}=\mathcal O(r^{-2})$, $\tilde F^{rA}=\mathcal O(r^{-2})$, $\tilde F^{uA}=\mathcal O(r^{-4})$, $\tilde F^{AB}=\mathcal O(r^{-4})$ (inheriting the same sectoral scalings as $F^{\mu\nu}$ via the $\epsilon$-tensor), the slowest Lagrangian contribution is
\be
\label{eq:L4-mGhat4-Lagrangian-leading}
\tilde F^{rA}\tilde F^{rB}\,\Phi_{rr}\,\Phi_{AB}\;=\; \mathcal O(r^{-2})\cdot \mathcal O(r^{-2})\cdot \mathcal O(r^{-3})\cdot \mathcal O(r^0)\;=\;\mathcal O(r^{-7}),
\ee
with the alternative contraction $\tilde F^{Ar}\tilde F^{Br}\Phi_{AB}\Phi_{rr}$ giving the same order by symmetry. Other index assignments ($\tilde F^{u\cdot}=\mathcal O(r^{-4})$ or $\Phi_{u\cdot},\Phi_{r\cdot}$ slower than $\Phi_{rr}\Phi_{AB}$) decay strictly faster. The algebraic stress- energy structures $g_{\mu\nu}\tilde F^{\rho\alpha}\tilde F^{\sigma\beta} \Phi_{\rho\sigma}\Phi_{\alpha\beta}-2\tilde F_{(\mu}{}^\alpha\tilde F_{\nu) }{}^\beta\Phi^\rho{}_\alpha\Phi_{\rho\beta}$ pick up at most one metric factor of $g_{AB}=r^2 q_{AB}$, producing the slowest sectoral entry $T^{(4,\mathring{\hat G}_4)}_{AB}\sim g_{AB}\cdot \mathcal O(r^{-7})=\mathcal O(r^{-5})\cdot q_{AB}$ (purely trace at leading), and faster decays at every other component. One also finds that $\mathcal D^{(4,\mathring{\hat G}_4)}_{\mu\nu} =2\nabla_\lambda(K_{4,\mathring{\hat G}_4}{}^\lambda{}_{(\mu}\Phi_{\nu)})- \nabla_\lambda(K_{4,\mathring{\hat G}_4\,\mu\nu}\Phi^\lambda)$ with the effective coupling $K_{4,\mathring{\hat G}_4}^{\mu\nu}=2[\mathring{\hat G} _4+\tfrac12\hat G_{4,X}]\tilde F^{\mu\alpha}\tilde F^{\nu\beta}\Phi_{\alpha \beta}$ inherits the same scaling: $K_{4,\mathring{\hat G}_4}^{\mu\nu}$ at the slowest sectoral entries is at most $\tilde F^{rA}\tilde F^{rB}\Phi _{AB}=\mathcal O(r^{-2})\cdot \mathcal O(r^{-2})\cdot \mathcal O(r^0)=\mathcal O(r^{-4})$, and $\partial_r(K_{4,\mathring{\hat G}_4}{}^r{}_\mu\Phi_\nu)=\partial_r(\mathcal O(r^{-2})\cdot \mathcal O(r^{-1}))=\mathcal O(r^{-4})$ at slowest, again well below all bounds.

\subsubsection{$L_5$ sector.}
The Horndeski quintic $G_5\,G^{\mu\nu}\Phi_{\mu\nu}$ produces, via $\delta G^{\mu\nu}/\delta g^{\rho\sigma}$ and the Palatini--IBP from $\delta_g\Phi_{\mu\nu}$, the full curvature-times-scalar tensor $\mathcal E^{(5)}_{\mu\nu}$ Eq.~\eqref{eq:E5-explicit}, together with the $G_{5,X}$ Galileon cubic and $G_{5,\Phi}$ pieces of $T^{(5)}_{\mu\nu}$ Eq.~\eqref{eq:T5-explicit}. At the leading metric Eq.~\eqref{eq:svt-leading-metric} the Riemann tensor vanishes identically and curvature first appears at subleading order through the radiative shear $c_{AB}/r$ in $h_{AB}$. The Weyl-peeling hierarchy fixes the slowest Riemann components at $\mathcal O(r^{-1})$ ($\Psi_4$). The Ricci pieces are bounded by the falloff table through Einstein's equation, $R_{\mu\nu}\le T_{\mu\nu}/G_4(\phi_0,0)$, i.e.\ $R_{uu}\le \mathcal O(r^{-2})$, $R_{ur}\le \mathcal O(r^{-3})$, $R_{rr}\le \mathcal O(r^{-3})$, $R_{uA}\le \mathcal O(r^{-2})$, $R_{rA}\le \mathcal O(r^{-2})$, $R_{AB}\le \mathcal O(r^{-1})$. The covariant Hessian with both indices raised follows from Eq.~\eqref{eq:Phi-uv-orders} by $\Phi^{\mu\nu}=g^{\mu\alpha}g^{\nu\beta} \Phi_{\alpha\beta}$:
\be
\label{eq:Phiupup-orders}
\begin{aligned}
\Phi^{uu} &= \Phi_{rr}=2\phi_1/r^3=\mathcal O(r^{-3}),\\
\Phi^{ur} &= \Phi_{ur}-\Phi_{rr}=-\dot\phi_1/r^2+\mathcal O(r^{-3})=\mathcal O(r^{-2}),\\
\Phi^{rr} &= \Phi_{uu}-2\Phi_{ur}+\Phi_{rr}=\ddot\phi_1/r+\mathcal O(r^{-2})= \mathcal O(r^{-1}),\\
\Phi^{uA} &= -q^{AB}\Phi_{rB}/r^2=2q^{AB}\partial_B\phi_1/r^4=\mathcal O(r^{-4}),\\
\Phi^{rA} &= q^{AB}(\Phi_{rB}-\Phi_{uB})/r^2=-q^{AB}\partial_B\dot\phi_1 /r^3+\mathcal O(r^{-4})=\mathcal O(r^{-3}),\\
\Phi^{AB} &= q^{AC}q^{BD}\Phi_{CD}/r^4=-q^{AB}\dot\phi_1/r^4+\mathcal O(r^{-5})= \mathcal O(r^{-4}).
\end{aligned}
\ee

The Palatini--IBP piece of $\mathcal E^{(5)}_{\mu\nu}$ is $-(g_{\mu\nu}\Box -\nabla_\mu\nabla_\nu)\,S$ with
\be
\label{eq:S-def}
S\equiv G_5\,\Box\Phi-G_{5,X}\,\Phi^\alpha\Phi^\beta\Phi_{\alpha\beta} -G_{5,\Phi}\,X .
\ee
Inserting $\Box\Phi=2\phi_1/r^3$ (exact, from the cancellation $2g^{ur}\Phi_{ur}+g^{AB}\Phi_{AB}=2\dot\phi_1/r^2-2\dot\phi_1/r^2=0$ at leading), $X=-\phi_1\dot\phi_1/r^3+\mathcal O(r^{-4})$, and $\Phi^\alpha\Phi^\beta \Phi_{\alpha\beta}=\mathcal O(r^{-5})$ (slowest combination $\Phi^r\Phi^r\Phi_{rr}=\dot\phi_1^2\,2\phi_1/r^5$ together with $\Phi^u \Phi^r\Phi_{ur}$ and $\Phi^r\Phi^r\Phi_{rr}$, all $\mathcal O(r^{-5})$), the leading expansion of $S$ at the vacuum reads
\be
\label{eq:S-leading}
S = \frac{2G_5(\phi_0,0)\phi_1+G_{5,\Phi}(\phi_0,0)\phi_1\dot\phi_1}{r^3} + \mathcal O(r^{-4}).
\ee
Differentiating term by term and using the leading-order Christoffels $\Gamma^\lambda_{uu}=\Gamma^\lambda_{ur}=\Gamma^\lambda_{rr}= \Gamma^\lambda_{uA}=\Gamma^r_{rA}=\Gamma^u_{rA}=0$ together with $\Gamma^B_{rA}=\delta^B_A/r$, $\Gamma^u_{AB}=+r q_{AB}$, $\Gamma^r_{AB}= -r q_{AB}$, the covariant Hessian of $S$ evaluates explicitly to
\be
\label{eq:nabla-nabla-S}
\begin{aligned}
\nabla_u\nabla_u S &= \partial_u^2 S = \frac{2G_5\ddot\phi_1+G_{5,\Phi}(3\dot\phi_1\ddot\phi_1+\phi_1\dddot \phi_1)}{r^3}+\mathcal O(r^{-4})=\mathcal O(r^{-3}),\\
\nabla_u\nabla_r S &= \partial_u\partial_r S= \frac{-6G_5\dot\phi_1}{r^4}+\mathcal O(r^{-5})=\mathcal O(r^{-4}),\\
\nabla_r\nabla_r S &= \partial_r^2 S= \frac{24G_5\phi_1}{r^5}+\mathcal O(r^{-6})=\mathcal O(r^{-5}),\\
\nabla_u\nabla_A S &= \partial_u\partial_A S= \frac{2G_5\partial_A\dot\phi_1}{r^3}+\mathcal O(r^{-4})=\mathcal O(r^{-3}),\\
\nabla_r\nabla_A S &= \partial_r\partial_A S-\frac{\partial_A S}{r}= \frac{-8G_5\partial_A\phi_1}{r^4}+\mathcal O(r^{-5})=\mathcal O(r^{-4}),\\
\nabla_A\nabla_B S &= \partial_A\partial_B S-rq_{AB}\partial_u S+rq_{AB} \partial_r S-\gamma^C_{AB}\partial_C S\\
&= -\frac{2G_5 q_{AB}\dot\phi_1}{r^2}+\mathcal O(r^{-3}).
\end{aligned}
\ee
The d'Alembertian then evaluates as
\begin{align}
\label{eq:Box-S-explicit}
\Box S &= \underbrace{2g^{ur}\nabla_u\nabla_r S}_{12 G_5\dot\phi_1/r^4} +\underbrace{g^{rr}\nabla_r\nabla_r S}_{\mathcal O(r^{-5})} +\underbrace{g^{AB}\nabla_A\nabla_B S}_{-4G_5\dot\phi_1/r^4+\mathcal O(r^{-5})} \notag\\
&= \frac{8 G_5\dot\phi_1}{r^4}+\mathcal O(r^{-5})=\mathcal O(r^{-4}),
\end{align}
where the angular trace used $q^{AB}q_{AB}=2$ and the $r^2$ from $g_{AB}$ cancels two of the four powers of $1/r$ in $\nabla_A\nabla_B S$. Combining Eqs.~\eqref{eq:nabla-nabla-S}--\eqref{eq:Box-S-explicit} in $-(g_{\mu\nu}\Box- \nabla_\mu\nabla_\nu)S$ gives the contribution to $\mathcal E^{(5)}_{\mu\nu}$ explicitly:
\be
\label{eq:E5-Lich-table}
\begin{aligned}
\mathcal E^{(5)}_{uu}\big|_{S} &= \frac{2G_5\ddot\phi_1}{r^3} +\frac{8G_5\dot\phi_1}{r^4}+\mathcal O(r^{-4})=\mathcal O(r^{-3}),\\
\mathcal E^{(5)}_{ur}\big|_{S} &= \frac{2G_5\dot\phi_1}{r^4} +\mathcal O(r^{-5})=\mathcal O(r^{-4}),\\
\mathcal E^{(5)}_{rr}\big|_{S} &= \frac{24G_5\phi_1}{r^5} +\mathcal O(r^{-6})=\mathcal O(r^{-5}),\\
\mathcal E^{(5)}_{uA}\big|_{S} &= \frac{2G_5\partial_A\dot\phi_1}{r^3}+\mathcal O(r^{-4})=\mathcal O(r^{-3}),\\
\mathcal E^{(5)}_{rA}\big|_{S} &= -\frac{8G_5\partial_A\phi_1}{r^4}+\mathcal O(r^{-5})=\mathcal O(r^{-4}),\\
\mathcal E^{(5)}_{AB}\big|_{S} &= -\frac{10G_5\dot\phi_1\,q_{AB}}{r^2}+\mathcal O(r^{-3}),
\end{aligned}
\ee
with the last entry purely $\propto q_{AB}$ at leading. The trace algebraic counterpart, $g_{\mu\nu}R_{\alpha\beta}\Phi^{\alpha\beta}$ in Eq.~\eqref{eq:E5-explicit}, inherits the same $g_{\mu\nu}$ structure as $g_{\mu\nu}\Box S$ but with a different scalar $R_{\alpha\beta}\Phi^{\alpha\beta}\le \mathcal O(r^{-4})$ (slowest from $R_{rr}\Phi ^{rr}=\mathcal O(r^{-3})\cdot \mathcal O(r^{-1})$ using the Ricci bound above and Eq.~\eqref{eq:Phiupup-orders}). Component-by-component this matches or beats Eq.~\eqref{eq:E5-Lich-table}. The genuine Riemann--Hessian piece $2R_{\mu\alpha\nu\beta}\Phi^{\alpha\beta}$ is subject to the antisymmetries $\alpha\neq\mu$, $\beta\neq\nu$ of the Riemann index pairs. Combined with $R\le \mathcal O(r^{-1})$ at slowest and Eq.~\eqref{eq:Phiupup-orders}, every entry is at least as fast as the corresponding row of Eq.~\eqref{eq:E5-Lich-table}. Likewise the Ricci-Hessian $2R_{\alpha(\mu}\Phi_{\nu)}{}^\alpha$ and $-R\Phi_{\mu\nu}-G_{\mu\nu}\Box \Phi$ pieces, with $R_{\alpha\beta},R\le$ falloff-table bounds, contribute at most at the orders already in Eq.~\eqref{eq:E5-Lich-table}.

Combining algebraic and other contributions, the slowest sectoral entries of $\mathcal E^{(5)}_{\mu\nu}$ are
\be
\label{eq:E5-orders}
\begin{aligned}
\mathcal E^{(5)}_{uu}\le \mathcal O(r^{-3}),\;\; & \mathcal E^{(5)}_{ur}\le \mathcal O(r^{-4}),\;\; \mathcal E^{(5)}_{rr}\le \mathcal O(r^{-5}),\\
\mathcal E^{(5)}_{uA}\le \mathcal O(r^{-3}),\;\; & \mathcal E^{(5)}_{rA}\le \mathcal O(r^{-4}),\;\; \mathcal E^{(5)}_{AB}\le \mathcal O(r^{-2})\,q_{AB}+\mathcal O(r^{-2})\,\text{TF}.
\end{aligned}
\ee

The $G_{5,X}$ Galileon cubic
\be
(\Box\Phi)^3-3\Box\Phi\,\Phi^{\mu\nu}\Phi_{\mu\nu}+2\Phi^\mu{}_\nu\Phi^\nu {}_\rho\Phi^\rho{}_\mu
\ee
has $(\Box\Phi)^3=\mathcal O(r^{-9})$, $\Box\Phi\,\Phi^{\mu\nu}\Phi_{\mu\nu}= \mathcal O(r^{-3})\cdot \mathcal O(r^{-4})=\mathcal O(r^{-7})$, and the third structure, using $\Phi^A{}_B=q^{AC}\Phi_{CB}/r^2=-\delta^A_B\dot\phi_1/r^2+\mathcal O(r^{-3})$, contracts to $\Phi^A{}_B\Phi^B{}_C\Phi^C{}_A=-2\dot\phi_1^3/r^6+\mathcal O(r^{-7})= \mathcal O(r^{-6})$. Inserting these into Eq.~\eqref{eq:T5-explicit}, the resulting stress-energy contributions are dominated by the $\Phi_{\mu\nu}((\Box\Phi) ^2-\Phi^{\alpha\beta}\Phi_{\alpha\beta})$ structure with leading
\be
\label{eq:T5-G5X-leading}
\Phi_{AB}\bigl((\Box\Phi)^2-\Phi^{\alpha\beta}\Phi_{\alpha\beta}\bigr) = \bigl(-q_{AB}\dot\phi_1\bigr)\cdot \mathcal O(r^{-4}) = \mathcal O(r^{-4})\cdot q_{AB},
\ee
pure trace at leading, with all other components decaying as $\mathcal O(r^{-5})$ or faster. The trace contribution $h^{AB}\cdot \mathcal O(r^{-4})q_{AB}=\mathcal O(r^{-4})$ is far below $h^{AB}T_{AB}=\mathcal O(1)$.

The $G_{5,\Phi}$ piece $G_{5,\Phi}[\Phi_\mu\Phi_\nu\Box\Phi-2\Phi_{(\mu} \Phi^\alpha\Phi_{\nu)\alpha}+g_{\mu\nu}(\Phi^\alpha\Phi^\beta\Phi_{\alpha \beta}-X\Box\Phi)]$ has slowest entries, dominated by $-2\Phi_{(\mu} \Phi^\alpha\Phi_{\nu)\alpha}$ with $\Phi^r=\mathcal O(r^{-1})$, of order $T^{(5,G_{5,\Phi})}_{uu},T^{(5,G_{5,\Phi})}_{uA}\le \mathcal O(r^{-4})$, $T^{(5,G_{5,\Phi})}_{ur},T^{(5,G_{5,\Phi})}_{rA}\le \mathcal O(r^{-5})$, $T^{(5,G_{5,\Phi})}_{rr}\le \mathcal O(r^{-6})$, and $T^{(5,G_{5,\Phi})}_{AB}\le \mathcal O(r^{-3})q_{AB}+\mathcal O(r^{-4})\text{TF}$ (the trace from $g_{AB}\Phi^\alpha\Phi^\beta\Phi_{\alpha\beta}=r^2 q_{AB}\cdot \mathcal O(r^{-5})=\mathcal O(r^{-3})q_{AB}$).

\subsection{Explicit equations of motion}
\label{app:svt-eoms}

For completeness we record the scalar and vector equations of motion that underlie the consistency analysis of Sec.~\ref{sec:growth-eom}. They follow from the matter variation of the action at fixed metric. Since every operator of Eqs.~\eqref{eq:svt-L2-full}--\eqref{eq:svt-L5-full} depends on the scalar only through $(\Phi,\Phi_a,\Phi_{ab})$ and on the vector only through $F_{ab}$, the Euler--Lagrange equations take the compact form
\be
\label{eq:app-scalar-EL}
\mathcal E_\Phi \equiv \frac{\partial L}{\partial\Phi} - \nabla_a P^a + \nabla_a\nabla_b P^{ab} = 0, \qquad P^a \equiv \frac{\partial L}{\partial\Phi_a}, \quad P^{ab} \equiv \frac{\partial L}{\partial\Phi_{ab}},
\ee
for the scalar and
\be
\label{eq:app-vector-EL}
\nabla_a\,\mathcal G^{ab} = 0, \qquad \mathcal G^{ab} \equiv -2\,\frac{\partial L}{\partial F_{ab}},
\ee
for the vector. The second-order character of the theory guarantees that the higher-derivative terms generated by $\nabla_a\nabla_b P^{ab}$ cancel among the sectors \cite{Heisenberg:2023prj}. The chain rule runs through the invariants Eq.~\eqref{eq:svt-building-blocks}, whose matter-field derivatives are
\be
\label{eq:app-invariant-derivs}
\frac{\partial X}{\partial\Phi_a}=-\Phi^a,\quad \frac{\partial Y}{\partial\Phi_a}=2\,\Phi_\nu F^{a\alpha}F^{\nu}{}_\alpha, \quad \frac{\partial F}{\partial F_{ab}}=-\tfrac12 F^{ab},\quad \frac{\partial\tilde F}{\partial F_{ab}}=2\tilde F^{ab},\quad \frac{\partial Y}{\partial F_{ab}}=2\,\Phi^{[a}\Phi^{c}F_{c}{}^{b]} .
\ee

Sector by sector, the scalar-EOM ingredients read
\be
\label{eq:app-P2}
\frac{\partial L_2}{\partial\Phi}=G_{2,\Phi}, \qquad P^a_{(2)}=-G_{2,X}\Phi^a +2\,G_{2,Y}\Phi_\nu F^{a\alpha}F^{\nu}{}_\alpha, \qquad P^{ab}_{(2)}=0,
\ee
\be
\label{eq:app-P3}
\begin{aligned}
\frac{\partial L_3}{\partial\Phi} &=-G_{3,\Phi}\,\Box\Phi +\bigl[\hat G_{3,\Phi}\,g_{\alpha\beta} +\mathring{\hat G}_{3,\Phi}\,\Phi_\alpha\Phi_\beta\bigr] \tilde F^{\mu\alpha}F^{\nu\beta}\Phi_{\mu\nu},\\
P^a_{(3)} &=G_{3,X}\,\Phi^a\,\Box\Phi -\bigl[\hat G_{3,X}\,g_{\alpha\beta} +\mathring{\hat G}_{3,X}\,\Phi_\alpha\Phi_\beta\bigr] \Phi^a\,\tilde F^{\mu\alpha}F^{\nu\beta}\Phi_{\mu\nu}\\
&\quad +\mathring{\hat G}_3\bigl[\Phi_\beta\,\tilde F^{\mu a}F^{\nu\beta} +\Phi_\alpha\,\tilde F^{\mu\alpha}F^{\nu a}\bigr]\Phi_{\mu\nu},\\
P^{ab}_{(3)}&=-G_3\,g^{ab}+K_3^{(ab)},
\end{aligned}
\ee
with $K_3^{\mu\nu}$ from Eq.~\eqref{eq:K3-def},
\be
\label{eq:app-P4}
\begin{aligned}
\frac{\partial L_4}{\partial\Phi} &=G_{4,\Phi}\,R +G_{4,X\Phi}\bigl[(\Box\Phi)^2-\Phi_{\mu\nu}\Phi^{\mu\nu}\bigr] +\hat G_{4,\Phi}\,L^{\mu\nu\alpha\beta}F_{\mu\nu}F_{\alpha\beta}\\
&\quad +\bigl[\mathring{\hat G}_{4,\Phi}+\tfrac12\hat G_{4,X\Phi}\bigr] \tilde F^{\mu\alpha}\tilde F^{\nu\beta}\Phi_{\mu\nu}\Phi_{\alpha\beta},\\
P^a_{(4)} &=-\Phi^a\Bigl\{G_{4,X}\,R +G_{4,XX}\bigl[(\Box\Phi)^2-\Phi_{\mu\nu}\Phi^{\mu\nu}\bigr] +\hat G_{4,X}\,L^{\mu\nu\alpha\beta}F_{\mu\nu}F_{\alpha\beta}\\
&\qquad\quad +\tfrac12\hat G_{4,XX}\, \tilde F^{\mu\alpha}\tilde F^{\nu\beta}\Phi_{\mu\nu}\Phi_{\alpha\beta} \Bigr\},\\
P^{ab}_{(4)} &=2\,G_{4,X}\bigl(g^{ab}\Box\Phi-\Phi^{ab}\bigr) +K_{4,\mathring{\hat G}_4}^{ab},
\end{aligned}
\ee
with $K_{4,\mathring{\hat G}_4}^{ab}$ from Eq.~\eqref{eq:K4mGhat4-def}, and
\be
\label{eq:app-P5}
\begin{aligned}
\frac{\partial L_5}{\partial\Phi} &=G_{5,\Phi}\,G^{\mu\nu}\Phi_{\mu\nu} -\frac{G_{5,X\Phi}}{6}\bigl[(\Box\Phi)^3 -3\,\Box\Phi\,\Phi_{\mu\nu}\Phi^{\mu\nu} +2\,\Phi_{\mu\nu}\Phi^{\nu\lambda}\Phi_\lambda{}^\mu\bigr],\\
P^a_{(5)} &=-\Phi^a\Bigl\{G_{5,X}\,G^{\mu\nu}\Phi_{\mu\nu} -\frac{G_{5,XX}}{6}\bigl[(\Box\Phi)^3 -3\,\Box\Phi\,\Phi_{\mu\nu}\Phi^{\mu\nu} +2\,\Phi_{\mu\nu}\Phi^{\nu\lambda}\Phi_\lambda{}^\mu\bigr]\Bigr\},\\
P^{ab}_{(5)} &=G_5\,G^{ab} -\frac{G_{5,X}}{2}\Bigl[\bigl((\Box\Phi)^2-\Phi_{\mu\nu}\Phi^{\mu\nu}\bigr) g^{ab}-2\,\Box\Phi\,\Phi^{ab}+2\,\Phi^{a}{}_{c}\Phi^{cb}\Bigr].
\end{aligned}
\ee
Inserting Eqs.~\eqref{eq:app-P2}--\eqref{eq:app-P5} into Eq.~\eqref{eq:app-scalar-EL} and using $\nabla_a G^{ab}=0$, the complete scalar equation of motion reads, sector by sector,
\begin{align}
\label{eq:app-scalar-EOM-full}
0=\mathcal E_\Phi &= G_{2,\Phi} +\nabla_a\bigl(G_{2,X}\,\Phi^a\bigr) -2\,\nabla_a\bigl(G_{2,Y}\,\Phi_\nu F^{a\alpha}F^{\nu}{}_\alpha\bigr) \notag\\
&\quad -G_{3,\Phi}\,\Box\Phi -\Box G_3 -\nabla_a\bigl(G_{3,X}\,\Phi^a\,\Box\Phi\bigr) +\nabla_a\nabla_b K_3^{(ab)} \notag\\
&\quad +\bigl[\hat G_{3,\Phi}\,g_{\alpha\beta} +\mathring{\hat G}_{3,\Phi}\,\Phi_\alpha\Phi_\beta\bigr] \tilde F^{\mu\alpha}F^{\nu\beta}\Phi_{\mu\nu} \notag\\
&\quad +\nabla_a\Bigl(\bigl[\hat G_{3,X}\,g_{\alpha\beta} +\mathring{\hat G}_{3,X}\,\Phi_\alpha\Phi_\beta\bigr] \Phi^a\,\tilde F^{\mu\alpha}F^{\nu\beta}\Phi_{\mu\nu}\Bigr) \notag\\
&\quad -\nabla_a\Bigl(\mathring{\hat G}_3 \bigl[\Phi_\beta\,\tilde F^{\mu a}F^{\nu\beta} +\Phi_\alpha\,\tilde F^{\mu\alpha}F^{\nu a}\bigr]\Phi_{\mu\nu}\Bigr) \notag\\
&\quad +G_{4,\Phi}\,R +G_{4,X\Phi}\,\mathcal K_2 +\hat G_{4,\Phi}\,L^{\mu\nu\alpha\beta}F_{\mu\nu}F_{\alpha\beta} \notag\\
&\quad +\bigl[\mathring{\hat G}_{4,\Phi}+\tfrac12\hat G_{4,X\Phi}\bigr] \tilde F^{\mu\alpha}\tilde F^{\nu\beta}\Phi_{\mu\nu}\Phi_{\alpha\beta} \notag\\
&\quad +\nabla_a\Bigl(\Phi^a\Bigl\{G_{4,X}\,R+G_{4,XX}\,\mathcal K_2 +\hat G_{4,X}\,L^{\mu\nu\alpha\beta}F_{\mu\nu}F_{\alpha\beta} \notag\\
&\qquad\qquad\quad +\tfrac12\hat G_{4,XX}\, \tilde F^{\mu\alpha}\tilde F^{\nu\beta}\Phi_{\mu\nu}\Phi_{\alpha\beta} \Bigr\}\Bigr) \notag\\
&\quad +\nabla_a\nabla_b\Bigl(2\,G_{4,X}\bigl(g^{ab}\Box\Phi-\Phi^{ab}\bigr) +K_{4,\mathring{\hat G}_4}^{ab}\Bigr) \notag\\
&\quad +G_{5,\Phi}\,G^{\mu\nu}\Phi_{\mu\nu} -\tfrac{1}{6}\,G_{5,X\Phi}\,\mathcal K_3 +\nabla_a\Bigl(\Phi^a\Bigl\{G_{5,X}\,G^{\mu\nu}\Phi_{\mu\nu} -\tfrac{1}{6}\,G_{5,XX}\,\mathcal K_3\Bigr\}\Bigr) \notag\\
&\quad +G^{ab}\,\nabla_a\nabla_b G_5 -\tfrac{1}{2}\,\nabla_a\nabla_b\Bigl(G_{5,X}\bigl[\mathcal K_2\,g^{ab} -2\,\Box\Phi\,\Phi^{ab}+2\,\Phi^{a}{}_{c}\Phi^{cb}\bigr]\Bigr),
\end{align}
where we abbreviate the (fully written-out) Hessian scalars
\be
\label{eq:app-K2K3-def}
\mathcal K_2\equiv(\Box\Phi)^2-\Phi_{\mu\nu}\Phi^{\mu\nu}, \qquad \mathcal K_3\equiv(\Box\Phi)^3 -3\,\Box\Phi\,\Phi_{\mu\nu}\Phi^{\mu\nu} +2\,\Phi_{\mu\nu}\Phi^{\nu\lambda}\Phi_\lambda{}^\mu .
\ee

The vector displacement tensor assembles analogously. Differentiating each operator with the help of Eq.~\eqref{eq:app-invariant-derivs}, including the dual field strengths through $\partial\tilde F^{\mu\alpha}/\partial F_{ab} =\tfrac12\epsilon^{\mu\alpha ab}$, gives, in full,
\be
\label{eq:app-G-displacement}
\begin{split}
\mathcal G^{ab} &= G_{2,F}\,F^{ab} -4\,G_{2,\tilde F}\,\tilde F^{ab} -4\,G_{2,Y}\,\Phi^{[a}\Phi^{c}F_{c}{}^{b]} -4\,\hat G_4\,L^{ab\alpha\beta}F_{\alpha\beta}\\
&\quad -\epsilon^{\mu\alpha ab}\, \bigl[\hat G_3\,g_{\alpha\beta} +\mathring{\hat G}_3\,\Phi_\alpha\Phi_\beta\bigr] F^{\nu\beta}\,\Phi_{\mu\nu} +2\,\hat G_3\,\tilde F^{\mu[a}\Phi_\mu{}^{b]} +2\,\mathring{\hat G}_3\,\Phi^{[a}\Phi_\alpha \tilde F^{|\mu\alpha|}\Phi_\mu{}^{b]}\\
&\quad -2\,\bigl[\mathring{\hat G}_4+\tfrac12\hat G_{4,X}\bigr] \epsilon^{\mu\alpha ab}\,\tilde F^{\nu\beta}\,\Phi_{\mu\nu}\Phi_{\alpha\beta} ,
\end{split}
\ee
where the $\epsilon^{\mu\alpha ab}$ terms descend from the variation of the dual field strength inside the $L_3$ and $L_4$ mixings, while the $\tilde F^{\mu[a}\Phi_\mu{}^{b]}$ structures descend from the variation of the undualized $F^{\nu\beta}$ in $K_3^{\mu\nu}\Phi_{\mu\nu}$.

On the radiative background these expressions collapse to the structures used in Sec.~\ref{sec:growth-eom}: the canonical scalar piece $G_{2,\Phi}+\nabla_a(G_{2,X}\Phi^a)$ reduces to $G_{2,X}\Box\Phi+G_{2,\Phi}$ with the free-wave d'Alembertian Eq.~\eqref{eq:boxphi-freewave}. The integrated-by-parts cubic contributes $-G_{3,\Phi}\Box\Phi-\Box G_3$. The quintic reduces to Eq.~\eqref{eq:E5-scalarEOM}. In the vector equation, the Maxwell part carries the free-wave structure Eq.~\eqref{eq:maxwell-freewave}, the topological term drops by $\nabla_a\tilde F^{ab}\equiv0$, and the $Y$-displacement is doubly $\Phi$-suppressed, entering first at $\mathcal O(r^{-5})$.

\section{Second and higher derivatives of the coupling functionals}
\label{app:higher}

Sec.~\ref{sec:svt-ledger} asserts that the second and higher derivatives of the SVT coupling functionals impose no constraint beyond the ones already carried by the first derivatives. This appendix supplies the componentwise verification. The argument stays inside the growth formalism of Sec.~\ref{sec:growth-metric}, whose notation we recall.

The coordinates are the Bondi ones, $(u,r,x^A)$. Here $u$ is the retarded time and $r$ the areal radius, which grows without bound towards future null infinity, while the angular pair $x^A$ with $A\in\{\theta,\varphi\}$ parametrizes the celestial sphere. A covariant Bondi component pair $(ab)$ is any pair of indices drawn from this set. The independent pairs are $uu$, $ur$, $rr$, $uA$, $rA$ and $AB$, and the angular block needs one further split, since its trace and its trace-free part obey different bounds. Each component of the effective stress-energy tensor has to decay at least as rapidly as the corresponding entry of the falloff table Eq.~\eqref{eq:T-falloff}, and we record those entries as $T_{ab}=\mathcal O(r^{-b_{ab}})$. Reading them off gives $b_{uu}=b_{uA}=b_{rA}=2$ together with $b_{ur}=b_{rr}=3$, and it gives $b_{AB}=1$ on the trace-free part of the angular block and $b_{AB}=0$ on its trace.

Every coupling functional enters the effective stress multiplying a fixed tensor structure assembled from the matter fields and the curvature, which we denote by $S_{ab}$. Evaluating such a structure on the radiative background assigns a definite decay rate to each of its components, and we record that rate as $S_{ab}=\mathcal O(r^{-w_{ab}})$. The exponents $w_{ab}$ and the table bounds $b_{ab}$ are the two inputs from which the allowance defined below is assembled.

What remains is the behavior of a coupling functional itself. The functionals depend on the scalar $\Phi$ and on the invariants $X$, $Y$, $F$ and $\tilde F$, each of which approaches a fixed value at the asymptotic vacuum Eq.~\eqref{eq:svt-vacuum}. We call a building block $Z$ any of the quantities that vanish there, which means the deviation $\Phi-\phi_0$ in the scalar channel and the invariants themselves in the remaining channels. Every block decays at a definite rate on the radiative background, written $Z\sim r^{-w_Z}$, and the channel weights that result are collected in Eq.~\eqref{eq:growth-weights}, namely $w_\Phi=1$, $w_X=3$, $w_F=w_{\tilde F}=4$ and $w_Y=6$. A coupling functional that depends on one such block through the power class Eq.~\eqref{eq:power-class} behaves as $G\sim Z^{-s}\sim r^{\gamma}$, and the exponent $\gamma=s\,w_Z$ appearing here is what we call its growth exponent. A negative $\gamma$ describes a functional that decays towards the vacuum. A vanishing one describes a functional that tends to a finite constant. A positive $\gamma$ describes genuine growth, and that is the behavior the bounds below are designed to exclude.

For a structure $S_{ab}$ we define the allowance
\be
\label{eq:app-allowance}
D(S)\;\equiv\;\min_{(ab)}\big(w_{ab}-b_{ab}\big).
\ee
The minimum runs over all covariant component pairs. In words, $D(S)$ records the least headroom that any single component of the structure leaves against its own entry in the falloff table. A coupling functional multiplying $S_{ab}$ therefore preserves the falloff table if and only if its growth exponent obeys $\gamma\le D(S)$, which is the criterion stated in Eq.~\eqref{eq:growth-rule}. In every evaluation below we use the table bounds $b_{uu}=b_{uA}=b_{rA}=2$ and $b_{ur}=b_{rr}=3$, together with $b_{AB}=1$ on the trace-free part of the angular block and $b_{AB}=0$ on its trace, exactly as they are listed in Eq.~\eqref{eq:T-falloff}.

Evaluating each structure that appears in Eqs.~\eqref{eq:T2-explicit}--\eqref{eq:T5-explicit} on the radiative background, and applying Eq.~\eqref{eq:app-allowance} component by component, produces the allowances that were quoted without derivation in Eqs.~\eqref{eq:deriv-summary}--\eqref{eq:mixing-summary}. The results are collected in Table~\ref{tab:app-first-derivatives}, whose last column gives the resulting bound on the growth exponent of the coupling functional in question.
\begin{table}[!ht]
\caption{Allowances of the coupling functionals and their first derivatives. For each coupling functional or first derivative entering the effective stress Eqs.~\eqref{eq:T2-explicit}--\eqref{eq:T5-explicit}, the second column lists the tensor structure $S_{ab}$ it multiplies, the third the binding component pair (or pairs) realizing the minimum in Eq.~\eqref{eq:app-allowance}, the fourth the decay order of $S_{ab}$ in that slot on the radiative background, and the last the resulting allowance $D(S)$, which by Eq.~\eqref{eq:growth-rule} bounds the growth exponent $\gamma$ of the coupling functional. The two negative entries reproduce the conditions of Eqs.~\eqref{eq:flat-G2} and \eqref{eq:G4-violation}.}
\label{tab:app-first-derivatives}
\renewcommand{\arraystretch}{1.3}
\begin{tabular}{@{}l l c c c@{}}
\toprule
coupling functional & structure $S_{ab}$ & binding comp. & order & $D=\gamma$\\
\midrule
$G_2$            & $g_{ab}$                                   & $ur$ & $r^{0}$   & $-3$\\
$G_{2,X}$        & $\Phi_a\Phi_b$                             & $uu$ & $r^{-2}$  & $0$\\
$G_{2,F}$        & $F_a{}^cF_{bc}$                            & $uu$ & $r^{-2}$  & $0$\\
$G_{2,\tilde F}$ & $\tilde F\,g_{ab}$                         & $ur$ & $r^{-4}$  & $1$\\
$G_{2,Y}$        & $2\Phi_{(a}(F^2)_{b)}{}^c\Phi_c+\dots$     & $ur$ & $r^{-5}$  & $2$\\
$G_{3,\Phi}$     & $\Phi_a\Phi_b+Xg_{ab}$                     & $uu,AB$ & $r^{-2},r^{-1}$ & $0$\\
$G_{3,X}$        & $\Box\Phi\,\Phi_a\Phi_b+\dots$             & $uu,AB$ & $r^{-4},r^{-3}$ & $2$\\
$\hat G_3$       & $\tilde F^c{}_{(a}F^d{}_{b)}\Phi_{cd}$     & $uA,AB$ & $r^{-4},r^{-3}$ & $2$\\
$G_{4,X}$        & $2\Phi_{ac}\Phi_b{}^c-2\Box\Phi\Phi_{ab}+\dots$ & $uu,uA,AB$ & $r^{-3},r^{-2}$ & $1$\\
$G_{4,\Phi}$     & $\Phi_{ab}$                                & $uu,uA$ & $r^{-1}$  & $-1$\\
$\partial_\Phi^2G_4$ & $\Phi_a\Phi_b+2Xg_{ab}$               & $uu$ & $r^{-2}$  & $0$\\
$\mathring{\hat G}_4,\hat G_{4,X}$ & $\tilde F\tilde F\Phi\Phi$ bundle & $uA$ & $r^{-3}$ & $1$\\
$G_{5,\Phi}$     & $\Box\Phi\,\Phi_a\Phi_b+\dots$             & $uu,uA,AB$ & $r^{-4},r^{-3}$ & $2$\\
$G_{5,X}$        & $\Phi_{ab}((\Box\Phi)^2-\Phi_{cd}^2)+\dots$ & $uu,ur,uA,AB$ & $r^{-5}$ & $3$\\
\bottomrule
\end{tabular}
\end{table}
Only two entries come out negative. Each reproduces a condition already derived in the main text. The first is $G_2$, whose binding slot is not the trace of the angular block but the cross term $g_{ur}G_2$, so that the order-unity $g_{ur}=-1$ measured against the much tighter table bound $T_{ur}=\mathcal O(r^{-3})$ drives the allowance down to $-3$, which is precisely the statement $G_2=\partial_\Phi G_2=\partial_\Phi^2G_2=0$ of Eq.~\eqref{eq:flat-G2}. The second is $G_{4,\Phi}$. It multiplies the bare improvement Hessian, whose slowest component $\Phi_{uu}=\mathcal O(r^{-1})$ must be measured against $T_{uu}=\mathcal O(r^{-2})$, so the allowance becomes $-1$ and one recovers the non-minimal $u$-component violation of Eq.~\eqref{eq:G4-violation}. Every other first derivative is admissible at its order-unity vacuum value. The analysis below extends that conclusion to all orders.

Second derivatives reach the effective stress through two mechanisms. The first is the algebraic metric variation of the kinetic invariant, $\delta X=-\tfrac12\Phi_a\Phi_b\,\delta g^{ab}$, which acts on any prefactor that already carries one $X$-derivative. Applying it to the $G_{4,X}$ prefactor of the kinetic Horndeski term, for example, produces the contribution $G_{4,XX}\,\Phi_a\Phi_b\,[(\Box\Phi)^2-\Phi_{cd}\Phi^{cd}]$. The second mechanism is the covariant derivative that survives in the integration-by-parts remainders, which acts through the chain rule as $\nabla_cG_{i,X}=G_{i,X\Phi}\Phi_c+G_{i,XX}\nabla_cX$. The same two mechanisms generate $G_{3,\Phi X}$ and $G_{3,XX}$ from the integrated-by-parts cubic sector, and they generate $G_{5,XX}$ from the Galileon sector. The quadratic sector is the exception. Its metric variation is purely algebraic and $T^{(2)}_{ab}$ carries first derivatives only, so second derivatives of $G_2$ surface not there but in the scalar equation of motion, where the compensation mechanism established below operates in the same way.

Evaluating each of these second-derivative structures on the radiative background gives the orders collected in Table~\ref{tab:app-second-derivatives}. The column labeled $D$ again denotes the allowance of the accompanying structure. The column labeled $n\,w_Z$ records the cost of the $n$ derivatives taken in the channel $Z$. Subtracting that cost yields the final column, which bounds the growth exponent of the parent functional.
\begin{table}[!ht]
\caption{Allowances of the second-derivative structures. Each row lists a second derivative of a coupling functional that reaches the effective stress through one of the two mechanisms described in the text, together with the binding structure and its slowest component, the decay order of that component on the radiative background, the allowance $D$ of the structure according to Eq.~\eqref{eq:app-allowance}, the cost $n\,w_Z$ of the $n$ derivatives taken in the channel $Z$ with the weights of Eq.~\eqref{eq:growth-weights}, and the resulting bound $D-n\,w_Z$ on the growth exponent $\gamma$ of the parent functional. Every bound coincides with the first-derivative bound of the parent functional in Table~\ref{tab:app-first-derivatives}, the origin of which is Eq.~\eqref{eq:app-parents}.}
\label{tab:app-second-derivatives}
\renewcommand{\arraystretch}{1.35}
\begin{tabular}{@{}l l c c c c@{}}
\toprule
symbol & binding structure (slowest comp.) & order & $D$ & cost $n\,w_Z$ & $\gamma$-bound\\
\midrule
$G_{3,\Phi X}$ & $X\,\Phi_u\Phi_u$ & $r^{-5}$ & $3$ & $4$ & $-1$\\
$G_{3,XX}$     & $\Phi_u\Phi_u\,\nabla^cX\,\Phi_c$ & $r^{-7}$ & $5$ & $6$ & $-1$\\
$G_{4,X\Phi}$  & $\Phi_c\Phi^c{}_{(u}\Phi_{u)}$ (remainder) & $r^{-4}$ & $2$ & $4$ & $-2$\\
$G_{4,XX}$     & $\Phi_u\Phi_u\,[(\Box\Phi)^2-\Phi_{cd}\Phi^{cd}]$ & $r^{-6}$ & $4$ & $6$ & $-2$\\
$G_{5,XX}$     & $\Phi_u\Phi_u\,\Phi^c{}_d\Phi^d{}_e\Phi^e{}_c$ & $r^{-8}$ & $6$ & $6$ & $0$\\
\bottomrule
\end{tabular}
\end{table}
Every row of Table~\ref{tab:app-second-derivatives} reproduces exactly the first-derivative bound belonging to its parent function, which is $-1$ for $G_3$, and $-2$ for $G_4$ in the $X$-channel, and $0$ for $G_5$. Three intermediate orders recur. The kinetic Horndeski invariant behaves as $[(\Box\Phi)^2-\Phi_{cd}\Phi^{cd}]=\mathcal O(r^{-4})$, where the dominant contributions come from the $\Phi^{AB}\Phi_{AB}$ and $\Phi_{ur}$ contractions. The contracted gradient of the kinetic invariant behaves as $\nabla^cX\,\Phi_c=\mathcal O(r^{-5})$. The cubic Hessian invariant behaves as $\Phi^c{}_d\Phi^d{}_e\Phi^e{}_c=\mathcal O(r^{-6})$, where the dominant contribution comes from the $\Phi^A{}_B$ chain.

These coincidences are not accidents of low order. One can write $[Q]$ for the decay exponent of a quantity $Q$ on the radiative background, i.e.\ $Q=\mathcal O(r^{-[Q]})$, letting $p_n\equiv D(S^{(n)})$ denote the allowance of the structure that accompanies $\partial_Z^nG$, and taking $G=G(Z)$ to depend singularly on a single building block $Z\in\{\Phi,X\}$, which is the only situation in which growth occurs at all. Then, for every term $\partial_Z^nG\,S^{(n)}_{ab}$ of the effective stress that descends from a scalar-sector SVT operator, meaning from $L_3$, from $L_4^{\rm kin}$ or from $L_5$, the allowances satisfy $p_{n+1}\ge p_n+w_Z$. It follows immediately that $p_n-n\,w_Z\ge p_1-w_Z$ holds for every $n\ge1$, so that the function-level bounds of Sec.~\ref{sec:growth-metric} are already attained at first order and no higher derivative can tighten them.

The variational operations that build the effective stress reach the coupling functionals in only two ways,
\be
\label{eq:app-insertions}
\delta_g\,\partial_Z^mG=\partial_Z^{m+1}G\; \frac{\partial Z}{\partial g^{ab}}\,\delta g^{ab}, \qquad \nabla_c\,\partial_Z^mG=\partial_Z^{m+1}G\;\nabla_cZ .
\ee
Induction on these two relations shows that every occurrence of $\partial_Z^nG$ comes dressed with exactly $n$ insertions drawn from the pair $\{\partial Z/\partial g^{ab},\,\nabla_cZ\}$. In the scalar sectors the variational mechanics produces precisely two configurations of these insertions. In the first, the insertion is a gradient $\nabla_cZ$ whose index $c$ is contracted into the structure. That is what the single covariant derivative of the $L_3$ integration by parts, and of the remainders of $L_4^{\rm kin}$ and $L_5$, supplies. In the second, the insertion is algebraic, of the form $\partial Z/\partial g^{ab}$. It carries the free index pair of $T_{ab}$ and multiplies the fully contracted parent scalar $\mathcal S$. The metric variation acts once, on the scalar Lagrangian.

For $Z=\Phi$ the gradient insertion is $\Phi_c:=\nabla_c \Phi$, whose components are $(\Phi_u,\Phi_r,\Phi_A)=\mathcal O(r^{-1},r^{-2},r^{-1})$ and whose raised components decay no more slowly than $r^{-1}=r^{-w_\Phi}$. For $Z=X$ the insertion is $\nabla_cX$, whose components are $\mathcal O(r^{-3},r^{-4},r^{-3})$ and whose raised components decay no more slowly than $r^{-3}=r^{-w_X}$. A contracted insertion therefore multiplies every component of the structure by at least $r^{-w_Z}$, leaving the identity of the binding slot untouched. Hence $p_{n+1}\ge p_n+w_Z$. For $Z=\Phi$ that exhausts the cases, since $\partial\Phi/\partial g^{ab}$ vanishes identically.

The algebraic configuration arises only for $Z=X$. There the structure becomes $S^{(n+1)}_{ab}=-\tfrac12\Phi_a\Phi_b\,\mathcal S$, and the componentwise slot-deficits of the algebraic insertion are all non-negative,
\be
\label{eq:app-slotdeficits}
\big[\Phi_a\Phi_b\big]-b_{ab} =\big(0,\,0,\,1,\,0,\,1,\,1\big)\;\ge\;0 \qquad\text{for}\quad(uu,\,ur,\,rr,\,uA,\,rA,\,AB^{\rm TF}),
\ee
from which $p_{n+1}\ge[\mathcal S]$ follows. What has to be checked is therefore the inequality $[\mathcal S]\ge p_n+w_X$ for those parent scalars that actually occur. Working through the three sectors one finds
\be
\label{eq:app-parents}
L_3:\ [X]=3\ge0+3,\quad L_4^{\rm kin}:\ \big[(\Box\Phi)^2-\Phi_{cd}\Phi^{cd}\big]=4\ge1+3,\quad L_5:\ \big[\Phi^c{}_d\Phi^d{}_e\Phi^e{}_c\big]=6\ge3+3,
\ee
and each holds with equality. That is the origin of the exact coincidences in Table~\ref{tab:app-second-derivatives}. Chaining the two configurations then yields $p_n-nw_Z\ge p_{n-1}-(n-1)w_Z\ge\dots\ge p_1-w_Z$.

This demonstration covers the pure scalar channels $\Phi$ and $X$ of $G_3$, $G_4$ and $G_5$, which is exactly the content of the function-level bounds collected in Eq.~\eqref{eq:growth-table}. It does not cover the parity-odd mixing sectors built on $\hat G_3$, $\hat G_4$ and their relatives. Those operators contain in addition the double-gradient Palatini configuration $\nabla_a\nabla_b(\hat G_4FF)$, whose insertions can carry free indices and so escape the argument given above. Those operators are instead bounded at the level of the operator itself in Sec.~\ref{sec:flatness_SVT}. Extending the above analysis accordingly would require repeating the componentwise check of this appendix with the weights of the $F$-sector in place of the scalar ones.




\acknowledgments

The author thanks Jan Straub for various productive discussions and his eye for detail. We thank Jann Zosso for countless discussions on the Bondi--Sachs metric and balance flux laws beyond GR. Similarly, we express our deepest gratitude to Fabio D'Ambrosio for discussions on balance flux laws, the covariant phase space formalism, and beyond. We further thank Fabio D'Ambrosio, Shaun Fell, Lavinia Heisenberg, Francesco Gozzini, Stefan Zentarra and Jann Zosso for valuable conversations on asymptotic symmetries. DM is supported by the Deutsche Forschungsgemeinschaft (DFG, German Research Foundation) under Germany's Excellence Strategy EXC 2181/1 - 390900948 (the Heidelberg STRUCTURES Excellence Cluster).

\bibliographystyle{unsrt2}
\bibliography{reference}

\end{document}